\documentclass[acmsmall]{acmart}

\usepackage{subcaption}
\usepackage{listings}
\usepackage{xspace}
\usepackage{tikz}
\usetikzlibrary{positioning, calc, fit, backgrounds, automata, tikzmark,arrows,shapes}
\usepackage{pgfmath}

\usepackage{amsmath}
\usepackage{amsfonts}

\usepackage{amssymb}
\usepackage[capitalize]{cleveref}
\usepackage{multirow}
\usepackage{stmaryrd}
\usepackage{wrapfig}
\usepackage{algorithm}
\usepackage{algpseudocode}
\usepackage{algorithmicx}
\usepackage{xcolor}
\algrenewcommand\algorithmicindent{0.5em}%

\newcommand{\sys}{\textsc{Bonsai}\xspace}

\newcommand{\sysIR}{\textsc{TTIR}\xspace}

\definecolor{codebackground}{RGB}{252,252,252}

\lstdefinelanguage{lang}{
  keywords={def, func, match, with, struct, elem, type, for, in, if, elif, else, elif, continue, yield, from, scan, update, upd, iter, Set, bool, set, float, f32, int, i32, vec, u32, let, tree, in, with, data, implements, mut},
  keywordstyle=\color{blue},
  identifierstyle=\color{black},
  sensitive=true,
  comment=[l]{//},
  morecomment=[s]{/*}{*/},
  commentstyle=\color{darkgray}\ttfamily,
  stringstyle=\color{blue}\ttfamily,
  morekeywords = [2]{true, false, null},
  keywordstyle = [2]\color{green},
  morekeywords = [2]{filter, reduce, product, map, argmin, argmax, min, max, any, all, count, sum, prod, avg},
  keywordstyle = [2]\color{purple},
  morekeywords = [3]{intersects, contains, overlaps, equals, covers, disjoint, touches, within, distmin, distmax, minb, maxb, argminb, argmaxb, abs, ite, ceil, floor, exp, ln, pow, round, sign, sqrt, trunc},
  keywordstyle = [3]\color{red},
  morekeywords = [4]{always, never, maybe},
  keywordstyle = [4]\color{red}\bfseries,
  morekeywords = [4]{Point, Rectangle, Triangle, Object, Ray},
  keywordstyle = [4]\color{cyan},
  morekeywords = [5]{ITree, Leaf, Interior, LargeLeaf, ExampleTree, TriBVH, MBVH, MinTree, MaxTree},
  keywordstyle = [5]\color{olive},
  morekeywords = [6]{rewrite, rule, cif, celse, has, celif},
  keywordstyle = [6]\color{cyan}\bfseries,
}

\lstdefinelanguage{ajcpp}{
  keywords={void, if, else},
  keywordstyle=\color{blue}\bfseries,
  identifierstyle=\color{black},
  sensitive=true,
  comment=[l]{//},
  morecomment=[s]{/*}{*/},
  commentstyle=\color{teal}\ttfamily,
  stringstyle=\color{blue}\ttfamily,
  morekeywords = [2]{Tree, u64},
  keywordstyle = [2]\color{purple},
  morekeywords = [3]{intersects, contains, is_leaf},
  keywordstyle = [3]\color{red},
}

\lstdefinelanguage{sql}{
  keywords={SELECT, WHERE, ORDER BY, AND, OR, FROM},
  keywordstyle=\color{blue},
  identifierstyle=\color{black},
  sensitive=true,
  comment=[l]{//},
  morecomment=[s]{/*}{*/},
  morecomment=[s]{'}{'},
  commentstyle=\color{darkgray}\ttfamily,
  stringstyle=\color{red}\ttfamily,
  morekeywords = [2]{min, max, abs, ite, ceil, floor, exp, ln, pow, round, sign, sqrt, trunc, DATEDIFF},
  keywordstyle = [2]\color{purple},
  morekeywords = [3]{intersects, contains, distance, month},
  keywordstyle = [3]\color{red},
}

\tikzstyle{lattice-point} = [rectangle, minimum width=0.75cm, minimum height=0.5cm, draw=gray, align=center, font=\scriptsize\ttfamily, rounded corners]

\definecolor{color0}{RGB}{255,255,217}
\definecolor{color1}{RGB}{237,248,177}
\definecolor{color2}{RGB}{199,233,180}
\definecolor{color3}{RGB}{127,205,187}
\definecolor{color4}{RGB}{65,182,196}
\definecolor{color5}{RGB}{29,145,192}
\definecolor{color6}{RGB}{34,94,168}
\definecolor{color7}{RGB}{37,52,148}
\definecolor{color8}{RGB}{8,29,88}

\newcommand{\code}[1]{\texttt{\lstinline[language=lang]|#1|}}

\newcommand{\Linear}{\textcolor{gray!80}{Linear}}

\newcommand{\Index}{\textbf{Index}}

\newcommand{\algcomment}[1]{\textcolor{darkgray}{#1}}

\AtBeginDocument{%
  }

\setcopyright{cc}
\setcctype{by}
\acmDOI{10.1145/3808256}
\acmYear{2026}
\acmJournal{PACMPL}
\acmVolume{10}
\acmNumber{PLDI}
\acmArticle{178}
\acmMonth{6}
\acmSubmissionID{pldi26main-p25-p}
\received{2025-11-07}
\received[accepted]{2026-04-03}

\begin{document}

\title{Bonsai: Compiling Queries to Pruned Tree Traversals}

\author{Alexander J Root}
\orcid{0000-0001-6221-1389}
\affiliation{%
  \department{Computer Science}
  \institution{Stanford University}
  \streetaddress{353 Jane Stanford Way}
  \city{Stanford}
  \state{CA}
  \country{USA}
  \postcode{94305}
}
\email{ajroot@cs.stanford.edu}

\author{Christophe Gyurgyik}
\orcid{0000-0001-8493-1133}
\affiliation{%
  \department{Computer Science}
  \institution{Stanford University}
  \streetaddress{353 Jane Stanford Way}
  \city{Stanford}
  \state{CA}
  \country{USA}
  \postcode{94305}
}
\email{cpg@cs.stanford.edu}

\author{Purvi Goel}
\orcid{0000-0003-2618-092X}
\affiliation{%
  \department{Computer Science}
  \institution{Stanford University}
  \streetaddress{353 Jane Stanford Way}
  \city{Stanford}
  \state{CA}
  \country{USA}
  \postcode{94305}
}
\email{pgoel2@cs.stanford.edu}


\author{Kayvon Fatahalian}
\orcid{0000-0001-8754-0429}
\affiliation{%
  \department{Computer Science}
  \institution{Stanford University}
  \streetaddress{353 Jane Stanford Way}
  \city{Stanford}
  \state{CA}
  \country{USA}
  \postcode{94305}
}
\email{kayvonf@cs.stanford.edu}

\author{Jonathan Ragan-Kelley}
\orcid{0000-0001-6243-9543}
\affiliation{%
  \institution{Massachusetts Institute of Technology}
  \city{Cambridge}
  \state{MA}
  \country{USA}
  \postcode{02139}
}
\email{jrk@mit.edu}

\author{Andrew Adams}
\orcid{0000-0002-6766-670X}
\affiliation{%
  \institution{Adobe Research}
  \city{San Francisco}
  \state{CA}
  \country{USA}
  \postcode{94103}
}
\email{andrew.b.adams@gmail.com}

\author{Fredrik Kjolstad}
\orcid{0000-0002-2267-903X}
\affiliation{%
  \department{Computer Science}
  \institution{Stanford University}
  \streetaddress{353 Jane Stanford Way}
  \city{Stanford}
  \state{CA}
  \country{USA}
  \postcode{94305}
}
\email{kjolstad@cs.stanford.edu}

\renewcommand{\shortauthors}{Root et al.}

\begin{abstract}

Trees can accelerate queries that search or aggregate values over large collections. They achieve this by storing metadata that enables quick pruning (or inclusion) of subtrees when predicates on that metadata can prove that none (or all) of the data in a subtree affect the query result.
Existing systems implement this pruning logic manually for each query predicate and data structure.
We generalize and mechanize this class of optimization.
Our method derives conditions for when subtrees can be pruned (or included wholesale), expressed in terms of the metadata available at each node.
We efficiently generate these conditions using symbolic interval analysis, extended with new rules to handle geometric predicates (e.g., intersection, containment).
Additionally, our compiler fuses compound queries (e.g., reductions on filters) into a single tree traversal.
These techniques enable the automatic derivation of generalized single-index and dual-index tree joins that support a wide class of join predicates beyond standard equality and range predicates.
The generated traversals match the behavior of expert-written code that implements query-specific traversals, and can asymptotically outperform the linear scans and nested-loop joins that existing systems fall back to when hand-written cases do not apply.

\end{abstract}




\maketitle

\section{Introduction}
\label{sec:intro}

Augmented tree data structures accelerate queries over large collections of data.
They achieve this by storing metadata, such as bounding boxes or aggregate values, at internal nodes that enable traversals to exclude or include entire subsets without examining each individual element. This mechanism, known as pruning or culling, is widely used: in database systems \textit{indexes} are used to accelerate range and point queries~\cite{guttman1984rtree, comer1979btree}, in graphics systems \textit{acceleration structures} are used to skip occluded geometry~\cite{kay1986, meister2021bvh} and to efficiently find collisions~\cite{ericson2004rtcd, pan2012fcl}, and in scientific computing \textit{spatial trees} are used to limit computation to relevant regions~\cite{barnes1986bhut, howard2019quantized}.

Despite their ubiquity, pruning logic is manually implemented. Traditional acceleration trees, such as Bounding Volume Hierarchies (BVHs)~\cite{kay1986, ericson2004rtcd, meister2021bvh}, B-trees~\cite{comer1979btree}, and R-Trees~\cite{guttman1984rtree}, encode pruning rules operationally through hand-written traversal logic. Generalized Search Trees~\cite{hellerstein1998gist} attempt to unify these structures under a common abstraction, but rely on user-specified search and consistency predicates. As a result, every new query requires a new hand-engineered traversal \textit{for every data structure}~\cite{samet2005multi, sawhney2020wos, sawhney2021fcpw, emre2025cones}. Modern query engines~\cite{neumann2011querycompilation, selinger1979systemr, tahboub2018architectquery} must therefore treat these tree queries as opaque, manually optimized operators specialized to particular queries, rather than reusable mechanisms for accelerating a broad class of queries.

This work introduces a new direction: automatically generating tree traversals from separate specifications of the query and the tree data structure.
Hand-written traversals across domains follow a single principle: traversals exploit \textit{necessary} and \textit{sufficient} conditions parameterized by tree metadata that determine whether a predicate is guaranteed to hold or fail for all elements in a subtree.
This principle extends naturally to reductions: for associative reductions (e.g., sum, product), subtree metadata can be used to include entire subsets without visiting individual elements; for reductions of idempotent operators (e.g., min, max), subtree metadata can be used to skip subtrees that cannot impact the running aggregate.
Our key insight is that these conditions can be automatically derived from high-level query predicates and annotations of tree metadata.

To enable the derivation of custom pruning logic, we adopt the perspective of \textit{data independence}~\cite{codd1970relational}: the query should be decoupled from the metadata used to accelerate it. Tree specifications provide this metadata as language-level annotations, allowing the compiler to derive pruning logic from the annotations and query operators.  Mechanizing both the derivation of pruning conditions and the use of metadata for reductions enables fully automated generation of fused traversal code, eliminating the need for hand-written traversals.


Our approach is built on two core steps:
First, a lowering algorithm fuses the operators of a high-level query into a tree traversal. The resulting traversal is expressed in terms of abstract necessary and sufficient predicates that guide pruning.
Second, we use symbolic analysis to derive the implementation of the concrete necessary and sufficient conditions specific to the given query and tree's metadata.
We use symbolic interval analysis to generate these conditions and provide a novel extension for analyzing spatial relations such as intersection and containment.
Together, these techniques enable generation of specialized traversals for a broad class of search and non-equijoin algorithms, extending beyond what traditional systems support. Our technical contributions are:

\begin{itemize}
    \item A lowering algorithm that fuses set filters and reductions into work-efficient tree traversals.

    \item A technique, termed \textit{predicate analysis}, that extends symbolic interval analysis with rules for geometric operators, to derive pruning conditions from query predicates and tree metadata.

    \item Two generalized tree-based non-equijoin algorithms, enabled by the above techniques.
\end{itemize}

\begin{wrapfigure}[5]{r}{0.49\textwidth}
  \centering
  \vspace{-1em} 
  \includegraphics[width=\linewidth]{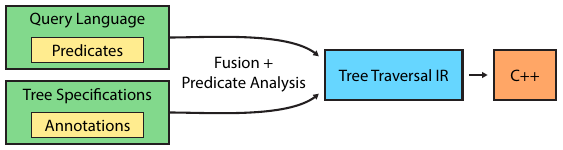}
  \label{fig:system_pldi}
\end{wrapfigure}
\noindent We implement these ideas in the \sys compiler, whose architecture is shown on the right. \sys compiles queries written in a simple functional query language (\Cref{sec:query-spec}). Trees are separately declared as ADTs with metadata annotations (\Cref{sec:tree-spec}). \sys's lowering algorithm (\Cref{sec:lowering}) fuses query operations into a tree traversal by recursively rewriting a tree traversal intermediate representation (\sysIR) defined in \Cref{sec:ir}. Lowering filters and idempotent reductions to pruning traversals requires the generation of pruning and inclusion functions, which \sys derives via predicate analysis (\Cref{sec:predicate-analysis}). Finally, we demonstrate that the fusion algorithm can also generate code for joins (filters of products) that coiterate multiple trees (\Cref{sec:joins}). Across a range of benchmarks, our generated traversals match hand-written code and asymptotically outperform the linear scans and nested-loop joins that systems fall back to when hand-written cases do not apply.

Although \sys is a set operator compiler that produces efficient implementations of range queries, spatial joins, and similar classes of operations, it lacks features needed for it to be used as a backend for a relational database management system (DBMS). In particular, it would need a query planner and associated cost model, as well as efficient non-tree operator implementations (e.g., sort-merge join, hash join, deduplication, etc.). \sys is most accurately described as an \textit{operator compiler}: it generates efficient implementations of individual and fused operators, but does not decide the order of the operators, which operators to fuse, or the allocation of temporaries. We discuss directions of future work for integrating our techniques into DBMSs in \Cref{sec:future-work}.

\section{The Structure of Accelerating Queries with Trees}
\label{sec:background}


To illustrate the basic concepts of how trees can accelerate queries, consider a range query: finding all 1D points in a collection whose \code{x} coordinate lies between \code{lo} and \code{hi} (inclusive):
\begin{minipage}{\linewidth}
\begin{lstlisting}[
  language=lang, 
  mathescape=true, 
  basicstyle=\scriptsize\ttfamily, 
  backgroundcolor=\color{codebackground}, 
  frame=single,
  xleftmargin=3.4pt,
  xrightmargin=3.4pt
]
type Point = { x : f32; id : i32; };
func rangeq(ps : Set<Point>, lo hi : f32) = filter(|p : Point| lo $\leq$ p.x && p.x $\leq$ hi, ps);
\end{lstlisting}
\end{minipage}

\noindent Building a tree over the points can accelerate filtering. Consider the following tree, built on two variants: a leaf node that stores a single point; and an interior node with two children augmented to hold the lower bound and upper bound (\code{xl}, \code{xh}) of the \code{x} coordinates of all points stored below it.
\begin{minipage}{\linewidth}
\begin{lstlisting}[language=lang, mathescape=true, basicstyle=\scriptsize\ttfamily, backgroundcolor=\color{codebackground}, frame=single,
xleftmargin=3.4pt,
xrightmargin=3.4pt
]
type ITree = Leaf(p : Point) | Interior(left right : ITree, xl xh : f32);
\end{lstlisting}
\end{minipage}

\noindent The range query can be efficiently implemented on this tree by considering three cases on an \code{Interior} node's metadata:
\vspace{-0.25em}
\begin{description}
    \item[Always] If the query range contains a subtree's range, all descendants are returned.

    \item[Maybe] If the query range overlaps the subtree's range, some descendants may be in the query range, so the children are recursively visited and their subresults are unioned.

    \item[Never] Otherwise, no descendants are included in the query, so the node is \textit{pruned}.
\end{description}
\vspace{-0.25em}

We illustrate these cases in the below tree traversal code, where \code{yield} returns a singleton set and \code{scan t} returns all points in subtree \code{t}. We illustrate an example tree on the right. A range query with \code{lo=-10} and \code{hi=10} descends to the root's children, pruning the left subtree (first dashed box). It then descends on the right subtree, scanning the dotted orange box because its interval is contained in the query interval, and pruning its sibling (second dashed box).

\newsavebox{\codebox}
\newlength{\toppad}
\newlength{\bottompad}

\setlength{\toppad}{0pt} 
\setlength{\bottompad}{3pt} 

\begin{lrbox}{\codebox}
\begin{minipage}[b]{0.6\textwidth} 
\begin{lstlisting}[language=lang, mathescape=true, basicstyle=\scriptsize\ttfamily, backgroundcolor=\color{codebackground}, frame=single,
xleftmargin=3.4pt,
xrightmargin=3.4pt
]
func _rangeq(t : ITree, lo hi : f32) =
 match t
 | Leaf(p) $\shortrightarrow$ if lo $\leq$ p.x && p.x $\leq$ hi: yield p
 | Interior(left, right, xl, xh) $\shortrightarrow$
    if lo $\leq$ xl && xh $\leq$ hi: scan t
    elif lo $\leq$ xh && xl $\leq$ hi: _rangeq(left, lo, hi) $\cup$
$\hspace{4.38cm}$_rangeq(right, lo, hi)
    else: {};
\end{lstlisting}
\end{minipage}
\end{lrbox}

\noindent
\usebox{\codebox}\hfill
\begin{minipage}[b]{0.38\textwidth}
\centering
\raisebox{\bottompad}{%
    \includegraphics[height=\dimexpr\ht\codebox+\dp\codebox-\toppad-\bottompad\relax]{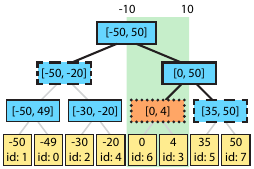}%
}
\end{minipage}

\vspace{-0.5em}
These cases correspond to necessary and sufficient conditions on the query predicate. 
The \textit{always} case requires a sufficient condition (one that guarantees the 
query predicate is true). The \textit{never} case 
prunes a subtree when the necessary condition is false (the query predicate cannot be true). If neither of these cases is satisfied, the query predicate is \textit{maybe} true, so the traversal must recurse.

Similar reasoning can be used to accelerate \textit{idempotent} and \textit{associative} reductions. To illustrate, we extend our range query example to compute the minimum \code{id} of any point within the range:
\begin{minipage}{\linewidth}
\begin{lstlisting}[language=lang, mathescape=true, basicstyle=\scriptsize\ttfamily, backgroundcolor=\color{codebackground}, frame=single,
xleftmargin=3.4pt,
xrightmargin=3.4pt
]
func minq(ps : Set<Point>, lo hi : f32) = min(|p : Point| p.id,
$\hspace{6.75cm}$filter(|p : Point| lo $\leq$ p.x && p.x $\leq$ hi, ps));
\end{lstlisting}
\end{minipage}

The \code{min} operation is both idempotent and associative.
These properties enable two complementary optimizations:
Associativity allows traversal to skip traversing fully-contained subtrees (in the \textit{always} case) by directly composing a precomputed minimum stored in the tree instead of scanning for it; the idempotent property enables value-based pruning: if a subtree's minimum 
cannot improve the current best, we can skip it entirely (another \textit{never} case).

To illustrate the use of these properties, we can extend the \code{ITree} to store the minimum \code{id} of its subtree as \code{idl}. This value can be used both to update the running minimum when a node is fully contained and in the pruning condition for the idempotent property. We provide the fused and lowered min-id-range query on the extended \code{ITree} below. When visiting the dotted orange box, the traversal uses its \code{idl} value to update the running minimum. The rightmost dashed box can then be pruned in one of two ways: it does not overlap the query interval, \textit{and} it has a greater \code{idl} than the best found so far.

\vspace{0.4em}
\noindent
\begin{minipage}[c]{0.675\textwidth}
\begin{lstlisting}[language=lang, mathescape=true, basicstyle=\scriptsize\ttfamily, backgroundcolor=\color{codebackground}, frame=single,
xleftmargin=3.4pt,
xrightmargin=3.4pt
]
func _minq_impl(t : ITree, lo hi : f32, best : mut i32) =
 match t
 | Leaf(p) $\shortrightarrow$ if lo $\leq$ p.x && p.x $\leq$ hi: best = min(p.id, best)
 | Interior(left, right, xl, xh, idl, idh) $\shortrightarrow$
    if lo $\leq$ xl && xh $\leq$ hi: best = min(idl, best) // inclusion
    elif lo $\leq$ xh && xl $\leq$ hi:
      if idl < best:                   // value-based pruning
        _minq_impl(left, lo, hi, best)
        _minq_impl(right, lo, hi, best);

func _minq(t : ITree, lo hi : f32) =
  let best : mut i32 = i32_max in _minq_impl(t, lo, hi, best);
  best;
\end{lstlisting}
\end{minipage}
\hfill
\begin{minipage}[c]{0.3\textwidth}
\centering
\includegraphics{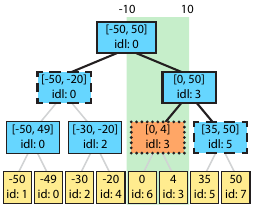}
\end{minipage}

This fusion provides asymptotic benefits by eliminating the allocation of an intermediate \textit{and} enabling value-based pruning via the running minimum. Without fusion, the range query must construct the entire filter output before computing the minimum \code{id}.

To make such optimizations systematic, we now describe how queries and data structures can be specified in a way that makes their properties explicit. The next sections introduce a simple language for describing high-level set queries, alongside a language for expressing tree structures annotated with the metadata they store.

\section{Query Language}
\label{sec:query-spec}

\sys's query specification language consists of queries over unordered finite sets and multisets. Their element type can be any primitive type (integer, float, enum, fixed-length vectors of primitive types) or (often) a product type of primitive types. Geometry types are product types extended with spatial relationships defined on them (e.g., \code{intersects} and \code{contains}).

\subsection{Set Operations}
\begin{figure*}[b]
\centering
\footnotesize
\begin{minipage}{0.58\linewidth}
\begin{tabular}{@{}l@{}}
\hline\\[-0.8em]
$q$ : Query \;::=\; $s$ \hfill \algcomment{set variable} \\
\phantom{$q$ : }\;|\; \code{filter}($P$, $q$) \hfill \algcomment{filtered set} \\
\phantom{$q$ : }\;|\; \code{map}($F$, $q$) \hfill \algcomment{mapped set} \\
\phantom{$q$ : }\;|\; \code{reduce}($e_r$, $\oplus$, $q$) \hfill \algcomment{aggregation} \\
\phantom{$q$ : }\;|\; \code{product}($q_1$, $q_2$) \hfill \algcomment{Cartesian product} \\
\phantom{$q$ : }\;|\; \code{min}($M$, $q$) \;|\; \code{max}($M$, $q$) \hfill \algcomment{metric optimization} \\
\phantom{$q$ : }\;|\; \code{argmin}($M$, $q$) \;|\; \code{argmax}($M$, $q$) \hfill \algcomment{arg optimization} \\
\phantom{$q$ : }\;|\; \code{any}($P$, $q$) \;|\; \code{all}($P$, $q$) \hfill \algcomment{conditional logic} \\[0.3em]
$P$ : Predicate \;::=\; $|x : T|\ e_b$ \hfill \algcomment{boolean lambda} \\[0.3em]
$M$ : Metric \;::=\; $|x : T|\ e_r$ \hfill \algcomment{real-valued lambda} \\[0.3em]
$F$ : Map \;::=\; $|x : T|\ e$ \hfill \algcomment{element transform} \\[0.3em]
$e_b$ : BoolExpr \;::=\; $e_r \odot e_r$ \hfill\;\; \algcomment{scalar comparison $\odot \in \{< \; \leq \; = \; \neq \; \geq \; >\}$} \\
\phantom{$e_b$ : }\;|\; $e_b \wedge e_b \mid e_b \vee e_b \mid \neg e_b$ \hfill \algcomment{logical connectives} \\
\phantom{$e_b$ : }\;|\; \code{contains}($e_g$, $e_g$) \;|\; \code{covers}($e_g$, $e_g$) \hfill \; \algcomment{topological predicates~\cite{egenhofer1990topological}} \\
\phantom{$e_b$ : }\;|\; \code{disjoint}($e_g$, $e_g$) \;|\; \code{intersects}($e_g$, $e_g$) \hfill \\
\phantom{$e_b$ : }\;|\; \code{equals}($e_g$, $e_g$) \;\;\;\,\;|\;\ \code{touches}($e_g$, $e_g$) \hfill \\
\phantom{$e_b$ : }\;|\; \code{within}($e_g$, $e_g$) \hfill \\
\phantom{$e_b$ : }\;|\; $e_g \leq_D e_g$ \;|\; $e_g <_D e_g$ \hfill \hfill \algcomment{ordering predicates on dimension $D$} \\[0.3em]

$e_r$ : RealExpr \;::=\; $x$ \;|\; $n$ \;|\; $e_r \odot e_r$ \hfill \algcomment{variables, literals, arithmetic} \\
\phantom{$e_r$ : }\;|\; \code{distmax}($e_g$, $e_g$) \;|\; \code{distmin}($e_g$, $e_g$) \hfill \algcomment{distance metrics} \\[0.3em]
$e_g$ : GeoExpr \;::=\; $x$ \;|\; $G(e^*)$ \hfill \algcomment{geometric variable or constructor} \\[0.3em]
$T$ : Type \;::=\; $t$ \;|\; $T \times n$ \hfill \algcomment{scalar and vector types} \\
\phantom{$T$ : }\;|\; $\code{Set<}T\code{>}$ \hfill \algcomment{set of elements} \\
\phantom{$T$ : }\;|\; $(T_1, \ldots, T_k)$ \hfill \algcomment{product type (including geometric types)} \\
\hline
\end{tabular}
\caption{Context-free grammar of \sys's query language. The symbol $s$ denotes a set variable, $x$ a variable, $n$ a numeric literal, $\oplus$ an associative and commutative binary operator, and $G$ a geometric type constructor (e.g., \code{Ray()}).}
\label{fig:query-grammar}
\end{minipage}
\hfill
\begin{minipage}{0.4\linewidth}
  \vspace{-6mm}
  \begin{minipage}[t]{\linewidth}
  \begin{lstlisting}[language=lang, mathescape=true,
      basicstyle=\scriptsize\ttfamily,
      backgroundcolor=\color{codebackground}, frame=single,
      xleftmargin=3.4pt,
    xrightmargin=3.4pt
]
filter($T \shortrightarrow$ bool, Set<$T$>) : Set<$T$>
map($T \shortrightarrow S$, Set<$T$>) : Set<$S$>
reduce($T$, $T \times T \shortrightarrow T$, Set<$T$>) : $T$
product(Set<$T$>, Set<$S$>) : Set<($T,S$)>
min($T \shortrightarrow \mathbb{R}$, Set<$T$>) : $\mathbb{R}$
max($T \shortrightarrow \mathbb{R}$, Set<$T$>) : $\mathbb{R}$
argmin($T \shortrightarrow \mathbb{R}$, Set<$T$>) : $T$
argmax($T \shortrightarrow \mathbb{R}$, Set<$T$>) : $T$
any($T \shortrightarrow$ bool, Set<$T$>) : bool
all($T \shortrightarrow$ bool, Set<$T$>) : bool
  \end{lstlisting}
  \captionof{figure}{Type signatures for \sys's set operations. $T$ and $S$ are primitive types (e.g., integers, floats, or product types).}
  \label{lst:setOps}
  \end{minipage}

  \vspace{1.5em}

  \begin{minipage}{\linewidth}
  \centering
  \includegraphics{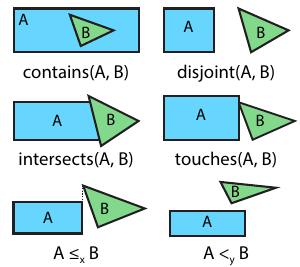}
  \captionof{figure}{Geometric predicate illustrations.}
  \label{fig:geo-preds}
  \end{minipage}
\end{minipage}

\end{figure*}

\sys supports operations on unordered sets and multisets: \code{filter}, \code{reduce}
\code{map}, and \code{product} (the Cartesian product).
In addition to these standard operations, \sys provides a number of idempotent reductions (e.g., \code{argmin}/\code{argmax}, \code{min}/\code{max}, and \code{any}/\code{all}). \sys operations have the same semantics for sets and multisets. For the sake of brevity, we refer to sets and multisets collectively as \textit{sets} throughout the rest of the paper. Standard aggregations such as \code{count} or \code{avg} can be implemented by a \code{map} followed by a \code{reduce}: \sys's fusion algorithm ensures that this decomposition does not introduce inefficiencies. Reductions and filters are the primary operations accelerated by tree data structures, though filters or reductions of products can also be accelerated, as we illustrate in \Cref{sec:joins}. We provide the grammar of \sys queries in \Cref{fig:query-grammar} and type signatures in \Cref{lst:setOps}.

The set operators \code{any}, \code{all}, and \code{filter} accept a predicate that maps the set elements to boolean values. We support a grammar of scalar operations (comparators, conjunctions, disjunctions, mathematical operators, etc.) in addition to geometric predicates. The latter is discussed below.

\subsection{Geometric Operations}

In \sys, the type of a geometric object is a product type along with defined spatial relationships between pairs of geometric object types. Rather than restricting users to a hardcoded set of shapes, \sys supports arbitrary geometric entities (e.g., volumes, polygons, rays, points) provided the type exposes spatial relationships.




Our spatial operators follow the semantics of~\citet{egenhofer1990topological}, and include topological operators (e.g., \code{intersects} and \code{contains}), ordering operators (e.g., $\leq_X$, which denotes ordering in the $X$ dimension), and metric operations (e.g., \code{distmin}). \Cref{fig:geo-preds} gives visual examples of the topological and ordering relationships of the spatial operators and \Cref{fig:query-grammar} lists their grammar, which is part of the predicate subset of the query languages. While not exhaustive, these operators are expressive enough to encompass a broad class of spatial queries, including ray tracing, collision detection, and spatial SQL.

\sys provides a user-extensible library of geometric object types (e.g., \texttt{Ray}, \texttt{Triangle}, and \texttt{AABB}) with their spatial relationships following the implementations of ~\citet{ericson2004rtcd}.


\subsection{Examples}
\sys's query language allows concise data-structure-agnostic representations of not just scalar queries, but also spatial queries, e.g., closest-hit ray tracing is expressible via:

\begin{lstlisting}[language=lang, mathescape=true, basicstyle=\scriptsize\ttfamily, backgroundcolor=\color{codebackground}, frame=single,
xleftmargin=3.4pt,
xrightmargin=3.4pt
]
closest(r : Ray, ts : Set<Triangle>) = argmin(|t : Triangle| distmin(r, t),
$\hspace{6.75cm}$filter(|t : Triangle| intersects(r, t), ts));
\end{lstlisting}
Shadow ray tracing, which only queries if there is a hit, is even more concise:
\begin{lstlisting}[language=lang, mathescape=true, basicstyle=\scriptsize\ttfamily, backgroundcolor=\color{codebackground}, frame=single,
xleftmargin=3.4pt,
xrightmargin=3.4pt
]
shadow(r : Ray, ts : Set<Triangle>) = any(|t : Triangle| intersects(r, t), ts);
\end{lstlisting}
Similarly, collision detection (a spatial join) is also concisely representable:
\begin{lstlisting}[language=lang, mathescape=true, basicstyle=\scriptsize\ttfamily, backgroundcolor=\color{codebackground}, frame=single,
xleftmargin=3.4pt,
xrightmargin=3.4pt
]
collisions(s0 : Set<Object>, s1 : Set<Object>) = filter(|a b : Object| intersects(a, b),
$\hspace{8.3cm}$product(s0, s1));
\end{lstlisting}


\section{Tree Specification Language}
\label{sec:tree-spec}
\sys provides a data modeling language based on Algebraic Data Types (ADTs) extended with augmentation \textit{annotations} that enable the optimization of queries. Augmentations are declarative annotations over recursive data structures that specify useful metadata invariants, allowing \sys to accelerate certain operations. We focus on three primary types of augmentations:

\begin{description}
    \item[Bounds augmentations] Capture geometric bounds (e.g., intervals in 1D, bounding volumes in higher dimensions) over a subtree, attached to primitives or fields of sum-typed primitives. These can be used to accelerate filters and idempotent reductions. 

    \item[Reduction augmentations] Store partial aggregates over a subtree. These can be used to accelerate associative reductions.

    \item[Data tags] Explicitly state that an ADT field should be interpreted as a member of the set that the tree implements.
\end{description}

Annotations are attached using a \code{with} clause on standard recursive ADT definitions. For example, the interval tree from \Cref{sec:background}, implementing a set of \code{Point}s with fields \code{x} and \code{id}, is encoded as:
\begin{lstlisting}[language=lang, mathescape=true, basicstyle=\scriptsize\ttfamily, backgroundcolor=\color{codebackground}, frame=single,
xleftmargin=3.4pt,
xrightmargin=3.4pt
]
tree ITree implements Set<Point> =
| Interior(left right : ITree, xl xh : f32, idl idh : i32)
 with x in [xl, xh]    // implicit forall expands to: forall p in subtree: p.x in [xl, xh]
 with id in [idl, idh] // implicit forall expands to: forall p in subtree: p.id in [idl, idh]
 with min(id) = idl    // implicit forall expands to: min(forall p in subtree: p.id) = idl
| Leaf(p : Point) with data = p;
\end{lstlisting}

These annotations directly correspond to the augmentations described above. The first and second \code{Interior} annotations are scalar bounds on the fields \code{x} and \code{id}, respectively, of all points stored in the subtree.
The third annotation is a reduction augmentation: it essentially marks the lower bound of \code{id} as \textit{tight}, meaning it can be used to accelerate a \code{min} operation over \code{id} on a subtree.
For most interval tree implementations, all bounds would be implemented as tight bounds; for brevity, we do not add these annotations to \code{x} and \code{idh} in the code snippet above.
The \code{Leaf} annotation simply tags \code{p} as a set element.

\sys allows geometric data to be labeled via the same notation as scalar data. Our specification language explicitly models \textit{object hierarchies}, where nodes bound discrete subsets of geometric primitives (e.g., BVHs~\cite{kay1986, ericson2004rtcd, meister2021bvh} or R-trees~\cite{guttman1984rtree}), rather than \textit{spatial hierarchies} that recursively partition the underlying space itself (e.g., k-d trees~\cite{bentley1975kdtree} or octrees). Because spatial partitions allow primitives to span multiple nodes, they yield weaker structural invariants (overlap rather than strict set containment) and require complex, query-specific deduplication techniques~\cite{eltabakh2007deduplication}, which we discuss further in \Cref{sec:future-work}. Just as \sys's query language supports any geometry-typed object (\Cref{sec:query-spec}), any geometry-typed object may serve as a bounding volume annotation.

Consider a standard binary Axis-Aligned Bounding Box (AABB) tree\footnote{Referred to as an R-tree~\cite{guttman1984rtree} in the database community.}~\cite{pbrt4}, which has a leaf and interior node that both store low and high vectors representing an AABB in 3D space.
\sys's data modeling language represents this structure as:
\begin{lstlisting}[language=lang,mathescape=true, basicstyle=\scriptsize\ttfamily,backgroundcolor=\color{codebackground}, frame=single,
xleftmargin=3.4pt,
xrightmargin=3.4pt
]
tree TriBVH implements Set<Triangle> =
| Leaf(low high : vec<f32, 3>, prims : Triangle[]) with data = prims with AABB(low, high)
| Interior(low high : vec<f32, 3>, left right : TriBVH) with AABB(low, high);
\end{lstlisting}
This bounds annotation states that all geometries beneath the node lie within the volume. Note that while AABBs can be decomposed into per-dimension scalar annotations, other bounding volumes~\cite{woop2014hair, ericson2004rtcd} cannot. This motivates \sys's support for geometry-typed bounds annotations.

We note that there are some tree data structures, such as \citet{benthin2018clbvh}'s, that store augmentations for the child nodes in the parent node. This subtly shifts where pruning occurs, but not in a meaningful way; for brevity of explanation, this paper describes lowering machinery only for the case where a node stores its own augmentations, not its children.

\section{Lowering Queries onto Trees}
\label{sec:lowering}

Our algorithm for fusing queries into tree traversals is a recursive bottom-up rewrite-based technique, in the spirit of StreamFusion~\cite{coutts2007streamfusion}.
The key observation is that a tree traversal is structured as a case analysis over node variants: leaves yield data and interior nodes recurse.
Each set operator applies uniformly over this structure: modifying how data is yielded, whether recursion proceeds, or how results are combined. This uniformity allows lowering to be expressed as local rewrites at leaves and recursive nodes, which naturally compose into a single fused traversal. To make these rewrites precise, we introduce a compact intermediate representation (\sysIR) that isolates exactly the constructs needed for tree traversals.
We first show how to lower and fuse set operations onto trees, abstractly assuming a technique for generating the pruning and inclusion functions \code{always}, \code{maybe}, and \code{never} for a given predicate; \Cref{sec:predicate-analysis} shows how we derive these functions.
\Cref{alg:lower} summarizes our lowering procedure, whose rewrite rules are presented and explained in the following subsections.

\subsection{Tree Traversal Intermediate Representation (TTIR)}
\label{sec:ir}

\sysIR is a small fusion calculus for tree traversals, analogous to StreamFusion's calculus for lists~\cite{coutts2007streamfusion}. It is not intended as a general-purpose language; its role is to expose the structure on which operators act: producing data at leaves, recursing at internal nodes, and aggregating results. To that end, \sysIR is a functional IR with \code{match} and \code{if} for control flow, extended by a handful of domain-specific constructs:

\begin{itemize}
    \item \code{yield} and \code{iter} return data items from leaves, for singletons and sub-collections (e.g., vectors, arrays) respectively.

    \item \code{scan} aggregates the results of subtrees; it performs a set union by default, but can apply other reductions (e.g., \code{sum}). It can also apply a function to subtree elements before aggregation.

    \item \code{from} recurses on subtrees.

    \item \code{upd} modifies the accumulator of a running reduction.
\end{itemize}

This design enables fusion: operator-specific rewrites can be defined locally on each construct and composed recursively. 
Note that this section discusses \code{scan} and \code{from} being applied to a single tree (recursing on its children). \Cref{sec:joins} illustrates an extension to multiple arguments, which is necessary for coiterating multiple trees, e.g., in lowering \code{product}. 

The base case of lowering produces a direct traversal of a tree by \code{yield}ing all singleton data, \code{iter}ating all sub-collection data, and \code{scan}ning all interior nodes. This is the \textsc{GenerateTreeIterator} method referenced on Line 5 of \Cref{alg:lower}. To illustrate, consider the lowering of iteration on the example tree on the left, into the traversal code on the right:

\noindent\begin{minipage}[t]{0.48\linewidth}
\vspace*{\fill}
\begin{lstlisting}[language=lang, mathescape=true, basicstyle=\scriptsize\ttfamily, backgroundcolor=\color{codebackground}, frame=single,
xleftmargin=3.4pt
]
tree ExampleTree implements Set<i32> =
| Leaf(i : i32) with data = i
| LargeLeaf(is : vec<i32, 4>) with data = is
| Interior(left right : ExampleTree);
\end{lstlisting}
\end{minipage}
\hspace{1em}
\begin{minipage}[t]{0.48\linewidth}
\begin{lstlisting}[language=lang, mathescape=true, basicstyle=\scriptsize\ttfamily, backgroundcolor=\color{codebackground}, frame=single,
xrightmargin=3.4pt
]
func traverse(t : ExampleTree) = match t
 | Leaf(i) $\shortrightarrow$ yield i
 | LargeLeaf(is) $\shortrightarrow$ iter is
 | Interior(left, right) $\shortrightarrow$ scan t;
\end{lstlisting}
\end{minipage}

The \code{scan} is applied to \code{t} itself rather than its children. This reflects the fact that augmentations are associated with the current node, and later compilation steps will exploit them. If instead augmentations were stored on the children (as in ~\citet{woop2014hair}), the traversal would \code{scan} the children directly, with their results implicitly unioned.


\subsection{Lowering Filters}
\label{sec:lower-filters}

Filtering refines which leaves yield an element, and whether recursion continues at interior nodes. \textsc{LowerFilter} in \Cref{alg:lower-filter-min} generates code with this behavior. At leaves, data is tested against the predicate $P$ before being yielded or iterated. Recursive nodes are scanned if $P$ is always true for the subtree, are recursed on if $P$ might be true, and are pruned otherwise.
\Cref{fig:filter_P} shows the lowered filter for predicate $P$ on \code{ExampleTree}.

This rewrite allows naturally fusing chained filters, as each simply inserts local conditions on \code{yield}, \code{iter}, and \code{scan} constructs. We show the result of such fusion in \Cref{fig:filter_PQ}, which \code{scan}s if both predicates $P$ and $Q$ can be proven always true and recursively evaluates otherwise.
Crucially, this means the algorithm does not attempt to enumerate all possible truth tables of compound predicates (e.g., for \code{filter(Q, filter(P, x))} considering every combination of \code{always(P)}, \code{always(Q)}, \code{maybe(P)}, \code{maybe(Q)}). This favors avoiding exponential code explosion and the need to produce filters specialized for simplified versions of predicates. It is possible that doing so could yield performance improvements; we simply avoid it to prevent exponential code blow-up. 

\begin{figure}[h]
  \centering
  \begin{subfigure}[t]{0.48\textwidth}
    \centering
\begin{lstlisting}[language=lang, mathescape=true, basicstyle=\scriptsize\ttfamily, backgroundcolor=\color{codebackground}, frame=single,
xleftmargin=3.4pt
]
func filter_P(t : ExampleTree) = match t
 | Leaf(i) $\shortrightarrow$ if $P$(i): yield i
 | LargeLeaf(is) $\shortrightarrow$ filter(|i| $P$(i), is)
 | Interior(left, right) $\shortrightarrow$
    if always($P$, t): scan t
    elif maybe($P$, t): from t
\end{lstlisting}
    \caption{Lowering a single filter $P$.}
    \label{fig:filter_P}
  \end{subfigure}
  \hfill
  \begin{subfigure}[t]{0.48\textwidth}
    \centering
\begin{lstlisting}[language=lang, mathescape=true, basicstyle=\scriptsize\ttfamily, backgroundcolor=\color{codebackground}, frame=single,
xrightmargin=3.4pt
]
func filter_PQ(t : ExampleTree) = match t ...
 | Interior(left, right) $\shortrightarrow$
    if always($P$, t):
    $\;\;$if always($Q$, t): scan t
    $\;\;$elif maybe($Q$, t): from t
    elif maybe($P$, t) && maybe($Q$, t): from t
\end{lstlisting}
\caption{Interior case for filters $P$ and $Q$}
    \label{fig:filter_PQ}
  \end{subfigure}

   \caption{Examples of lowered \code{filter}s on the \code{ExampleTree}.}
  \label{fig:filter_examples}
  \vspace{-1em}
\end{figure}

\begin{figure}[h]
\vspace{-1.5em}
\noindent\begin{minipage}[t]{0.48\textwidth}
\begin{algorithm}[H]
\caption{Recursive Query Lowering}
\tiny
\label{alg:lower}
\begin{algorithmic}[1]
\State \textbf{Input}: Query expression $Q$
\State \textbf{Output}: TTIR that iterates the query result
\Function{Lower}{$Q$}
    \State \textbf{match} $Q$ \textbf{with}
    \State \hspace{0.1em} \texttt{|} \textsc{s} $\Rightarrow$ \textsc{GenerateTreeIterator}(\textsc{s})
    \State \Comment{Apply \code{yield} and \code{iter} to tagged data, and \code{scan} to nodes.}

    \State \hspace{0.1em} \texttt{|} \code{filter}$(P, S)$ $\Rightarrow$ \textsc{LowerFilter}($P, S$) \Comment{\Cref{alg:lower-filter-min}}

    \State \hspace{0.1em} \texttt{|} \code{map}$(F, S)$ $\Rightarrow$
    \State \hspace{1.5em} \textbf{rewrite} \textsc{Lower}($S$) \textbf{with}
    \State \hspace{1.5em} \texttt{|} \code{yield x} $\Rightarrow$ \code{yield} $F$\code{(x)}
    \State \hspace{1.5em} \texttt{|} \code{iter xs} $\Rightarrow$ \code{iter map(}$F$\code{, xs)}
    \State \hspace{1.5em} \texttt{|} \code{scan tr} $\Rightarrow$ \code{scan} $F$\code{(tr)}

    \State \hspace{0.1em} \texttt{|} \code{reduce}$(id, C, S)$ $\Rightarrow$ \Comment{Lower associative reductions}
    \State \hspace{1em} \textsc{WrapWithAccumulator}(\code{a}, $id$, 
    \State \hspace{1.5em} \textbf{rewrite} \textsc{Lower}($S$) \textbf{with}
    \State \hspace{1.5em} \texttt{|} \code{yield x} $\Rightarrow$ \code{upd a}  $C$(\code{a, x)}
    \State \hspace{1.5em} \texttt{|} \code{iter xs} $\Rightarrow$ \code{upd a} \code{reduce(}$a, C$, \code{xs)}
    \State \hspace{1.5em} \texttt{|} \code{scan tr} $\Rightarrow$ \textbf{if} \code{tr} \textbf{has} $C$\code{(tr)} \textbf{then}
    \State \hspace{8.25em} \code{upd a}  $C$\code{(a, tr.}$C$\code{(tr))} 
    \State \hspace{7.25em} \textbf{else}
    \State \hspace{8.25em} \code{scan<}$C$\code{> tr} \hspace{1em}) \Comment{End wrapper}

    \State \hspace{0.1em} \texttt{|} \code{product}$(S_0, S_1)$ $\Rightarrow$ \textsc{LowerProd}($S_0, S_1$) \Comment{\Cref{alg:lower-product}}

    \State \hspace{0.1em} \texttt{|} \code{min}$(M, S)$ $\Rightarrow$ \textsc{LowerMin}($M, S$) \Comment{\Cref{alg:lower-filter-min}}
    \State \hspace{0.1em} \texttt{|} \code{max}$(M, S)$ $\Rightarrow$ \textsc{LowerMax}($M, S$) 

    \State \hspace{0.1em} \texttt{|} \code{argmin}$(M, S)$ $\Rightarrow$ \textsc{LowerArgMin}($M, S$) \Comment{\Cref{alg:lower-argmin}}
    \State \hspace{0.1em} \texttt{|} \code{argmax}$(M, S)$ $\Rightarrow$ \textsc{LowerArgMax}($M, S$) 

    \State \hspace{0.1em} \texttt{|} \code{any}$(P, S)$ $\Rightarrow$ \textsc{LowerAny}($P, S$) \Comment{\Cref{alg:lower-any}}
    \State \hspace{0.1em} \texttt{|} \code{all}$(P, S)$ $\Rightarrow$ \textsc{LowerAll}($P, S$) 
\EndFunction
\end{algorithmic}
\end{algorithm}
\end{minipage}
\hfill
\begin{minipage}[t]{0.48\textwidth}
\begin{algorithm}[H]
\caption{Filter and Min Lowering}
\tiny
\label{alg:lower-filter-min}
\begin{algorithmic}[1]
\State \textbf{Input}: Query predicate $P$ and set expression $S$
\State \textbf{Output}: TTIR that iterates the query result
\Function{LowerFilter}{$P$, $S$}
    \State \textbf{rewrite} \textsc{Lower}($S$) \textbf{with}

    \State \hspace{0.1em} \texttt{|} \code{yield x} $\Rightarrow$ \code{if} $P$\code{(x): yield x}

    \State \hspace{0.1em} \texttt{|} \code{iter xs} $\Rightarrow$ \code{iter filter(}$P$\code{, xs)}

    \State \hspace{0.1em} \texttt{|} \code{scan tr} $\Rightarrow$ \code{if always(}$P$\code{, tr): scan tr}
    \State \hspace{6em} \code{elif maybe(}$P$\code{, tr): from tr}

    \State \hspace{0.1em} \texttt{|} \code{from tr} $\Rightarrow$ \code{if maybe(}$P$\code{, tr): from tr}
\EndFunction
\vspace{0.55em}
\State \textbf{Input}: Query metric $M$ and set expression $S$
\State \textbf{Output}: TTIR that stores the result in accumulator \code{a}
\Function{LowerMin}{$M$, $S$}
    \State \textsc{WrapWithAccumulator}(\code{a}, $\infty$, 
    \State \hspace{0.3em} \textbf{rewrite} \textsc{Lower}($S$) \textbf{with}

    \State \hspace{0.35em} \texttt{|} \code{yield x} $\Rightarrow$ \code{upd a minb(a, }$M$\code{(x))}

    \State \hspace{0.35em} \texttt{|} \code{iter xs} $\Rightarrow$ \code{upd a minb(a, min(}$M$\code{, xs))}

    \State \hspace{0.35em} \texttt{|} \code{scan tr} $\Rightarrow$ \textbf{if} \code{tr} \textbf{has} \code{min(}$M$\code{, tr)} \textbf{then}
    \State \hspace{6.75em} \code{upd a minb(a, min(}$M$\code{, tr))}
    \State \hspace{6.25em} \textbf{else if} \code{tr} \textbf{has} \code{max(}$M$\code{, tr)} \textbf{then}
    \State \hspace{6.75em} \code{upd a minb(a, max(}$M$\code{, tr))}
    \State \hspace{6.75em} \code{if maybe(min(}$M$\code{(tr)) < a):}
    \State \hspace{7.25em} \code{from tr}
    \State \hspace{6.25em} \textbf{else}
    \State \hspace{6.75em} \code{if maybe(min(}$M$\code{(tr)) < a):}
    \State \hspace{7.25em} \code{from tr}

    \State \hspace{0.35em} \texttt{|} \code{from tr} $\Rightarrow$ \code{if maybe(min(}$M$\code{(tr)) < a):}
    \State \hspace{6.75em} \code{from tr} \hspace{1em} ) \Comment{End accumulator wrapper}
\EndFunction
\end{algorithmic}
\end{algorithm}
\end{minipage}
\vspace{-.5em}
\end{figure}







\subsection{Lowering Associative Reductions}
\label{sec:lower-cred}

Associative reductions can be computed hierarchically, allowing reuse of intermediate results across subtrees. Each is defined by an idempotent identity value and a commutative,\footnote{Commutativity is required because our sets are unordered.} associative binary operator for combining subresults. These algebraic properties make them amenable to acceleration via tree augmentations that store precomputed values over subsets/subtrees.

Associative reductions on their own can be evaluated directly from augmentations stored at the root node, but also integrate naturally with filters. A reduction that wraps a filter can be fused into an efficient traversal that uses precomputed values only when the filter predicate is proven always true, recursively evaluating the query predicate otherwise, as illustrated in \Cref{sec:background}. Such fusion is asymptotically useful, avoiding the need to store the filter result before aggregation.

Lowering an associative reduction is also done through local rewrites on \sysIR constructs, illustrated in Lines 15--22 of \Cref{alg:lower}: \code{yield} and \code{iter} apply the operator on the running accumulator and the yielded set elements, \code{scan} incorporates subtree augmentations when available, and \code{from} continues to recurse. For \code{scan}s, there are two (compile-time) cases: Lines 19--20 handle the case that a tree node stores the subresult, and Line 22 applies a reduction scan if the tree node does not. Note that associative reductions that are also idempotent (e.g., \code{min}/\code{max} and \code{any}/\code{all}) exhibit extra pruning potential; they incorporate value-based pruning when possible.

\subsection{Lowering Idempotent Reductions}
\label{sec:lower-ired}

Idempotent reductions, including \code{min}, \code{max}, \code{argmin}, \code{argmax}, \code{any}, and \code{all}, form reductions over a lattice structure. They allow subtree pruning whenever it can be proven that a subtree cannot affect the final result. For example, consider the \code{min}-\code{id} query from \Cref{sec:background}. During traversal, if it can be determined that no value in a subtree has a smaller \code{id} than the current best, the entire subtree can be skipped. This observation generalizes to all idempotent reductions: whenever it can be proven that a subtree does not affect the final result, it need not be visited.

\textsc{LowerMin} in \Cref{alg:lower-filter-min} illustrates our lowering rewrite for \code{min} with a metric applied. \code{yield} and \code{iter} simply reduce on the leaf data, and \code{scan} and \code{from} are rewritten locally to exploit subtree metrics. Note that \code{scan} lowering first applies pruning via the associative reduction property, using stored metadata if available, but otherwise falls back to value-based pruning that the idempotent property enables. This lowering leverages minimum-value metadata if available, but can otherwise use maximum-value metadata to prune the search space conservatively. Such optimizations are in line with state-of-the-art minimum-distance queries~\cite{sawhney2021fcpw, fan2024gDist}.

To illustrate how different stored metadata enable distinct optimizations, consider a min-reduction on a filtered set with filter predicate $P$. When lowered on a tree that stores the minimum metric (\code{MinTree} traversal on the left, below), that stored subtree's value can be used to update the accumulator any time the predicate $P$ is proven true (inclusion-based skipping). When $P$ is \textit{maybe} true, the value can still aid further pruning: if it exceeds the running minimum, the subtree cannot contribute to the result. The same query on a tree that only stores the maximum metric (\code{MaxTree} traversal on the right, below) offers no inclusion-based skipping, but the stored maximum can still induce a tighter bound on the minimum value.

\noindent\begin{minipage}[t]{0.48\linewidth}
\vspace*{\fill}
\begin{lstlisting}[language=lang, mathescape=true, basicstyle=\scriptsize\ttfamily, backgroundcolor=\color{codebackground}, frame=single,
xleftmargin=3.4pt
]
func min_wmin(t : MinTree, acc : i32&) =
 match t
 | Leaf(i) $\shortrightarrow$ if $P$(i): upd acc minb(acc, i)
 | Interior(left, right, min_i) $\shortrightarrow$
    if always($P$, t): upd acc minb(acc, min_i)
    elif maybe($P$, t):
      if min_i < acc: from t
\end{lstlisting}
\end{minipage}
\hspace{1em}
\begin{minipage}[t]{0.48\linewidth}
\begin{lstlisting}[language=lang, mathescape=true, basicstyle=\scriptsize\ttfamily, backgroundcolor=\color{codebackground}, frame=single,
xrightmargin=3.4pt
]
func min_wmax(t : MaxTree, acc : i32&) =
 match t
 | Leaf(i) $\shortrightarrow$ if $P$(i): upd acc minb(acc, i)
 | Interior(left, right, max_i) $\shortrightarrow$
    if always($P$, t):
      upd acc minb(acc, max_i); from t
    elif maybe($P$, t): from t
\end{lstlisting}
\end{minipage}

\noindent Lowerings for \code{argmin} and \code{argmax} mirror those for \code{min} and \code{max}, but also track the element achieving the extremum (\Cref{alg:lower-argmin}). \code{any} and \code{all} can early-return when the predicate is proven always or never true on a subtree, as shown in \Cref{alg:lower-any}.

\vspace{-1.5em}
\noindent\begin{minipage}[t]{0.48\textwidth}
\begin{algorithm}[H]
\caption{Argmin Lowering}
\tiny
\label{alg:lower-argmin}
\begin{algorithmic}[1]
\State \textbf{Input}: Query metric $M$ and set expression $S$
\State \textbf{Output}: TTIR that stores the result in accumulator \code{a}
\Function{LowerArgMin}{$M$, $S$}
    \State \textsc{WrapWithAccumulator}(\code{a}, \{$\infty$, \textsc{null}\}, 
    \State \hspace{0.3em} \textbf{rewrite} \textsc{Lower}($S$) \textbf{with}

    \State \hspace{0.35em} \texttt{|} \code{yield x} $\Rightarrow$ \code{upd a argminb(a, \{}$M$\code{(x), x\})}

    \State \hspace{0.35em} \texttt{|} \code{iter xs} $\Rightarrow$ \code{upd a argminb(a, argmin(}$M$\code{, xs))}

    \State \hspace{0.35em} \texttt{|} \code{scan tr} $\Rightarrow$ \textbf{if} \code{tr} \textbf{has} \code{max(}$M$\code{, tr)} \textbf{then}
    \State \hspace{6.75em} \code{upd a minb(a, max(}$M$\code{, tr))} 
    \State \hspace{6.5em} \code{if maybe(min(}$M$\code{(tr)) < a): from tr}
    \State \hspace{6.1em} \textbf{else}
    \State \hspace{6.75em} \code{if maybe(min(}$M$\code{(tr)) < a): from tr}

    \State \hspace{0.1em} \texttt{|} \code{from tr} $\Rightarrow$ \code{if maybe(min(}$M$\code{(tr)) < a): from tr} \hspace{0.5em} )
\EndFunction
\end{algorithmic}
\end{algorithm}
\end{minipage}
\hfill
\begin{minipage}[t]{0.48\textwidth}
\begin{algorithm}[H]
\caption{Any Lowering}
\tiny
\label{alg:lower-any}
\begin{algorithmic}[1]
\State \textbf{Input}: Query predicate $P$ and set expression $S$
\State \textbf{Output}: TTIR that stores the result in accumulator \code{a}
\Function{LowerAny}{$P$, $S$}
    \State \textsc{WrapWithAccumulator}(\code{a}, \code{false}, 
    \State \hspace{0.3em} \textbf{rewrite} \textsc{Lower}($S$) \textbf{with}

    \State \hspace{0.35em} \texttt{|} \code{yield x} $\Rightarrow$ \code{upd a (a}$\;\lor\;P$\code{(x))}

    \State \hspace{0.35em} \texttt{|} \code{iter xs} $\Rightarrow$ \code{upd a (a}$\;\lor\;$\code{any(}$P$\code{, x))}

    \State \hspace{0.35em} \texttt{|} \code{scan tr} $\Rightarrow$ \code{if always(}$P$\code{, tr):}
    \State \hspace{6.75em} \code{upd a true} 
    \State \hspace{6em} \code{elif} $\neg$\code{a} $\land$ \code{ maybe(}$P$\code{, tr):}
    \State \hspace{6.5em} \code{from tr}

    \State \hspace{0.1em} \texttt{|} \code{from tr} $\Rightarrow$ \code{if} $\neg$\code{a} $\land$ \code{ maybe(}$P$\code{, tr):}
    \State \hspace{6.75em} \code{from tr} \hspace{1em} ) \Comment{End accumulator wrapper}
\EndFunction
\end{algorithmic}
\end{algorithm}
\end{minipage}

\section{Predicate Analysis}
\label{sec:predicate-analysis}

Lowering filters and idempotent reductions requires generating necessary (\code{maybe}) and sufficient (\code{always}) conditions, a process we call \textit{predicate analysis}. Our implementation uses symbolic interval analysis.
A key insight is that symbolic interval analysis~\cite{jrk2012halide} is sufficient for deriving necessary and sufficient conditions in linear time through a simple AST traversal of the query predicate, in contrast to general necessary/sufficient condition synthesis (e.g., inductive invariant generation~\cite{reynolds2020sygus, dillig2013invariant}), which often requires exponential search. Our evaluation shows that interval-derived conditions are tight enough for real-world applications and match the state-of-the-art systems that manually implement query-specific pruning. Symbolic analysis thus provides a practical and efficient way to derive pruning conditions directly from the structure of query predicates.

To clarify the relation between interval analysis and predicate analysis:
a necessary condition (\code{maybe}) is a condition implied by the predicate, and is thus an \textit{upper bound} of the predicate. Likewise, a sufficient condition (\code{always}) implies the predicate, and is thus a \textit{lower bound} of the predicate. For scalar expressions, it is therefore sufficient to derive these conditions by applying standard symbolic interval analysis to generate the bounds on a boolean expression. Many geometric relationships can be similarly bounded by necessary and sufficient conditions (see \Cref{sec:geometric-analysis}).

Notably, pruning works best when bounds are \textit{tight}: this means that a lower bound should be the weakest (most frequently true) sufficient condition, and the upper bound should be the strongest (least frequently true) necessary condition. These correspond to scanning as often as possible and pruning as often as possible, respectively. This is the ideal goal of generating such bounds, but we make no guarantees that our algorithm derives the weakest and strongest bounds (notably, for some correlated expressions, e.g., $x - x$, interval analysis is known to produce non-tight bounds).

We first describe our notation, then provide minor background on scalar interval analysis, and lastly illustrate our extension to handle geometric predicates such as intersection and containment.



\subsection{Notation and Terminology}

We denote the lower bound of an expression $E$ (either boolean or numeric) as $\lfloor E \rfloor$, and the upper bound as $\lceil E \rceil$. Upper bounds and lower bounds are equivalent to \code{always} and \code{maybe} by:
\[
    \code{always}(E) = \lfloor E \rfloor
    \qquad
    \code{maybe}(E) = \lceil E \rceil
\]

In interval analysis, it is important to note the difference between \textit{varying} parameters and \textit{uniform} parameters: \textit{varying} parameters are values in a predicate that can take multiple values, bounded by either an interval (in the scalar case) or a volume (in the geometric case); \textit{uniform} parameters are constant with respect to the queried data. For example, in a standard range query that searches for all $x$ such that $low \leq x \leq high$, $x$ is a varying parameter and $low$ and $high$ are uniform parameters. Interval analysis rules (including ours in \Cref{sec:geometric-analysis}) are frequently defined differently depending on which operands of an expression are varying or uniform.

If an expression cannot be bounded, its bounds default to the limits of its type, e.g., $[\texttt{false}, \texttt{true}]$ for boolean expressions, $[0, \texttt{UINTMAX}]$ for unsigned integers, and $[-\infty, \infty]$ for floats.

\subsection{Background: Scalar Interval Analysis}
\label{sec:scalar-interval}

Symbolic interval analysis derives bounds recursively in a bottom-up traversal of the predicate's AST. Varying parameters are replaced with their intervals, and operators are evaluated on the intervals themselves by considering the monotonicity of an operator. We illustrate the reasoning behind comparisons and boolean combinators here; additional operators are described in Appendix~\ref{sec:appendix-interval-analysis}. 

\paragraph{Comparisons} The comparison of two numbers can be bounded by a comparison of the ranges that bound each number. Consider the expression $x < y$: if $x$ and $y$ are varying, then the expression $x < y$ \textit{may} be true if $x$'s lower bound is less than $y$'s upper bound (otherwise, all values that $y$ can be are less than all values that $x$ can be). Conversely, $x < y$ is \textit{always} true if the upper bound of $x$ is less than the lower bound of $y$. This reasoning produces the following bounds:
\[
    \lceil x < y \rceil \mapsto \lfloor x \rfloor < \lceil y \rceil
    \qquad
    \lfloor x < y \rfloor \mapsto \lceil x \rceil < \lfloor y \rfloor
\]
The $\leq$ operator has the same monotonicity of operands as $<$, and can be bounded in the same way.
Likewise, similar reasoning applies to equality, though the lower bound requires that the intervals of the arguments each contain a single value:
\[
    \lceil x = y \rceil \mapsto \lfloor x \rfloor \leq \lceil y \rceil \land \lfloor y \rfloor \leq \lceil x \rceil
    \qquad
    \lfloor x = y \rfloor \mapsto \lfloor x \rfloor = \lfloor y \rfloor \land \lceil x \rceil = \lceil y \rceil \land \lfloor x \rfloor = \lceil x\rceil
\]

\paragraph{Boolean combinators} Boolean \textit{and}, $\land$, and boolean \textit{or}, $\lor$, are both monotonically increasing in their arguments, and are therefore bounded by the bounds of their arguments. Boolean \textit{negation}, $\neg$, is monotonically decreasing, and therefore is upper bounded by the negation of the lower bound of its argument, and lower bounded by the negation of the argument's upper bound.
\[
    \footnotesize
    \lceil a \land b \rceil \mapsto \lceil a \rceil \land \lceil b \rceil
    \qquad
    \lceil a \lor b \rceil \mapsto \lceil a \rceil \lor \lceil b \rceil
    \qquad
    \lceil \neg a \rceil \mapsto \neg \lfloor a \rfloor
\]
\[
    \footnotesize
    \lfloor a \land b \rfloor \mapsto \lfloor a \rfloor \land \lfloor b \rfloor
    \qquad
    \lfloor a \lor b \rfloor \mapsto \lfloor a \rfloor \lor \lfloor b \rfloor
    \qquad
    \lfloor \neg a \rfloor \mapsto \neg \lceil a \rceil
\]

While such bounds are well-established~\cite{snyder1992interval}, applying them to generate tree-pruning functions is, to our knowledge, novel. We further show that this reasoning naturally extends to spatial operators.

\subsection{Geometric Bounds}
\label{sec:geometric-analysis}

In the same way that scalar boolean operators are bounded by their necessary and sufficient conditions parameterized by their bounding intervals, geometric boolean operators can be bounded by necessary and sufficient conditions parameterized by their bounding volumes.

While this analogy is conceptually straightforward, applying scalar interval analysis directly to implementations of geometric predicates such as \code{intersects} or \code{contains} is ineffective in practice. Such implementations are typically hundreds of lines of specialized geometric code~\cite{pbrt4}, and interval propagation through this code rarely exposes the high-level spatial relationships necessary for pruning. Instead, we aim to derive these bounds directly from the semantics of each predicate.

Each geometric predicate in ~\Cref{fig:query-grammar} is binary, and can be analyzed by considering three cases: when the first argument is varying (contained by a bounding volume) and the second is uniform, when the first argument is uniform and the second is varying, and when both arguments are varying. Note that symmetric operators like \code{intersects} only have two cases, and in the case that both arguments are uniform, the expression itself is uniform and is a singular-valued interval.

An upper bound is implied by the predicate, and a lower bound implies the predicate. Thus, the upper bound and lower bound of a predicate $P$ satisfy:
\[
    P(X) \land \code{bounded}(X) \rightarrow \lceil P \rceil
    \qquad
    \lfloor P \rfloor \land \code{bounded}(X) \rightarrow P(X)
\]
where \code{bounded}$(X)$ asserts that all varying parameters are bounded by their respective bounding volumes, and both the upper and lower bounds refer only to the uniform parameters of $P$ and the bounding volumes of the varying parameters.

As a guiding example, consider searching for all objects contained within a query sphere.
If a subtree is ever fully contained within the query sphere, then all objects in the subtree must be contained within the query sphere. Alternatively, if the subtree's bounding volume intersects the query sphere, it is possible that some objects in the subtree may be contained within the query sphere. These give rise to the bounds on the containment predicate with a uniform first parameter, $u$ (\code{sphere}), and a varying second parameter, $v$ bounded by a bounding volume $V_v$:
\[
    \lceil \code{contains}(u, v) \rceil \mapsto \code{intersects}(u, V_v)
    \qquad
    \lfloor \code{contains}(u, v) \rfloor \mapsto \code{contains}(u, V_v)
\]

When the first argument is varying and the second is uniform (e.g., in a query which searches for all objects that \textit{contain} a query object), the predicate can be upper-bounded, but a bounding volume is not enough to prove objects in a bounding volume \textit{always} contain a query object, so there is no lower bound based on this augmentation.
\[
    \lceil \code{contains}(v, u) \rceil \mapsto \code{intersects}(V_v, u)
    \qquad
    \lfloor \code{contains}(v, u) \rfloor \mapsto \code{false}
\]

Likewise, when both arguments are varying, there is no lower bound, and the upper bound is simply the intersection of the two bounding volumes (if the bounding volumes do not intersect, it is impossible for any object in one to contain any object in the other).

We generally see this pattern in geometric predicates: almost all can be upper bounded, which means they benefit from pruning subtrees where the predicate can never be true, but many do not have lower bounds, so cannot often be proven to be \textit{always} true in a given subtree. We provide the rest of our lower bound and upper bound rules for geometric predicates in \Cref{alg:geom-upper} and \Cref{alg:geom-lower-metric} below. Note that distances and ordering predicates are monotonic functions, and therefore are bounded by simply replacing any varying arguments with their bounding volumes. $u$ denotes uniform geometric values, $v$, $v_0$, and $v_1$ denote varying geometric values with corresponding bounding volumes $V_v$, $V_0$, and $V_1$. Ordering relationships (e.g., $\leq$ and $<$) are monotonic and are therefore bounded by rewriting varying arguments to their bounding volumes (not included for brevity). Metric relationships, \code{distmin} and \code{distmax}, are always lower bounded by \code{distmin} and upper bounded by \code{distmax} applied to the same arguments but replacing varying arguments with their bounding volumes (uniform parameters are bounded by themselves for the sake of brevity).

\vspace{-1.5em}
\noindent\begin{minipage}[t]{0.48\textwidth}
\begin{algorithm}[H]
\caption{Geometric Upper Bounds}
\tiny
\label{alg:geom-upper}
\begin{algorithmic}[1]
\State \textbf{Input}: Geometric predicate $E$
\State \textbf{Output}: Upper bound of $E$
\State \textbf{function} $\lceil E \rceil$
\State \hspace{0.1em} \textbf{match} $E$ \textbf{with}
\State \hspace{0.5em} \texttt{|} \code{contains(}$u$, $v$\code{)} $\mapsto$ \code{intersects(}$u$, $V_v$\code{)}
\State \hspace{0.5em} \texttt{|} \code{contains(}$v$, $u$\code{)} $\mapsto$ \code{intersects(}$V_v$, $u$\code{)}
\State \hspace{0.5em} \texttt{|} \code{contains(}$v_0$, $v_1$\code{)} $\mapsto$ \code{intersects(}$V_0$, $V_1$\code{)}

\State \hspace{0.5em} \texttt{|} \code{covers(}$u$, $v$\code{)} $\mapsto$ \code{intersects(}$u$, $V_v$\code{)}
\State \hspace{0.5em} \texttt{|} \code{covers(}$v$, $u$\code{)} $\mapsto$ \code{covers(}$V_v$, $u$\code{)}
\State \hspace{0.15em} \texttt{|} \code{covers(}$v_0$, $v_1$\code{)} $\mapsto$ \code{intersects(}$V_0$, $V_1$\code{)}

\State \hspace{0.15em} \texttt{|} \code{disjoint(}$u$, $v$\code{)} $\mapsto$ $\neg$\code{contains(}$u$, $V_v$\code{)}
\State \hspace{0.15em} \texttt{|} \code{disjoint(}$v$, $u$\code{)} $\mapsto$ $\neg$\code{within(}$V_v$, $u$\code{)}

\State \hspace{0.15em} \texttt{|} \code{within(}$u$, $v$\code{)} $\mapsto$ \code{within(}$u$, $V_v$\code{)}
\State \hspace{0.15em} \texttt{|} \code{within(}$v$, $u$\code{)} $\mapsto$ \code{intersects(}$V_v$, $u$\code{)}
\State \hspace{0.15em} \texttt{|} \code{within(}$v_0$, $v_1$\code{)} $\mapsto$ \code{intersects(}$V_0$, $V_1$\code{)}

\State \hspace{0.15em} \texttt{|} \code{equals(}$u$, $v$\code{)} $\mapsto$ \code{within(}$u$, $V_v$\code{)}
\State \hspace{0.15em} \texttt{|} \code{equals(}$v$, $u$\code{)} $\mapsto$ \code{contains(}$V_v$, $u$\code{)}
\State \hspace{0.15em} \texttt{|} \code{equals(}$v_0$, $v_1$\code{)} $\mapsto$ \code{intersects(}$V_0$, $V_1$\code{)}

\State \hspace{0.15em} \texttt{|} \code{intersects(}$u$, $v$\code{)} $\mapsto$ \code{intersects(}$u$, $V_v$\code{)}
\State \hspace{0.15em} \texttt{|} \code{intersects(}$v$, $u$\code{)} $\mapsto$ \code{intersects(}$V_v$, $u$\code{)}
\State \hspace{0.15em} \texttt{|} \code{intersects(}$v_0$, $v_1$\code{)} $\mapsto$ \code{intersects(}$V_0$, $V_1$\code{)}

\State \hspace{0.15em} \texttt{|} \code{touches(}$u$, $v$\code{)} $\mapsto$ \code{intersects(}$u$, $V_v$\code{)}
\State \hspace{0.15em} \texttt{|} \code{touches(}$v$, $u$\code{)} $\mapsto$ \code{intersects(}$V_v$, $u$\code{)}
\State \hspace{0.15em} \texttt{|} \code{touches(}$v_0$, $v_1$\code{)} $\mapsto$ \code{intersects(}$V_0$, $V_1$\code{)}

\State \hspace{0.15em} \texttt{|} \code{_} $\mapsto$ \code{true}
 
\State \textbf{end function}
\end{algorithmic}
\end{algorithm}
\end{minipage}
\hfill
\begin{minipage}[t]{0.48\textwidth}
\begin{algorithm}[H]
\caption{Lower Bounds and Metric Bounds}
\tiny
\label{alg:geom-lower-metric}
\begin{algorithmic}[1]
\State \textbf{Input}: Geometric predicate $E$
\State \textbf{Output}: Lower bound of $E$
\State \textbf{function} $\lfloor E \rfloor$
\State \hspace{0.1em} \textbf{match} $E$ \textbf{with}
\State \hspace{0.5em} \texttt{|} \code{contains(}$u$, $v$\code{)} $\mapsto$ \code{contains(}$u$, $V_v$\code{)}

\State \hspace{0.5em} \texttt{|} \code{covers(}$u$, $v$\code{)} $\mapsto$ \code{covers(}$u$, $V_v$\code{)}

\State \hspace{0.5em} \texttt{|} \code{disjoint(}$u$, $v$\code{)} $\mapsto$ \code{disjoint(}$u$, $V_v$\code{)}
\State \hspace{0.5em} \texttt{|} \code{disjoint(}$v$, $u$\code{)} $\mapsto$ \code{disjoint(}$V_v$, $u$\code{)}
\State \hspace{0.5em} \texttt{|} \code{disjoint(}$v_0$, $v_1$\code{)} $\mapsto$ \code{disjoint(}$V_0$, $V_1$\code{)}

\State \hspace{0.15em} \texttt{|} \code{intersects(}$u$, $v$\code{)} $\mapsto$ \code{contains(}$u$, $V_v$\code{)}
\State \hspace{0.15em} \texttt{|} \code{intersects(}$v$, $u$\code{)} $\mapsto$ \code{within(}$V_v$, $u$\code{)}

\State \hspace{0.15em} \texttt{|} \code{within(}$v$, $u$\code{)} $\mapsto$ \code{within(}$V_v$, $u$\code{)}

\State \hspace{0.15em} \texttt{|} \code{_} $\mapsto$ \code{false}

\State \textbf{end function}

\State \textbf{Input}: Geometric metric $E$ $\qquad$ \textbf{Output}: Lower bound on $E$
\State \textbf{function} $\lfloor E \rfloor$
\State \hspace{0.1em} \textbf{match} $E$ \textbf{with}
\State \hspace{0.5em} \texttt{|} \code{distmin(}$v_0$, $v_1$\code{)} $\mapsto$ \code{distmin(}$V_0$, $V_1$\code{)}
\State \hspace{0.5em} \texttt{|} \code{distmax(}$v_0$, $v_1$\code{)} $\mapsto$ \code{distmin(}$V_0$, $V_1$\code{)}
\State \textbf{end function}

\State \textbf{Input}: Geometric metric $E$ $\qquad$ \textbf{Output}: Upper bound on $E$
\State \textbf{function} $\lceil E \rceil$
\State \hspace{0.1em} \textbf{match} $E$ \textbf{with}
\State \hspace{0.5em} \texttt{|} \code{distmin(}$v_0$, $v_1$\code{)} $\mapsto$ \code{distmax(}$V_0$, $V_1$\code{)}
\State \hspace{0.5em} \texttt{|} \code{distmax(}$v_0$, $v_1$\code{)} $\mapsto$ \code{distmax(}$V_0$, $V_1$\code{)}
\State \textbf{end function}

\end{algorithmic}
\end{algorithm}
\end{minipage}



\section{Joins as Tree Traversals}
\label{sec:joins}

The previous two sections introduced the machinery required to compile filters and reductions over sets. Now we show how this machinery can also be used to generate efficient code for non-equijoins, including spatial joins. The key insight is that a join predicate can be analyzed like a filter predicate with multiple varying parameters, enabling pruning of both sides of the join.

Traditional database systems implement joins using strategies like hash join (build a hash table on one side, probe from the other) or sort-merge join (sort both sides, then merge). These strategies work well for equijoins (equality predicates), but generally do not extend to non-equijoins with arbitrary predicates, particularly geometric predicates like intersection or containment. Non-equijoins are often lowered to a \textit{nested join}, i.e., quadratic enumeration of all pairs followed by predicate evaluation.

Spatial databases~\cite{guttman1984rtree} and collision detection algorithms~\cite{ericson2004rtcd} have long used tree-based indexes to accelerate spatial queries. Our compilation approach generalizes these techniques to arbitrary predicates by treating join predicates the same way we treat filter predicates: analyzing them to generate pruning conditions and compile efficient tree traversals. We extend two well-known strategies: a single-index join analogous to a hash join, and a dual-index join analogous to sort-merge join and inspired by dual-tree traversals in collision detection and spatial joins.

\subsection{Single-Index Join}

A straightforward non-equijoin implementation with our techniques is an iterate-locate pattern akin to hash-join: a tree (index) is built on one side; the other side iterates through its elements, locating values in the tree that satisfy the join predicate. In functional notation, this is simply:
\begin{lstlisting}[language=lang, mathescape=true, basicstyle=\scriptsize\ttfamily, backgroundcolor=\color{codebackground}, frame=single,
xleftmargin=3.4pt,
xrightmargin=3.4pt
]
single(set0 : Set<T>, set1 : Set<S>) = map(|t : T| (t, filter(|s : S| $P$(t, s), set1)), set0);
\end{lstlisting}
This expression iterates over each element \code{t} in \code{set0} (the outer loop), and for each \code{t}, filters \code{set1} to find all \code{s} that satisfy the join predicate $P$\code{(t, s)}. Note that for the purpose of predicate analysis, \code{t} is a uniform parameter and \code{s} is a varying parameter.


\textit{Output.}
The return type of this function is a set of tuples where each tuple contains an element of type \code{T} and a \textit{set} of elements of type \code{S}: \code{Set<(T, Set<S>)>}. This represents an implicit groupby operation, grouping matching \code{S} elements by their corresponding \code{T} element. If the desired output is a flat list of pairs \code{Set<(T, S)>}, a final flattening step is required.

\textit{Performance Characteristics}
When $n$ is the size of the outer set and $m$ is the size of the inner set, the worst-case runtime is $\Theta(n \cdot m)$, when no (or limited) pruning is possible, and the tree traversal takes linear time. For highly-selective filters that turn the filter logarithmic (e.g., a range predicate), the complexity is $O(n \cdot \log|m| + m \cdot \log|m|)$, where the first term corresponds to $n$ tree traversals and the second term corresponds to the tree build complexity (assuming standard tree construction algorithms~\cite{kay1986}). As in databases, the choice of which side to index can significantly impact performance: indexing the smaller set minimizes build and materialization costs, but indexing the set with better spatial locality may enable more effective pruning. The outer loop is also trivially parallelizable.

\subsection{Dual-Index Join}

In some cases, we can further improve performance with a coiterative join strategy akin to sort-merge join, inspired by the dual-tree traversals in collision detection~\cite{ericson2004rtcd} and spatial joins~\cite{guttman1984rtree}. By building trees on \textit{both} sides of the join, we can prune subtrees from both input sets simultaneously. Conceptually, a dual-index join computes a filtered Cartesian product:
\begin{lstlisting}[language=lang, mathescape=true, basicstyle=\scriptsize\ttfamily, backgroundcolor=\color{codebackground}, frame=single,
xleftmargin=3.4pt,
xrightmargin=3.4pt
]
dual(set0 : Set<T>, set1 : Set<S>) = filter(|t : T, s : S| $P$(t, s), product(set0, set1));
\end{lstlisting}

\begin{wrapfigure}[23]{r}{0.45\textwidth} 
\vspace{-2.5em} 
\begin{minipage}{0.45\textwidth}
\begin{algorithm}[H]
\caption{Product Lowering}
\tiny
\label{alg:lower-product}
\begin{algorithmic}[1]
\State \textbf{Input}: Set expressions $S_0$ and $S_1$
\State \textbf{Output}: TTIR that iterates the product of the two sets
\Function{LowerProduct}{$S_0$, $S_1$}
    \State \textbf{rewrite} \textsc{Lower}($S_0$) \textbf{with}

    \State \texttt{|} \code{yield x} $\Rightarrow$
    \State \hspace{1em} \textbf{rewrite} \textsc{Lower}($S_1$) \textbf{with}
    \State \hspace{1em} \texttt{|} \code{yield y} $\Rightarrow$ \code{yield (x, y)}
    \State \hspace{1em} \texttt{|} \code{iter ys} $\Rightarrow$ \code{iter map(}|\code{y}| \code{(x, y), ys)}
    \State \hspace{1em} \texttt{|} \code{scan t1} $\Rightarrow$ \code{scan \{x\} t1}
    \State \hspace{0.65em} \texttt{|} \code{from t1} $\Rightarrow$ \code{from \{x\} t1}

    \State \texttt{|} \code{iter xs} $\Rightarrow$
    \State \hspace{1em} \textbf{rewrite} \textsc{Lower}($S_1$) \textbf{with}
    \State \hspace{1em} \texttt{|} \code{yield y} $\Rightarrow$ \code{iter map(}|\code{x}| \code{(x, y), xs)}
    \State \hspace{1em} \texttt{|} \code{iter ys} $\Rightarrow$ \code{iter product(xs, ys)}
    \State \hspace{1em} \texttt{|} \code{scan t1} $\Rightarrow$ \code{scan \{xs\} t1}
    \State \hspace{1em} \texttt{|} \code{from t1} $\Rightarrow$ \code{from \{xs\} t1}

    \State \texttt{|} \code{scan t0} $\Rightarrow$
    \State \hspace{1em} \textbf{rewrite} \textsc{Lower}($S_1$) \textbf{with}
    \State \hspace{1em} \texttt{|} \code{yield y} $\Rightarrow$ \code{scan t0 \{y\}}
    \State \hspace{1em} \texttt{|} \code{iter ys} $\Rightarrow$  \code{scan t0 \{ys\}}
    \State \hspace{1em} \texttt{|} \code{scan t1} $\Rightarrow$ \code{scan t0 t1}
    \State \hspace{1em} \texttt{|} \code{from t1} $\Rightarrow$ \code{from t0 t1}

    \State \texttt{|} \code{from t0} $\Rightarrow$
    \State \hspace{1em} \textbf{rewrite} \textsc{Lower}($S_1$) \textbf{with}
    \State \hspace{1em} \texttt{|} \code{yield y} $\Rightarrow$ \code{from t0 \{y\}}
    \State \hspace{1em} \texttt{|} \code{iter ys} $\Rightarrow$  \code{from t0 \{ys\}}
    \State \hspace{1em} \texttt{|} \code{scan t1} $\Rightarrow$ \code{from t0 t1}
    \State \hspace{1em} \texttt{|} \code{from t1} $\Rightarrow$ \code{from t0 t1}
\EndFunction
\end{algorithmic}
\end{algorithm}
\end{minipage}
\end{wrapfigure}

\textit{Algorithm.}
The dual tree traversal computes the filtered Cartesian product by recursing on a pair of nodes, one from each side of the join, and checking if the join predicate can be true in those two subsets. Abstractly, the traversal is lowered to the traversal in \Cref{fig:dual-abstract}, which is the result of applying the lowering rules for \code{product}, provided in \Cref{alg:lower-product}, composed with the \code{filter} lowering in \Cref{alg:lower-filter-min}, onto a simple binary tree with \code{Leaf} and \code{Interior} variants.

\textit{Example.} For collision detection (where the filter predicate is \code{intersects}), this abstract code is lowered to \Cref{fig:dual-intersects}, which matches the standard dual tree collision detection algorithms~\cite{ericson2004rtcd}.

\begin{figure}[t]
  \centering
  \begin{subfigure}[t]{0.48\textwidth}
    \centering
\begin{lstlisting}[language=lang, mathescape=true, basicstyle=\scriptsize\ttfamily, backgroundcolor=\color{codebackground}, frame=single,
xleftmargin=3.4pt
]
func dual_traversal(t0 : Tree, t1 : Tree) =
 match t0, t1 with
 | Leaf(d0), Leaf(d1) $\shortrightarrow$
  if $P$(d0, d1): yield (d0, d1)
 | Leaf(d0), Interior(l, r) $\shortrightarrow$
  if always($P$, d0, t1): scan t0, t1
  elif maybe($P$, d0, t1): from t0, t1
  | Interior(l, r), Leaf(d1) $\shortrightarrow$
  if always($P$, t0, d1): scan t0, t1
  elif maybe($P$, t0, d1): from t0, t1
  | Interior(l0, r0), Interior(l1, r1) $\shortrightarrow$
  if always($P$, t0, t1): scan t0, t1
  elif maybe($P$, t0, t1): from t0, t1
\end{lstlisting}
    \caption{Generic dual traversal join of two trees under predicate $P$. \code{scan} and \code{from} recurse on each pair of children (the product of children).
    }
    \label{fig:dual-abstract}
  \end{subfigure}
  \hfill
  \begin{subfigure}[t]{0.48\textwidth}
    \centering
\begin{lstlisting}[language=lang, mathescape=true, basicstyle=\scriptsize\ttfamily, backgroundcolor=\color{codebackground}, frame=single,
xrightmargin=3.4pt
]
func collisions(t0 : Tree, t1 : Tree) =
 match t0, t1 with
 | Leaf(d0), Leaf(d1) $\shortrightarrow$
  if intersects(d0, d1): yield (d0, d1)
 | Leaf(d0), Interior(l, r) $\shortrightarrow$
  if contains(d0, t1): scan t0, t1
  elif intersects(d0, t1): from t0, t1
  | Interior(l, r), Leaf(d1) $\shortrightarrow$
  if contains(d1, t0): scan t0, t1
  elif intersects(t0, d1): from t0, t1
  | Interior(l0, r0), Interior(l1, r1) $\shortrightarrow$
  if intersects(t0, t1): from t0, t1
$\vspace{1em}$
\end{lstlisting}
    \caption{Dual tree traversal with an \code{intersects} join predicate. When both arguments vary, no sufficient condition exists, so the final \code{always} is false.}
    \label{fig:dual-intersects}
  \end{subfigure}

  \caption{Dual-tree traversal specialization. The generic dual traversal \textbf{(a)} lowers to collision detection \textbf{(b)} by instantiating predicate $P$ as \code{intersects}. \code{always} and \code{maybe} specialize to \code{contains} and \code{intersects}.}
  \label{fig:dual}
\end{figure}

\textit{Output.}
The return type is a set of tuple pairs.

\textit{Performance Characteristics.}
Let $n$ and $m$ be the sizes of the outer and inner sets. The worst-case runtime is $\Theta(n \cdot m)$ when no pruning occurs. The runtime for this class of algorithms depends heavily on the predicate and the data distributions.
For example, if the root nodes are disjoint w.r.t the query predicate, the traversal terminates immediately in $O(1)$ time.
In general, it is not possible to give a meaningful asymptotic bound without a specific predicate or data distribution, but runtime is always lower-bounded by construction of both trees: $\Omega(n \cdot \log|n| + m \cdot \log|m|)$.
Parallelization is nontrivial (there is no trivially parallelizable outer loop), but the recursive traversal can be parallelized via work stealing~\cite{blumofe1999workstealing}.

\section{Code Generation and Implementation}
\label{sec:codegen}

\sys compiles to C++.
Tree layouts are specified using \textsc{Scion}~\cite{gyurgyik2026scion} to provide compact layouts for self-comparisons and match other systems in cross-system comparisons in ~\Cref{sec:eval}.

\textit{Outputs.}
Every \sys tree traversal produces some accumulated value, the output type. Every generated function has a reference to an accumulator of this type as the output parameter.

\textit{Generating sets.}
If the output type of a query is a set, then the accumulator is our custom C++ \code{Set<T>} type: \code{yield}s become appends and \code{iter}s become grouped appends.

\textit{Producing scalars.}
If the output type of a query is a scalar (e.g., \code{min}), then the accumulator is just the C++ version of the scalar type, and \code{upd} is lowered to a mutation.

\textit{Lowering \code{from}.}
\code{from} is always lowered into recursive calls applied to all children of the argument type. For \code{from}s applied to multiple parameters (from lowering a \code{product}), these are lowered to the Cartesian product of children nodes, as in standard dual tree traversals~\cite{ericson2004rtcd}.

\textit{Lowering \code{scan}.}
Every scan is lowered into the standard tree traversal on the base tree type, with any \code{map} function applied to the leaf primitives, and any reduction operator lowered last. Note that in the case of multiple tree parameters (from lowering a \code{product}), \code{scan}s again become Cartesian products, and aggregate functions are applied on the products.

\textit{Example.}
Consider lowering a collision detection join that counts\footnote{A \code{count} aggregate is a map that maps all elements to $1$ followed by a sum reduction.} the number of collisions:
\begin{lstlisting}[language=lang, mathescape=true, basicstyle=\scriptsize\ttfamily, backgroundcolor=\color{codebackground}, frame=single,
xleftmargin=3.4pt,
xrightmargin=3.4pt
]
f(s0 : Set<T>, s1 : Set<S>) = count(filter(|t : T, s : S| intersects(t, s), product(s0, s)));
\end{lstlisting}
\sys fuses the \code{count} into the traversal in \Cref{fig:dual-intersects}. The lowered C++ (\Cref{fig:codegen_example}) passes the accumulator as the final parameter to both traversal functions (querying, scanning). While specialized scans could be generated (e.g., when the first parameter is always a leaf), we avoid this to prevent a combinatorial number of specialized variants. Such specialization could be profitable.

\begin{figure}
  \centering
  \begin{minipage}[t]{0.48\textwidth}
    \centering
    \begin{subfigure}[t]{\textwidth}
      \centering
\begin{lstlisting}[language=ajcpp, mathescape=true, basicstyle=\scriptsize\ttfamily, keywordstyle=\color{blue}\bfseries, backgroundcolor=\color{codebackground}, frame=single,
xleftmargin=3.4pt
]
void f(Tree t0, Tree t1, u64 &c) {
 if (is_leaf(t0)) {
  if (is_leaf(t1)) {
    if (intersects(t0.prim, t1.prim)) c++;
  } else { // t1 is interior
   if (contains(t0.prim, t1.vol)) {
    f_scan(t0, t1, c);
   } else if (intersects(t0.prim, t1.vol)) {
    f(t0, t1.left, c);
    f(t0, t1.right, c);
   }
  }
 } else { // t0 is interior
  if (is_leaf(t1)) {
   // flip the t0 leaf t1 interior case above
  } else { // t1 is interior
  if (intersects(t0.vol, t1.vol)) {
   f(t0.left, t1.left, c);
   f(t0.right, t1.left, c);
   f(t0.left, t1.right, c);
   f(t0.right, t1.right, c); }}}}
\end{lstlisting}
      \caption{The count of collisions, lowered to C++.}
      \label{fig:codegen_count_cd}
    \end{subfigure}
  \end{minipage}
  \hfill
  \begin{minipage}[t]{0.48\textwidth}
    \centering
    \begin{subfigure}[t]{\textwidth}
      \centering
\begin{lstlisting}[language=ajcpp, mathescape=true, basicstyle=\scriptsize\ttfamily, keywordstyle=\color{blue}\bfseries, backgroundcolor=\color{codebackground}, frame=single,
xrightmargin=3.4pt
]
void f_scan(Tree t0, Tree t1, u64 &c) {
 if (is_leaf(t0)) {
  if (is_leaf(t1)) {
   c++;
  } else {
   // t1 is interior
   f_scan(t0, t1.left, c);
   f_scan(t0, t1.right, c);
  }
 } else {
  // t0 is interior
  if (is_leaf(t1)) {
   f_scan(t0.left, t1, c);
   f_scan(t0.right, t1, c);
  } else {
   // t1 is interior
   f_scan(t0.left, t1.left, c);
   f_scan(t0.right, t1.left, c);
   f_scan(t0.left, t1.right, c);
   f_scan(t0.right, t1.right, c);
}}}
\end{lstlisting}
      \caption{Lowered \code{scan} with a count accumulator.}
      \label{fig:codegen_count_cd_scan}
    \end{subfigure}
  \end{minipage}

  \caption{Fused, final lowered C++ of a coiterating two trees, and counting collisions over leaf and interior variants. The user supplies intersects/contains code (in C++ or in \sys's kernel language).}
  \label{fig:codegen_example}
\end{figure}

\section{Evaluation}
\label{sec:eval}

We evaluate two primary claims for both regular filters and our generalized join algorithms:
\begin{enumerate}
    \item \sys achieves pruning efficiency and runtime performance comparable to hand-written tree traversals; and
    \item \sys can generate pruned traversals that existing systems are missing, resulting in improved performance for certain queries.
\end{enumerate}
We also present an ablation study quantifying the impact of fusion on compound filter-reduction queries, and analyze how data distribution impacts the performance of different join strategies.

\subsection{Methodology}

We evaluate on an Intel Core i9-14900K (3.2 GHz, 24 cores) with 196 GB DDR4 RAM, 896 KiB of L1 data cache, 32 MiB of L2 cache, and 36 MiB of L3 cache.
Benchmarks run single-threaded to isolate asymptotic behavior and bound to performance cores via \texttt{numactl}.
Generated kernels are compiled with \texttt{clang++ 21.1.3}.
Each benchmark runs 7 times; we discard the fastest and slowest and report the mean of the remaining 5 runs. Runs timeout after 30s.



For graphics queries, we compare against Fastest Closest Points in the West (FCPW)~\cite{sawhney2021fcpw} and the Flexible Collision Library (FCL)~\cite{pan2012fcl}. 
For scalar queries, we compare to SQLite~\cite{sqlite} and DuckDB~\cite{raasveldt2019duckdb}, both configured to run entirely in memory. 
Graphics models are drawn from the McGuire archive~\cite{McGuire2017Data}; scalar benchmarks use synthetic uniformly random data, except the salary join, which follows \citet{khayyat2015iejoin}. 
All generated data uses a fixed seed (42) for reproducibility. 

Unless otherwise stated, trees are built using a standard recursive spatial median-split algorithm~\cite{kay1986}. Optimizing tree construction is not a focus of this work; reported runtimes for joins include both with and without tree build times for transparency.

\subsection{Compile Times}
Compilation times are interactive, i.e.,  1--5\,ms per query for all queries. Compiling the generated C++ code with \code{-O3} takes approximately 50\,ms per query. This could be reduced with lighter compiler optimizations or less template metaprogramming in our benchmarking framework.

\subsection{Comparison to Hand-Written Traversals}

\subsubsection{Graphics Queries}
To demonstrate that \sys's geometric pruning and lowering match state-of-the-art performance, we compare to three representative and optimized graphics queries: closest point queries, ray tracing, and collision detection. 
Closest point queries (\code{min} over \code{distmin}) and ray tracing (\code{argmin} on \code{filter}) use a single-index join, iterating over an array of points or rays and querying a tree built on the scene geometry (e.g., triangles). We directly copy FCPW's tree topology and layout for fair comparison. For collision detection, we compare our lowered \code{filter} of a \code{product} against FCL’s hand-written dual-index traversal, matching FCL's tree layout, and made a best effort at duplicating their build algorithm.

\begin{figure}[b]
  \vspace{-1em}
  \centering
  \begin{subfigure}[b]{0.32\textwidth}
    \centering
    \includegraphics[width=\textwidth]{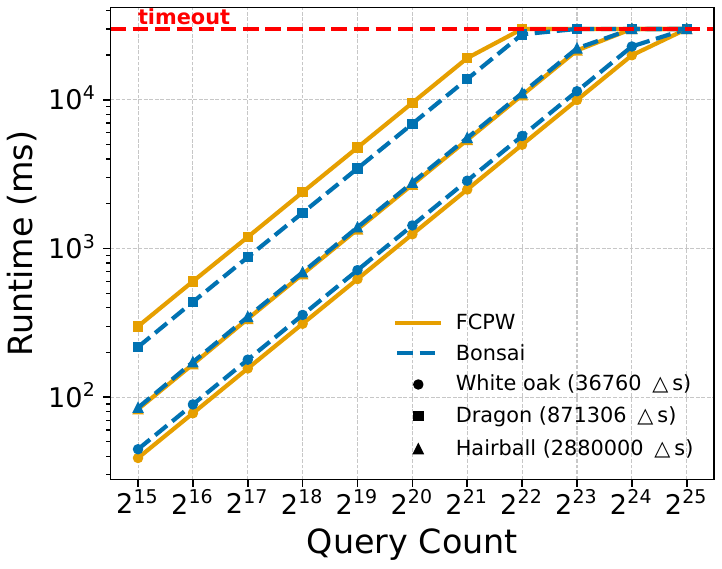}
    \caption{Closest point queries.}
    \label{fig:cp-speedup}
  \end{subfigure}
  \hfill
  \begin{subfigure}[b]{0.32\textwidth}
    \centering
    \includegraphics[width=\textwidth]{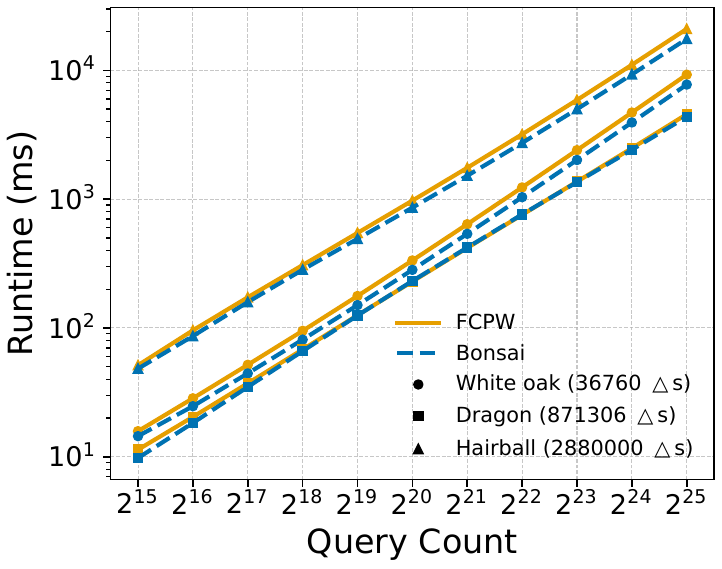}
    \caption{Ray tracing (primary rays).}
    \label{fig:rt-speedup}
  \end{subfigure}
  \hfill
  \begin{subfigure}[b]{0.32\textwidth}
    \centering
    \includegraphics[width=\textwidth]{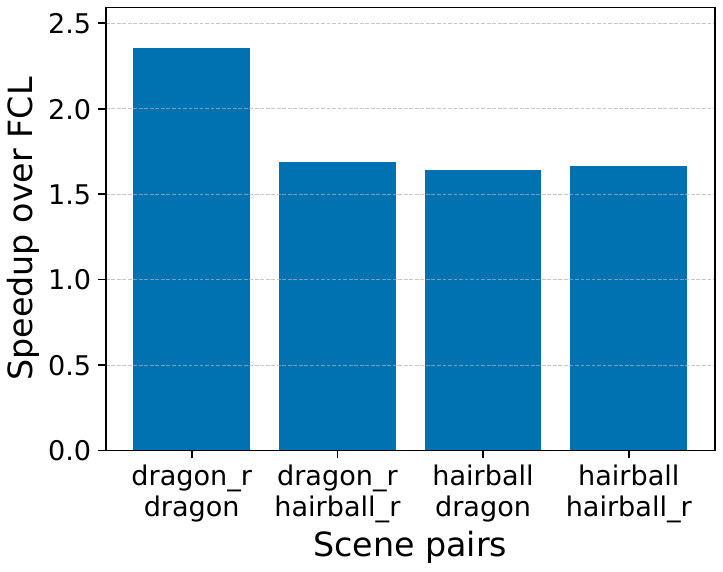}
    \caption{Collision detection.}
    \label{fig:cd-speedup}
  \end{subfigure}

  \caption{Comparison of \sys versus SotA closest point queries and ray tracing in FCPW~\cite{sawhney2021fcpw}, and collision detection versus FCL~\cite{pan2012fcl}. Lower is better for runtimes \textbf{(a)} and \textbf{(b)}, and higher is better for speedups \textbf{(c)}.}
  \label{fig:graphics-comparison}
\end{figure}

\textit{Closest point queries}:
\Cref{fig:cp-speedup} shows runtimes of \sys and FCPW~\cite{sawhney2021fcpw} closest point queries over three scenes and varying numbers of randomly generated query points within each scene’s bounding box.
\sys is on par with FCPW: it is on average (arithmetic mean) $0.87\times$ the throughput of FCPW on the White Oak scene; $1.38\times$ on the Dragon scene; and $0.97\times$ on the Hairball scene.
We see two differences between FCPW's code and \sys-generated code that could explain the performance differences: although FCPW has better vector instruction usage (via the Eigen library~\cite{eigenweb}), it records information beyond just the closest point to the query point. We believe the speedup on Dragon could be due to \sys's query specialization removing such metadata.

\textit{Closest hit ray tracing}: \Cref{fig:rt-speedup} compares \sys{} with FCPW on ray tracing performance. Again, \sys is on par with FCPW: the range of speedups for the White Oak scene is $1.09\times$--$1.19\times$ (avg. of $1.17\times$); for the Dragon scene: $0.99\times$--$1.16\times$ (avg. of $1.04\times$); and the Hairball scene: $1.06\times$--$1.19\times$ (avg. of $1.13\times$). As before, we believe any speedup comes from query specialization, as manual inspection confirms the tree traversals generated by \sys{} are identical to FCPW's hand-written traversals. Because ray-bounding box and ray-triangle intersection are less vectorizable than the point-box and point-triangle distance queries used for closest point queries, FCPW's improved vector instruction support offers less advantage here. However, extending \sys{} with improved vectorization support remains a valuable future direction.

\textit{Collision detection}: \Cref{fig:cd-speedup} shows a speedup plot of \sys versus FCL~\cite{pan2012fcl} across scene pairs from FCL's benchmarking suite. \sys-generated code is consistently faster---not due to improved pruning, but because FCL relies heavily on virtual function dispatch for geometry intersection and includes profiling hooks that cannot be disabled.
\sys achieves a $2.36\times$ speedup on dragon--dragon rotated scenes (8,688 collisions);
$1.69\times$ on dragon rotated--hairball rotated (48,238 collisions);
$1.64\times$ on hairball--dragon (123,055 collisions);
and $1.66\times$ on hairball--hairball rotated (5,118,441 collisions).
These results indicate that \sys matches the pruning efficiency of hand-written collision systems, while the speedup shows the benefit of specializing code at compile time instead of relying on runtime dynamic dispatch.

\subsubsection{Range and Inequality Joins}

\begin{figure}[t]
  \vspace{-1em}
  \centering
  \begin{subfigure}[b]{0.48\textwidth}
    \centering
    \includegraphics[width=\textwidth]{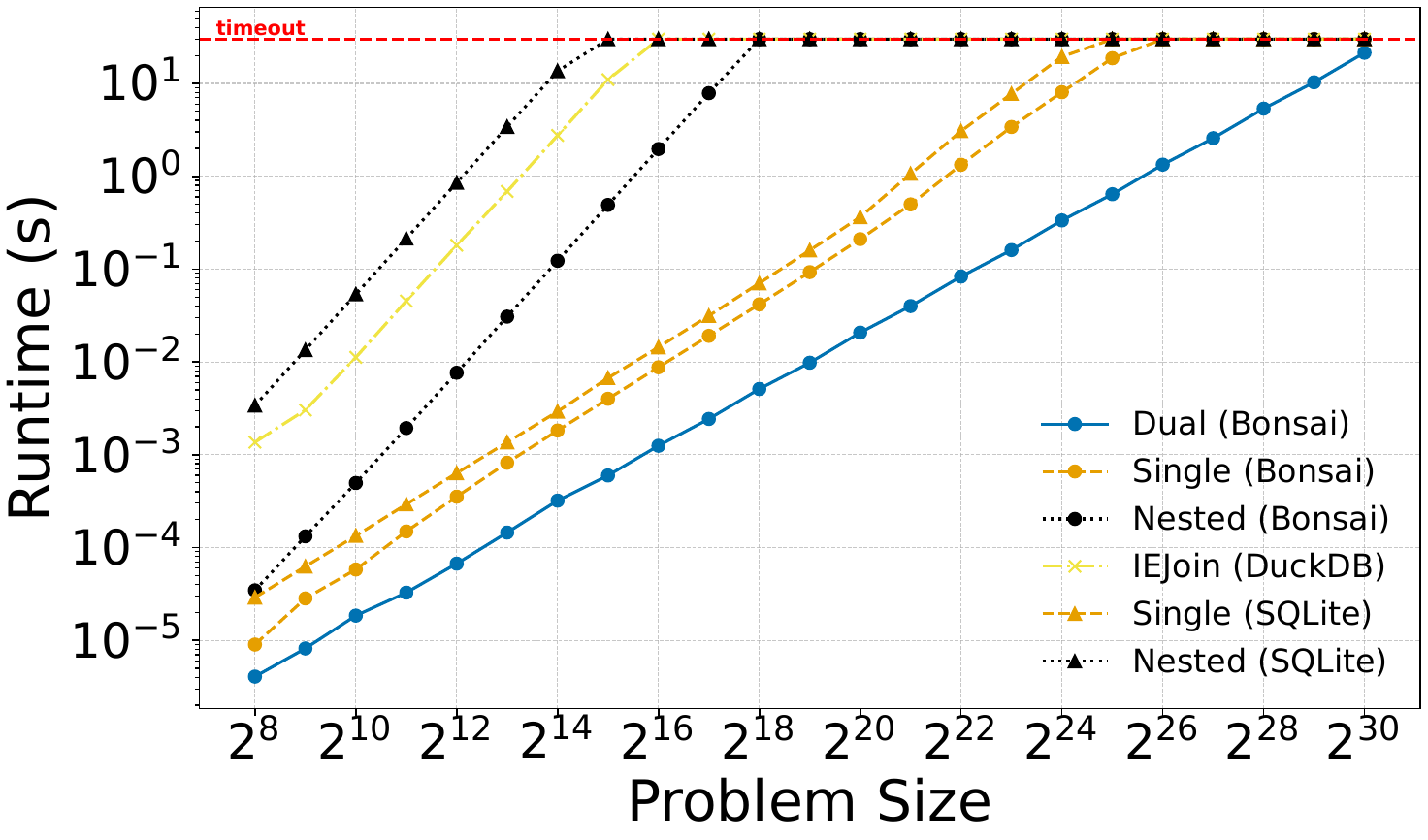}
    \caption{Range Join}
    \label{fig:chebyshev-sota}
  \end{subfigure}
  \hfill
  \begin{subfigure}[b]{0.48\textwidth}
    \centering
    \includegraphics[width=\textwidth]{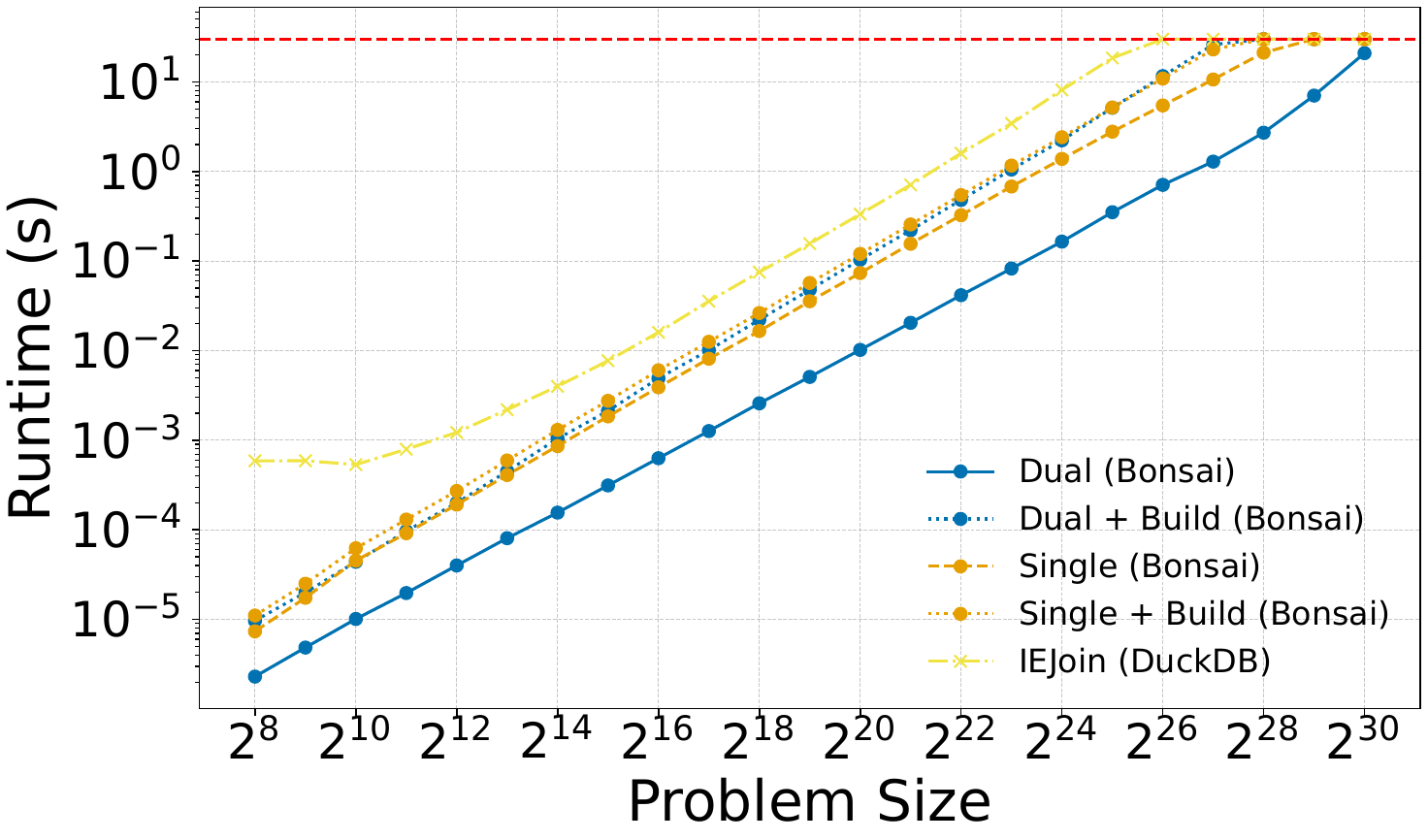}
    \caption{Salary Join}
    \label{fig:salary-sota}
  \end{subfigure}
  \vspace{-0.5em}

  \caption{Runtime comparisons with state of the art (SotA) database management systems. Lower is better. In \textbf{(a)}, we plot the runtimes of a 2D range join (join predicate $x_0 \in [x_1 - k, x_1 + k] \land y_0 \in [y_1 - k, y_1 + k]$). Without an index, SQLite produces a nested join, but with an index on $(x_1, y_1)$, it will perform a single-index join; DuckDB performs an IEJoin regardless of the presence of an index. \sys{}'s nested and single-index joins are on-par with SQLite, but the dual index join out-performs all of them. In \textbf{(b)} we directly compare to DuckDB's IEJoin on a benchmark from the original paper~\cite{khayyat2015iejoin}, counting the number of employees in a database who make less money but pay higher taxes than a peer. Even incorporating \sys{}'s unoptimized tree construction times, both \sys{} generated joins out-perform DuckDB's custom join.}
  \label{fig:sota-joins}

  \vspace{-1.5em}
\end{figure}

Relational DBMSs like SQLite~\cite{sqlite} use hand-optimized tree traversals (e.g., B-trees) to accelerate range queries of the form \texttt{x in [low, high]}. To evaluate \sys in this context, we test \textit{range joins}, where each row in one table performs a range query over the other.
\Cref{fig:chebyshev-sota} shows \sys’s single-index join matches SQLite’s native traversal, while its dual-index join is faster by leveraging both indexes. Our nested joins also outperform SQLite’s, confirming them as valid baselines. \sys{}'s gains stem from SQLite’s interpreter overhead.

There also exist specialized (non-tree-based) algorithms for \textit{inequality joins}~\cite{khayyat2015iejoin}, a class of open-ended range predicates. We evaluate \sys-generated code against DuckDB’s IEJoin on one of its benchmarks (\Cref{fig:salary-sota}), and find that it outperforms the native implementation, even when including index build time. Overall, these findings establish that \sys can reproduce the performance of expert-written join algorithms while generalizing to new join types, making it a strong foundation for exploring advanced query operators beyond current database capabilities.

\subsection{Comparison to Non-Pruning Code}

We evaluate \sys on queries for which state-of-the-art systems perform linear or quadratic scans. We include filters, reductions, and joins, highlighting cases where \sys generates pruned
\begin{wraptable}{r}{0.5\textwidth} 
    \vspace{-0.5em}
    \caption{Filter predicates used for our evaluation.}
    \vspace{-0.5em}
    \centering
    \tiny
    \begin{tabular}{l|ccccc}
    \toprule
    Query Predicate & Postgres & MySQL & DuckDB & SQLite & \sys \\
    \midrule
    $x \in [-10, 10]$ & \Index & \Index & \Linear & \Index & \Index \\
    $x = 42$ & \Index & \Index & \Index & \Index & \Index \\
    $|x| \leq 10$ & \Linear & \Linear & \Linear & \Linear & \Index \\
    $x^2 \leq 100$ & \Linear & \Linear & \Linear & \Linear & \Index \\
    $round(x) = 10$ & \Linear & \Linear & \Linear & \Linear & \Index \\
    $x^2 - 4x + 3 \leq 0$ & \Linear & \Linear & \Linear & \Linear & \Index \\
    $\sqrt{|x|} < \sqrt{10}$ & \Linear & \Linear & \Linear & \Linear & \Index \\
    $|x - u| > s$ & \Linear & \Linear & \Linear & \Linear & \Index \\
    \midrule
    $|x - y| < 1$ & \Linear & \Linear & \Linear & \Linear & \Index \\
    $|x| + |y| < 1$ & \Linear & \Linear & \Linear & \Linear & \Index \\
    $x^2 + y^2 < 10$ & \Linear & \Linear & \Linear & \Linear & \Index \\
    \bottomrule
    \end{tabular}
    \label{fig:linear-queries}
    \vspace{-3em}
\end{wraptable}
traversals that traditional systems do not map to tree queries.

\subsubsection{Filters and Reductions}

\Cref{fig:linear-queries} lists 12 linear queries that we use to highlight the asymptotic benefits of tree traversals. The first two queries (a range and a point query) are accelerated by most database systems we examined. None of the systems perform index scans for the remaining ten, despite seven being algebraically reducible to standard range queries that they already support.


\begin{figure}[t]
  \centering
  \begin{subfigure}[b]{0.48\textwidth}
    \centering
    \includegraphics[width=\textwidth]{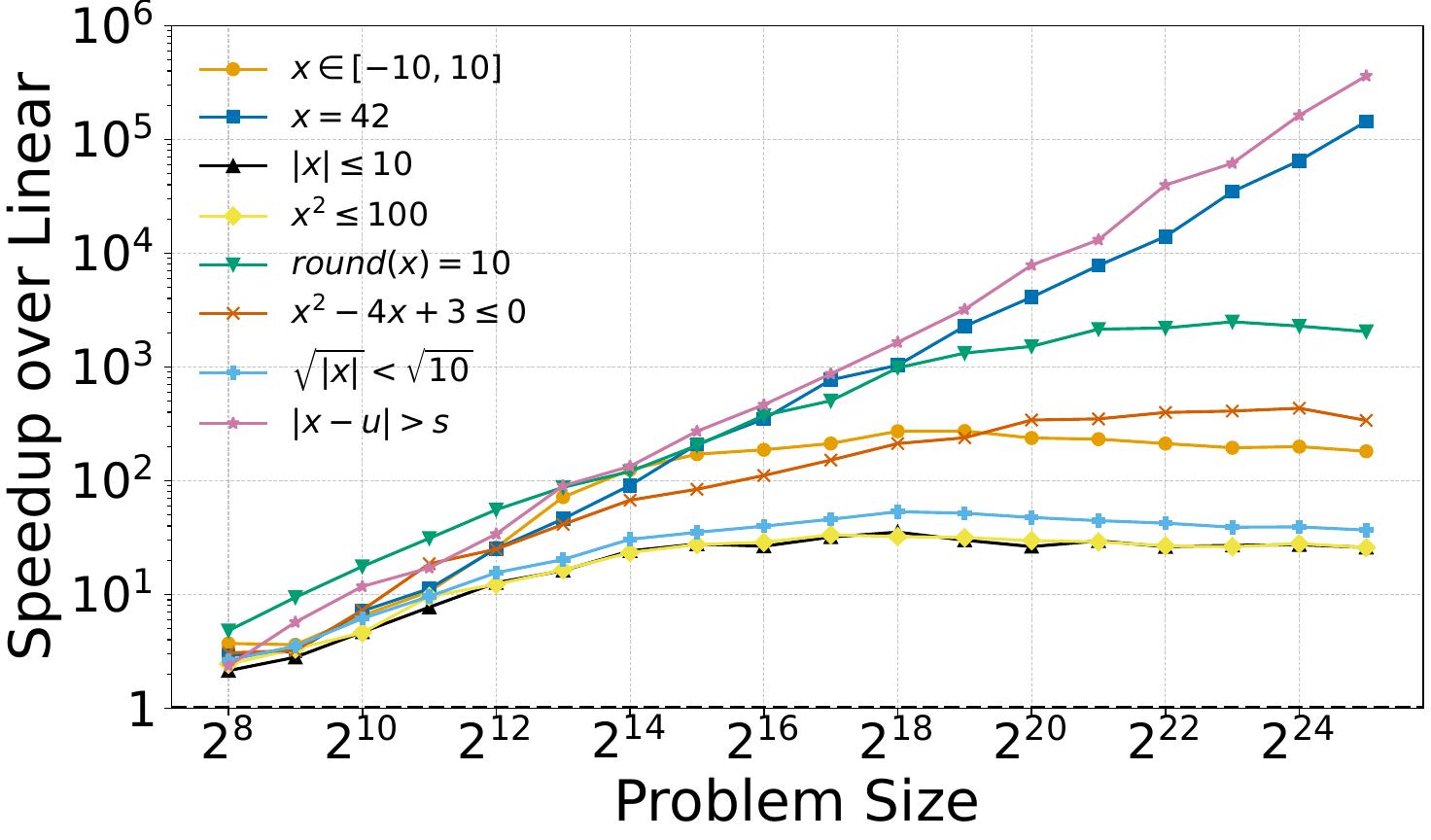}
    \caption{Accelerated filters on interval trees.}
    \label{fig:uniform-filters}
  \end{subfigure}
  \hfill
  \begin{subfigure}[b]{0.48\textwidth}
    \centering
    \includegraphics[width=\textwidth]{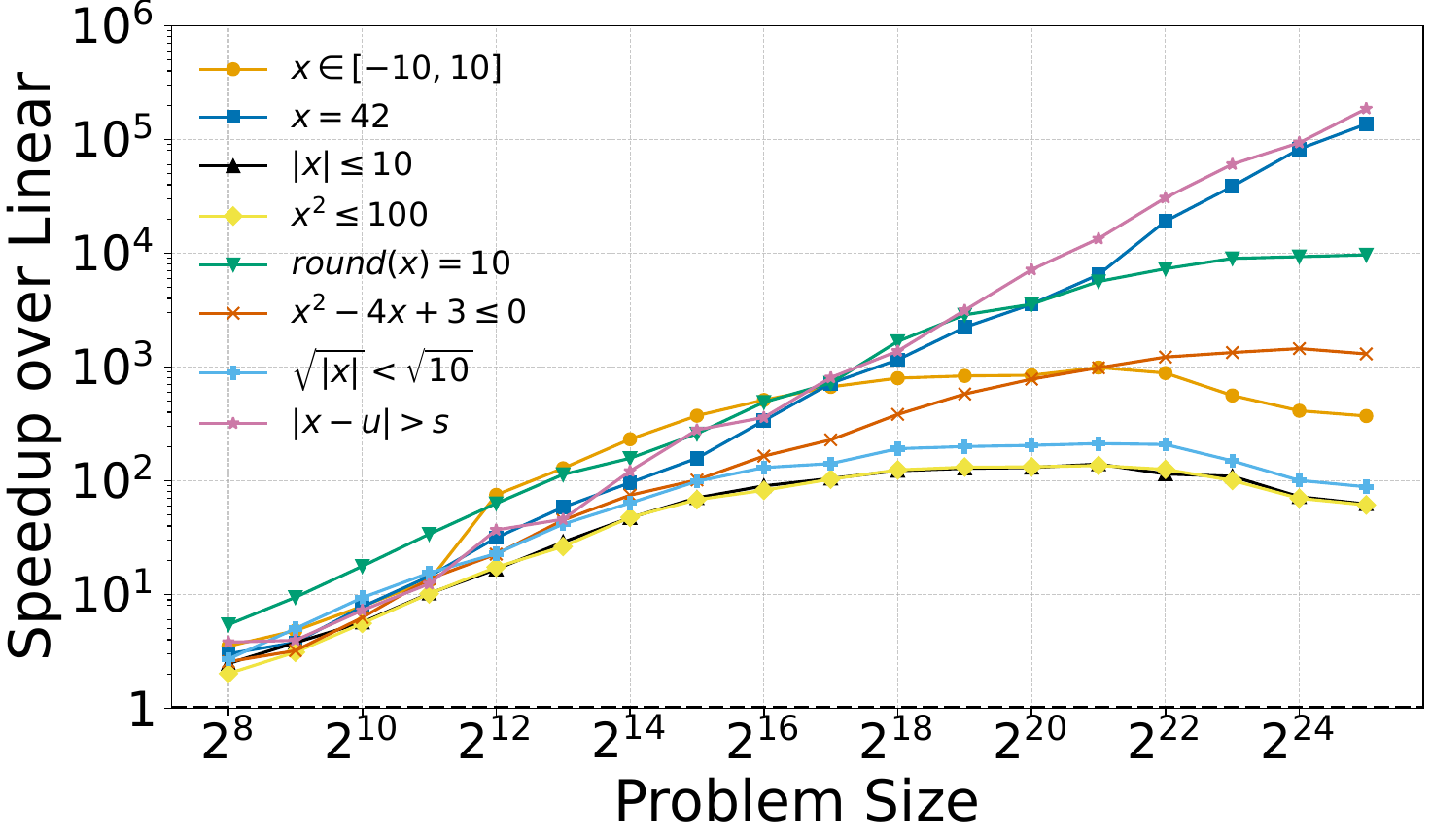}
    \caption{Accelerated aggregation (count) of filters.}
    \label{fig:uniform-count}
  \end{subfigure}

  \vspace{1em}

  \begin{subfigure}[b]{0.48\textwidth}
    \centering
    \includegraphics[width=\textwidth]{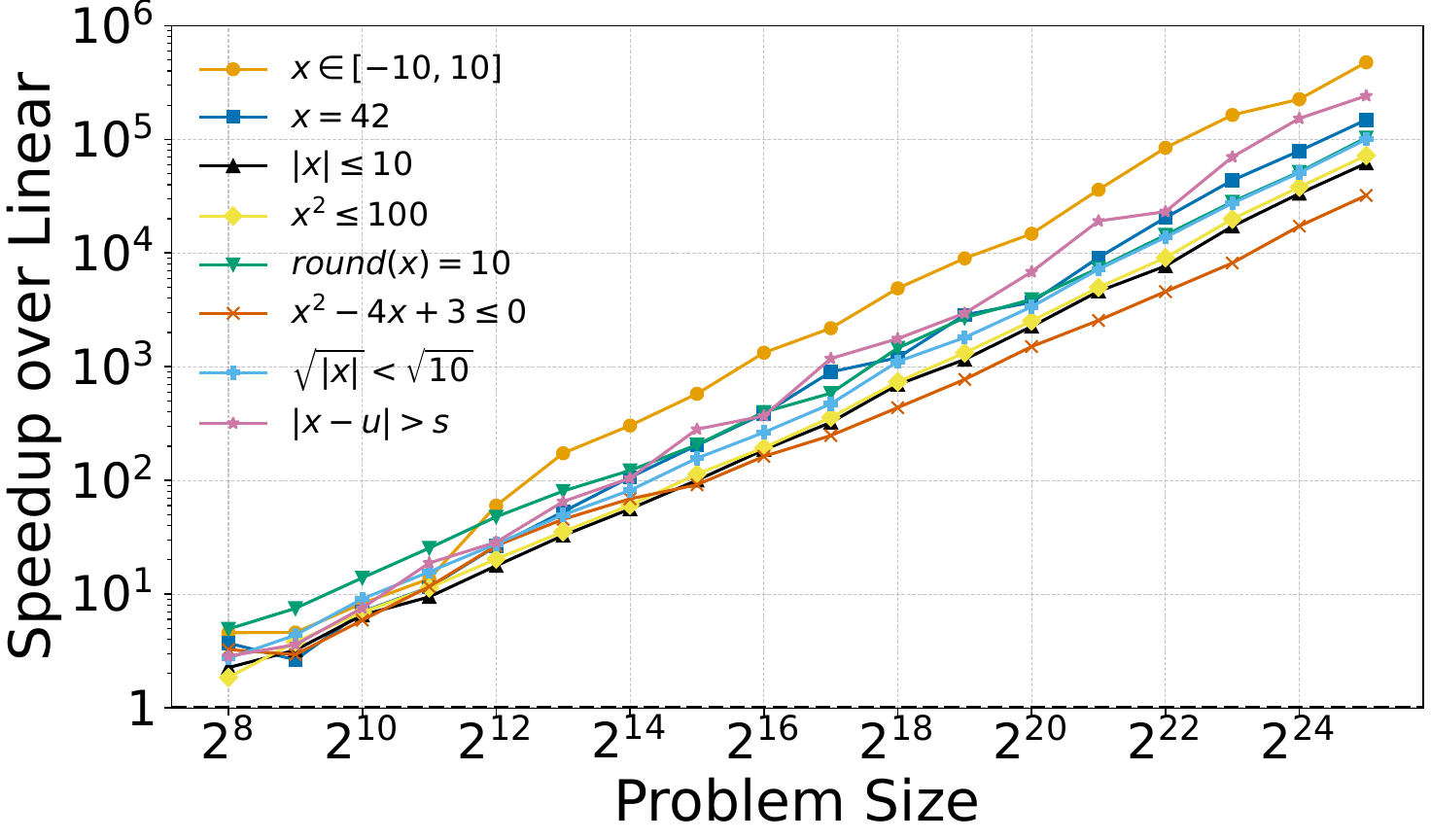}
    \caption{Accelerated aggregation with count aggregation.}
    \label{fig:uniform-count-agg}
  \end{subfigure}
  \hfill
  \begin{subfigure}[b]{0.48\textwidth}
    \centering
    \includegraphics[width=\textwidth]{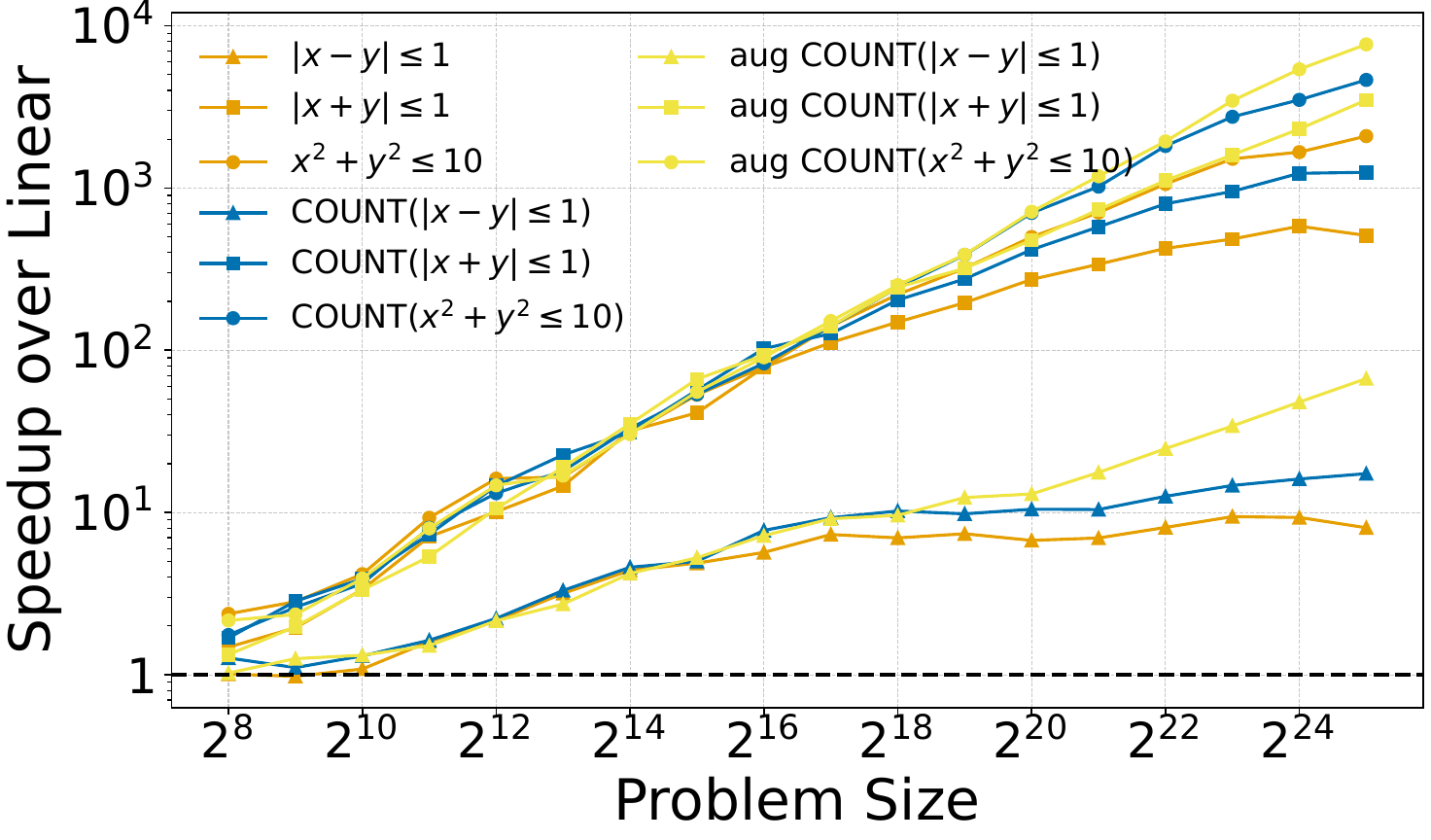}
    \caption{Multidimensional filter predicates.}
    \label{fig:uniform-2D}
  \end{subfigure}

  \caption{Speed-up plots over default linear scans for uniformly sampled data in $[-1000, 1000]$. Higher is better.
  }
  \label{fig:uniform-comparison}
  \vspace{-1em}
\end{figure}

These seven contain predicates on a single variable and  highlight the benefits of using tree traversals. \Cref{fig:uniform-filters} illustrates performance gains achieved by computing these filters as tree traversals. However, many queries plateau when they become write-bound, except for the highly selective point and standard deviation queries. \Cref{fig:uniform-count} extends these filters with a \code{count} aggregation. These traversals are now read-bound, and many simply perform scans to aggregate the count. We therefore extend the queried tree with a count augmentation in \Cref{fig:uniform-count-agg}, resulting in massive asymptotic improvements.

Although a sufficiently powerful rewrite system could convert some of these queries into range queries, others, particularly those with multi-variable predicates, cannot. The final three filter predicates in \Cref{fig:linear-queries} fall into this category: a diagonal band, a diamond-shaped region, and a circular filter.
\Cref{fig:uniform-2D} plots speedups for filters (orange), a fused count-filter reduction (blue), and the same reduction on a tree augmented with subtree count metadata (yellow). All outperform linear scans; more selective queries (diamond and circle) achieve the largest speedups.

\subsubsection{Fusion Ablation}

\begin{wrapfigure}{r}{0.4\textwidth}
    \vspace{-1.5em}
    \centering
    \includegraphics[width=\linewidth]{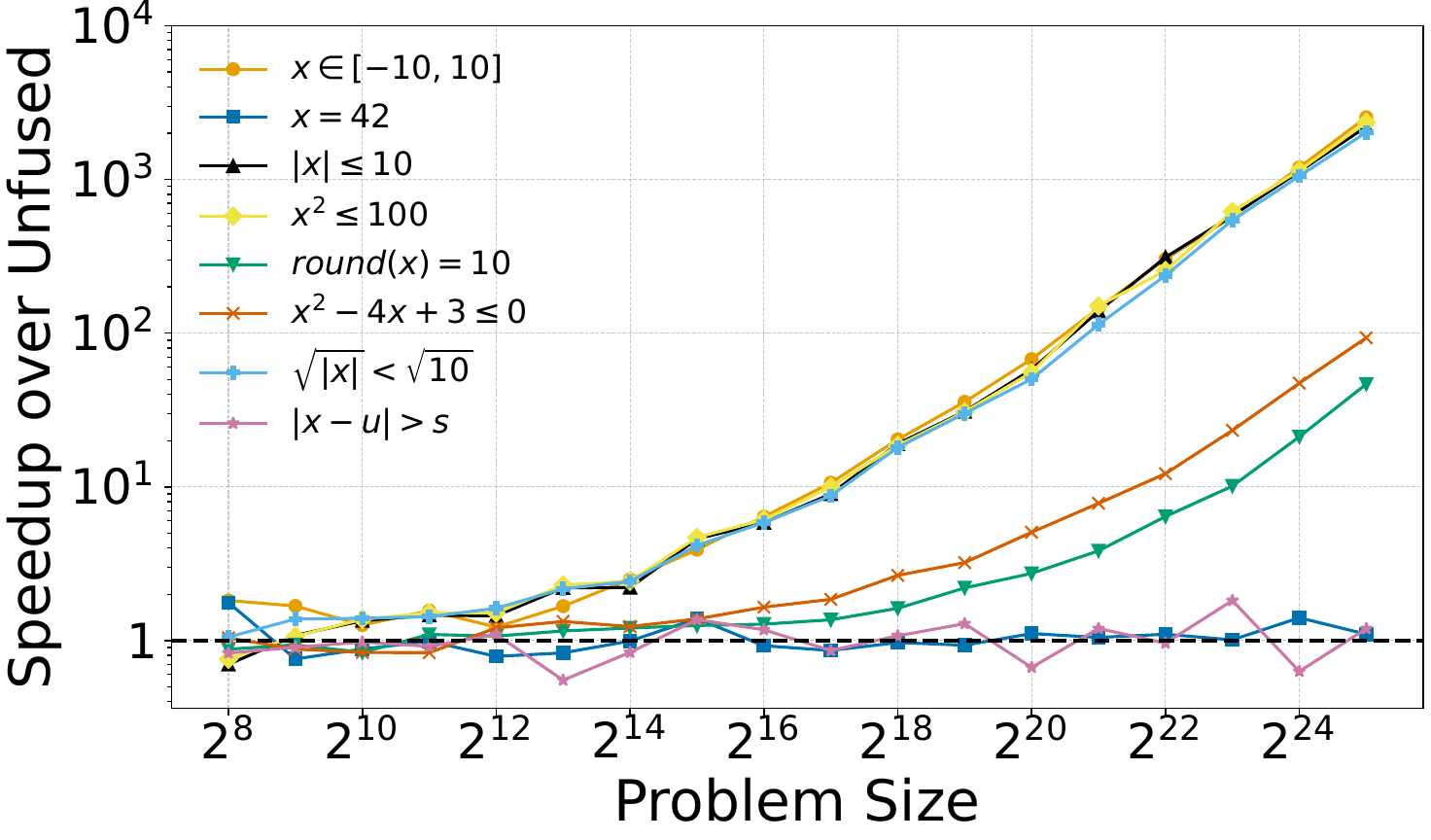}
    \vspace{-2em}
    \caption{Ablation of count-filter fusion.}
    \label{fig:uniform-count-ablation}
    \vspace{-2em}
\end{wrapfigure}

Fusion is particularly beneficial when filters are not very selective, as fusion of a reduction on a filter \textit{both} removes the need for an expensive intermediate data structure, and leverages reduction metadata available in the tree. To highlight these benefits, \Cref{fig:uniform-count-ablation} compares fused \code{count(filter())} queries to their unfused counterparts. Unfused variants perform a tree traversal to compute the filtered set, and then return its size. Because the reduction is an $O(1)$ operation on the intermediate data structure, any slowdown is a direct result of either (a) allocation and deallocation of the intermediate or (b) scanning subtrees that could otherwise return the count augmentation. In line with this observation, \Cref{fig:uniform-count-ablation} shows that fusion yields the greatest speedups for less selective queries, which benefit from reduced reading and writing overhead.

\subsubsection{Joins}

\begin{figure}[t]
  \centering
  \begin{subfigure}[b]{0.32\textwidth}
    \centering
    \includegraphics[width=\textwidth]{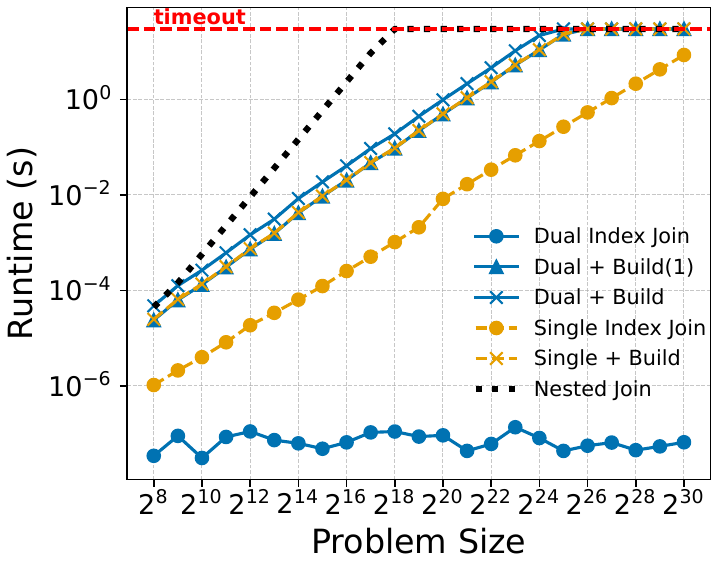}
    \caption{Radius $1$.}
    \label{fig:circle1-torus}
  \end{subfigure}
  \hfill
  \begin{subfigure}[b]{0.32\textwidth}
    \centering
    \includegraphics[width=\textwidth]{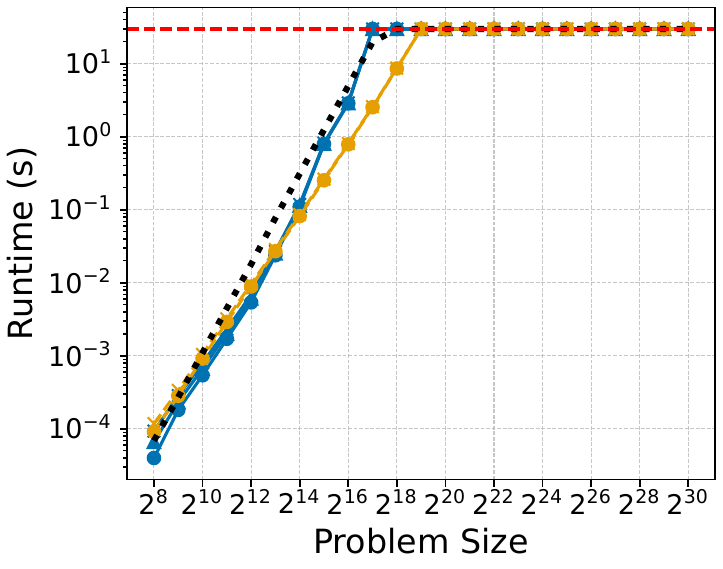}
    \caption{Radius $100$.}
    \label{fig:circle100-torus}
  \end{subfigure}
  \hfill
  \begin{subfigure}[b]{0.32\textwidth}
    \centering
    \includegraphics[width=\textwidth]{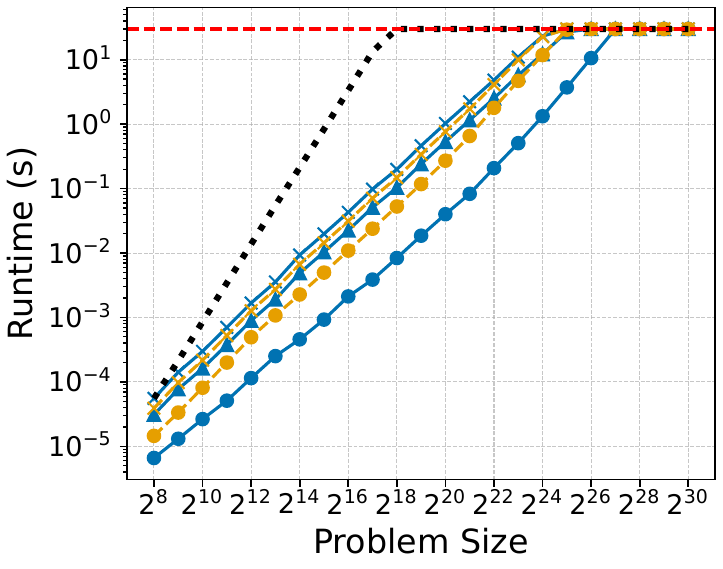}
    \caption{Radius $100000$.}
    \label{fig:circle10000-torus}
  \end{subfigure}

  \caption{Torus join results, with join predicate $\sqrt{(x_0-x_1)^2+(y_0-y_1)^2} \in [10, 20]$. Both tables are uniformly randomly sampled within circles of varying radii, to illustrate the impact of data distribution on join choice.}
  \label{fig:torus-join}
\end{figure}

Beyond range and salary joins (\Cref{fig:chebyshev-sota,fig:salary-sota}), we evaluate a \textit{torus join}, a hybrid of range and distance joins that retrieves all pairs of points within a distance range. No database we tested, including spatial systems, accelerated this join. \Cref{fig:torus-join} shows its performance under varying data distributions. Despite substantial differences in absolute performance, both single and dual-index joins scale better than the nested join, even when accounting for tree build time.

In \Cref{fig:circle1-torus}, points are sampled from a tight circle; \textit{no} pairs satisfy the join predicate. This highlights the strength of dual joins: the algorithm terminates almost immediately after detecting that the trees are disjoint w.r.t the predicate. When tree-build costs are included, the single-index join outperforms the dual join, as the latter must build both trees. Thus, for unindexed tables, a single-index join is preferable, but dual joins are asymptotically superior when both inputs are indexed. In \Cref{fig:circle100-torus}, sampling from a larger circle yields little pruning for either join type. Within the measured range (before timeout), the single-index join (with or without build times) appears to scale better than the dual-index join. \Cref{fig:circle10000-torus} considers an even larger circle. Pruning improves over \Cref{fig:circle100-torus} but is less extreme than in \Cref{fig:circle1-torus}, as many distant subtrees are pruned.

Overall, \sys{}-generated joins can outperform nested joins. The best join strategy depends on the predicate and data distribution; we leave cost models to automate this choice for future work.

\section{Related Work}
\label{sec:related-work}


Significant research has been conducted on the design of tree data structures. For graphics, we refer the reader to ~\citet{ericson2004rtcd} and ~\citet{meister2021bvh} for in-depth surveys of spatial acceleration structures in collision detection and ray tracing, respectively. For databases, we refer the reader to the spatio-temporal access methods surveys~\cite{mokbel2003survey, dinh2010survey, mahmood2019survey} and indexing survey~\cite{gani2016survey}.

\textit{Generalized Search Trees} (GiSTs)~\cite{hellerstein1998gist} attempt to unify database index structures under a common abstraction, but are not a compilation technique: they still require users to write the pruning function (termed "Consistent" in their model). GiSTs also do not support a notion of \code{always} functions (for scanning entire subtrees), nor subtree aggregates. We believe our model could serve as a useful extension of GiSTs to allow them to function for a larger variety of queries.


\textit{Term Rewriting Systems.}
Rewriting systems aim to canonicalize queries into a small number of queries for which pruning is possible, e.g., a range query. Rewriting systems are challenged by local minima and non-termination, motivating both extensive efforts to synthesize them from real-world data~\cite{newcomb2020trs, root2023pitchfork}, and alternatives like e-graphs~\cite{willsey2021egg, nandi2021ruler, yang2021equality, zhang2022relational, zhang2024dbt}. Our technique simplifies the task of targeting tree traversals by greatly expanding the space of valid targets: rather than rewriting solely to a limited set of query operators, a system needs only to produce predicates with derivable necessary or sufficient conditions. While no rewriting was needed for our benchmarks, such systems remain valuable in practice for simplifying predicates before bounds analysis (e.g., by removing correlated terms that yield suboptimal interval bounds).

\textit{Query Compilation.}
Query compilers~\cite{neumann2011querycompilation, tahboub2018architectquery, selinger1979systemr} largely treat index queries and tree traversals as external black-box operators, relying on pattern matching to invoke hand-optimized implementations when a cost model deems them profitable. The specific filters or database join algorithms (e.g., hash, sort-merge, and nested-loop joins~\cite{graefe1993query} and their spatial or multidimensional variants~\cite{samet2005multi}) generally depend on specialized traversal code for specific predicates or data types. These techniques accelerate common queries but do not generalize to arbitrary predicates or reductions. Our work complements these systems by automatically generating predicate-aware tree traversals, enabling efficient filters and joins without requiring hand-written specialization.



\textit{Interval Analysis in Computer Graphics.}
While our algorithm for generating pruning functions was inspired by Halide's symbolic interval analysis~\cite{jrk2012halide, jrk2013halide}, there has long been use of symbolic and numeric interval analysis in computer graphics~\cite{velasquez2009boundingshaders, mitchell1991interval, snyder1992interval, tricard2024intervalshading, keeter2020fidgit}. Interval analysis enables hierarchical reasoning about whether equations have solutions over an interval, allowing optimizers to prune solution spaces~\cite{snyder1992interval}. We extend this idea to compile tree queries and to handle spatial operators.

\textit{Pruning in Databases.}
Similarly, the databases community has long used pruning to reject partitions of data that cannot satisfy a query predicate~\cite{graefe2009zones, sun2014dataskip, sudhir2023pando}. This idea resembles our generation of the \code{maybe} pruning function, though it is typically used at runtime, not compile time. Recently, ~\citet{zimmerer2025snowflake} proposed computing an equivalent to the \code{always} function via the relationship $\code{always}(P) = \neg\code{maybe}(\neg P)$, but is limited to accelerating \code{LIMIT} queries, though it could extend to a wider range of query types, especially when partition metadata can be used to produce aggregates on scans without iterating over the data. Our formalization of pruning functions could strengthen such systems, while our spatial extension could support efficient pruning of spatial predicates, and our reduction-handling techniques could improve filtered reductions over partitioned data.


\textit{Compilers for Irregular Traversal Patterns.} Multiple domain-specific languages for other domains (and classes of data structures) have used data structure properties in the compilation of asymptotically-efficient code, e.g., sparse tensor algebra~\cite{kjolstad2017taco, chou2018formats, chou2022dynamic} with extensions to more general sparse array processing~\cite{henry2021, root2024burrito, sundram2024recuma, liu2024stencils}, sparse grids~\cite{hu2019taichi} with extensions to meshes~\cite{yu2022mesh}, and graphs~\cite{zhang2018graphit}. We provide a similar framework for tree data structures and pruning traversals. Though the graphics community has explored languages for collision detection~\cite{bernstein2019thesis} and ray tracing~\cite{perard2017ratrace, perard2019rodent}, each relies on hand-written traversal routines. In contrast, our higher-level representation (\Cref{sec:query-spec}) broadens support for more general spatial queries.

\textit{Parallelizing Traversals.} Significant work has explored parallelizing recursive programs and tree traversals outside of domain-specific languages~\cite{ren2019simdrec, koparkar2021pargibbon, singhal2024orchard, schardl2019tapir, frigo1998cilk5, sakka2019finegraintree, chen2022traversalsynthesis}. These works are largely orthogonal to ours, as they seek to efficiently parallelize code patterns such as those generated by our compiler.

\section{Conclusion and Future Work}
\label{sec:future-work}

We describe the first technique for compiling high-level queries into pruning tree traversals. Our technique is enabled by the key insight that hand-written tree traversals are based on necessary and sufficient conditions as functions of node metadata, which imply or disprove query predicates without iterating all data in the subtree. This technique is further enabled by an efficient derivation procedure for generating these conditions based on standard symbolic interval analysis with a novel extension to spatial operators. Our results demonstrate that accelerated filters and joins do not need to rely on bespoke data-structure-specific code, but can instead be generated by a compiler. This suggests a path toward query engines that treat tree-based acceleration as a reusable, derivable optimization rather than a small set of inflexible primitives. We envision many directions for future work, including:

\begin{description}
    \item[DBMS Integration.] Integrating our techniques into a database management system (DBMS) requires both implementing physical relational operators for tree queries (e.g., a generalized Index Scan) and developing cost models to drive operator selection. Such cost models must consider both query predicates and data distributions to assess pruning potential. For example, a DBMS observing frequent long-running queries that filter or aggregate on a column could automatically build bounds or reduction annotations for that column, respectively. However, this may only be profitable given certain data distributions on the column, as shown in ~\Cref{fig:torus-join}. Determining when a profile warrants index construction requires accurate modeling of compute gains against the memory overhead of storing the index, which we believe requires further research on cost models.


    \item[Tree Design.] Different choices of stored metadata induce different asymptotics of tree traversals, but also impact memory usage; synthesizing a tree design (i.e., the choice of metadata) automatically for a query or set of queries would enable automatic optimization of queries.

    \item[Space Partitioning Trees.] Space partitioning trees such as k-d trees~\cite{bentley1975kdtree} (as opposed to bounding volume hierarchies) offer weaker invariants for non-point data: their implicit bounding volumes only promise overlap with primitives beneath a node, and geometry often needs to be duplicated in the tree~\cite{eltabakh2007deduplication}. Nevertheless, they offer benefits for some classes of queries (e.g., in-order all-hit ray tracing). Developing predicate analysis for overlap volumes and deriving deduplication techniques could enable further query acceleration.

    \item[Scheduling.] The ray tracing community has performed extensive research on improving the performance of tree traversals on parallel hardware (e.g., packet tracing~\cite{wald2001interactive}, techniques for improving coherence on GPUs~\cite{aila2009understanding}, and wavefront traversals~\cite{aila2010treelets}). Supporting these optimizations for a larger class of queries via program transformations could be useful.

    \item[Construction Algorithms.] It is well-known that the construction of a tree on geometry has a significant impact on query performance~\cite{lauterbach2009fast, pantaleoni2010hlbvh, karras2012maximizing, gu2013efficient, apetrei2014fast, meister2017parallel, benthin2024hploc}. Exploring the design space of construction time versus tree quality is important work that could be made easier by compiler techniques.
\end{description}

Together, these directions broaden the scope of pruning-based optimization and move us toward compiler-driven methods for accelerated query processing. We hope this work serves as a foundation for research in tree traversal design and derivation.

\section*{Data-Availability Statement}
Performance results were generated with a publicly available artifact~\cite{root2026zenodo} containing all benchmarking code and scripts, as well as instructions for reproducibility. The \sys compiler is also available at \href{https://github.com/rootjalex/bonsai}{https://github.com/rootjalex/bonsai}. Benchmarking results may very based on the hardware used.

\begin{acks}
We thank our reviewers for their valuable feedback. We also thank
Amanda Liu,
Bala Vinaithirthan,
Ben Driscoll,
Bobby Yan,
Brennan Shacklett,
Devanshu Ladsaria,
Genghan Zhang,
Ishita Gupta,
James Dong,
Katherine Mohr,
Liza Pertseva,
Marco Siracusa,
Matt Pharr,
Nestan Tsiskaridze,
Olivia Hsu,
Rohan Sawhney,
Rohan Yadav,
Rubens Lacouture,
Sai Gautham Ravipati,
Scott Kovach,
and Zander Majercik
for their helpful feedback on drafts of this work.
Alexander and Christophe were supported by the Qualcomm Innovation Fellowship during part of this work; Alexander was also supported by the NSF Graduate Research Fellowship and took part in this work while interning at Adobe Research under Andrew.
\end{acks}

\bibliographystyle{ACM-Reference-Format}
\bibliography{references}

@inproceedings{egenhofer1990topological,
  author    = {Max J. Egenhofer and John Herring},
  title     = {A Mathematical Framework for the Definition of Topological Relationships},
  booktitle = {Proceedings of the Fourth International Symposium on Spatial Data Handling},
  pages     = {803--813},
  year      = {1990},
  publisher = {International Geographical Union},
  address   = {Zurich, Switzerland},
  url       = {https://web.archive.org/web/20100614161335/http://www.spatial.maine.edu/~max/MJEJRH-SDH1990.pdf},
  note      = {Accessed: 2025-05-01}
}

@inproceedings{frigo1998cilk5,
author = {Frigo, Matteo and Leiserson, Charles E. and Randall, Keith H.},
title = {The implementation of the Cilk-5 multithreaded language},
year = {1998},
isbn = {0897919874},
publisher = {Association for Computing Machinery},
address = {New York, NY, USA},
url = {https://doi.org/10.1145/277650.277725},
doi = {10.1145/277650.277725},
booktitle = {Proceedings of the ACM SIGPLAN 1998 Conference on Programming Language Design and Implementation},
pages = {212–223},
numpages = {12},
location = {Montreal, Quebec, Canada},
series = {PLDI '98}
}

@book{ericson2004rtcd,
    author = {Ericson, Christer},
    title = {Real-Time Collision Detection},
    year = {2004},
    isbn = {1558607323},
    publisher = {CRC Press, Inc.},
    address = {USA},
}

@article{jrk2012halide,
    author = {Ragan-Kelley, Jonathan and Adams, Andrew and Paris, Sylvain and Levoy, Marc and Amarasinghe, Saman and Durand, Fr\'{e}do},
    title = {Decoupling algorithms from schedules for easy optimization of image processing pipelines},
    year = {2012},
    issue_date = {July 2012},
    publisher = {Association for Computing Machinery},
    address = {New York, NY, USA},
    volume = {31},
    number = {4},
    issn = {0730-0301},
    url = {https://doi.org/10.1145/2185520.2185528},
    doi = {10.1145/2185520.2185528},
    journal = {ACM Trans. Graph.},
    month = {jul},
    articleno = {32},
    numpages = {12}
}

@article{jrk2013halide,
author = {Ragan-Kelley, Jonathan and Barnes, Connelly and Adams, Andrew and Paris, Sylvain and Durand, Fr\'{e}do and Amarasinghe, Saman},
title = {Halide: a language and compiler for optimizing parallelism, locality, and recomputation in image processing pipelines},
year = {2013},
issue_date = {June 2013},
publisher = {Association for Computing Machinery},
address = {New York, NY, USA},
volume = {48},
number = {6},
issn = {0362-1340},
url = {https://doi.org/10.1145/2499370.2462176},
doi = {10.1145/2499370.2462176},
journal = {SIGPLAN Not.},
month = jun,
pages = {519–530},
numpages = {12}
}

@article{fpbounds2017joldes,
    author = {Joldes, Mioara and Muller, Jean-Michel and Popescu, Valentina},
    title = {Tight and Rigorous Error Bounds for Basic Building Blocks of Double-Word Arithmetic},
    year = {2017},
    issue_date = {June 2018},
    publisher = {Association for Computing Machinery},
    address = {New York, NY, USA},
    volume = {44},
    number = {2},
    issn = {0098-3500},
    url = {https://doi.org/10.1145/3121432},
    doi = {10.1145/3121432},
    journal = {ACM Trans. Math. Softw.},
    month = {oct},
    articleno = {15res},
    numpages = {27}
}

@book{pbrt4,
  author = {Pharr, Matt and Jakob, Wenzel and Humphreys, Greg},
  title = {Physically Based Rendering: From Theory to Implementation},
  edition = {4},
  publisher = {MIT Press},
  year = {2023},
  isbn = {9780262048026},
  url = {https://mitpress.mit.edu/9780262048026/physically-based-rendering/}
}

@inproceedings{woop2014hair,
author = {Woop, Sven and Benthin, Carsten and Wald, Ingo and Johnson, Gregory S. and Tabellion, Eric},
title = {Exploiting local orientation similarity for efficient ray traversal of hair and fur},
year = {2014},
publisher = {Eurographics Association},
address = {Goslar, DEU},
booktitle = {Proceedings of High Performance Graphics},
pages = {41–49},
numpages = {9},
location = {Lyon, France},
series = {HPG '14}
}

@inproceedings{benthin2018clbvh,
author = {Benthin, Carsten and Wald, Ingo and Woop, Sven and \'{A}fra, Attila T.},
title = {Compressed-leaf bounding volume hierarchies},
year = {2018},
isbn = {9781450358965},
publisher = {Association for Computing Machinery},
address = {New York, NY, USA},
url = {https://doi.org/10.1145/3231578.3231581},
doi = {10.1145/3231578.3231581},
booktitle = {Proceedings of the Conference on High-Performance Graphics},
articleno = {6},
numpages = {4},
location = {Vancouver, British Columbia, Canada},
series = {HPG '18}
}

@inproceedings{coutts2007streamfusion,
author = {Coutts, Duncan and Leshchinskiy, Roman and Stewart, Don},
title = {Stream fusion: from lists to streams to nothing at all},
year = {2007},
isbn = {9781595938152},
publisher = {Association for Computing Machinery},
address = {New York, NY, USA},
url = {https://doi.org/10.1145/1291151.1291199},
doi = {10.1145/1291151.1291199},
booktitle = {Proceedings of the 12th ACM SIGPLAN International Conference on Functional Programming},
pages = {315–326},
numpages = {12},
location = {Freiburg, Germany},
series = {ICFP '07}
}

@software{sawhney2021fcpw,
author = {Sawhney, Rohan},
title = {FCPW: Fastest Closest Points in the West},
version = {1.0},
year = {2021}
}

@inproceedings{fan2024gDist,
author = {Fan, Peng and Wang, Wei and Tong, Ruofeng and Li, Hailong and Tang, Min},
title = {gDist: Efficient Distance Computation between 3D Meshes on GPU},
year = {2024},
isbn = {9798400711312},
publisher = {Association for Computing Machinery},
address = {New York, NY, USA},
url = {https://doi.org/10.1145/3680528.3687619},
doi = {10.1145/3680528.3687619},
booktitle = {SIGGRAPH Asia 2024 Conference Papers},
articleno = {71},
numpages = {11},
location = {Tokyo, Japan},
series = {SA '24}
}

@inproceedings{reynolds2020sygus,
author = {Reynolds, Andrew and Barbosa, Haniel and Larraz, Daniel and Tinelli, Cesare},
title = {Scalable Algorithms for Abduction via Enumerative Syntax-Guided Synthesis},
year = {2020},
isbn = {978-3-030-51073-2},
publisher = {Springer-Verlag},
address = {Berlin, Heidelberg},
url = {https://doi.org/10.1007/978-3-030-51074-9_9},
doi = {10.1007/978-3-030-51074-9_9},
booktitle = {Automated Reasoning: 10th International Joint Conference, IJCAR 2020, Paris, France, July 1–4, 2020, Proceedings, Part I},
pages = {141–160},
numpages = {20},
location = {Paris, France}
}

@inproceedings{dillig2013invariant,
author = {Dillig, Isil and Dillig, Thomas and Li, Boyang and McMillan, Ken},
title = {Inductive invariant generation via abductive inference},
year = {2013},
isbn = {9781450323741},
publisher = {Association for Computing Machinery},
address = {New York, NY, USA},
url = {https://doi.org/10.1145/2509136.2509511},
doi = {10.1145/2509136.2509511},
booktitle = {Proceedings of the 2013 ACM SIGPLAN International Conference on Object Oriented Programming Systems Languages \&amp; Applications},
pages = {443–456},
numpages = {14},
location = {Indianapolis, Indiana, USA},
series = {OOPSLA '13}
}

@book{date1989sql,
author = {Date, C. J.},
title = {A guide to the SQL standard (2nd ed.)},
year = {1989},
isbn = {0201502097},
publisher = {Addison-Wesley Longman Publishing Co., Inc.},
address = {USA}
}

@article{blumofe1999workstealing,
author = {Blumofe, Robert D. and Leiserson, Charles E.},
title = {Scheduling multithreaded computations by work stealing},
year = {1999},
issue_date = {Sept. 1999},
publisher = {Association for Computing Machinery},
address = {New York, NY, USA},
volume = {46},
number = {5},
issn = {0004-5411},
url = {https://doi.org/10.1145/324133.324234},
doi = {10.1145/324133.324234},
journal = {J. ACM},
month = sep,
pages = {720–748},
numpages = {29}
}

@article{kay1986,
author = {Kay, Timothy L. and Kajiya, James T.},
title = {Ray tracing complex scenes},
year = {1986},
issue_date = {Aug. 1986},
publisher = {Association for Computing Machinery},
address = {New York, NY, USA},
volume = {20},
number = {4},
issn = {0097-8930},
url = {https://doi.org/10.1145/15886.15916},
doi = {10.1145/15886.15916},
journal = {SIGGRAPH Comput. Graph.},
month = aug,
pages = {269–278},
numpages = {10}
}

@article{comer1979btree,
author = {Comer, Douglas},
title = {Ubiquitous B-Tree},
year = {1979},
issue_date = {June 1979},
publisher = {Association for Computing Machinery},
address = {New York, NY, USA},
volume = {11},
number = {2},
issn = {0360-0300},
url = {https://doi.org/10.1145/356770.356776},
doi = {10.1145/356770.356776},
journal = {ACM Comput. Surv.},
month = jun,
pages = {121–137},
numpages = {17}
}

@inproceedings{guttman1984rtree,
author = {Guttman, Antonin},
title = {R-trees: a dynamic index structure for spatial searching},
year = {1984},
isbn = {0897911288},
publisher = {Association for Computing Machinery},
address = {New York, NY, USA},
url = {https://doi.org/10.1145/602259.602266},
doi = {10.1145/602259.602266},
booktitle = {Proceedings of the 1984 ACM SIGMOD International Conference on Management of Data},
pages = {47–57},
numpages = {11},
location = {Boston, Massachusetts},
series = {SIGMOD '84}
}

@article{barnes1986bhut,
	author = {Barnes, Josh and Hut, Piet},
	date = {1986/12/01},
	doi = {10.1038/324446a0},
	id = {Barnes1986},
	isbn = {1476-4687},
	journal = {Nature},
	number = {6096},
	pages = {446--449},
	title = {A hierarchical O(N log N) force-calculation algorithm},
	url = {https://doi.org/10.1038/324446a0},
	volume = {324},
	year = {1986}}

@article{neumann2011querycompilation,
author = {Neumann, Thomas},
title = {Efficiently compiling efficient query plans for modern hardware},
year = {2011},
issue_date = {June 2011},
publisher = {VLDB Endowment},
volume = {4},
number = {9},
issn = {2150-8097},
url = {https://doi.org/10.14778/2002938.2002940},
doi = {10.14778/2002938.2002940},
journal = {Proc. VLDB Endow.},
month = jun,
pages = {539–550},
numpages = {12}
}

@inproceedings{tahboub2018architectquery,
author = {Tahboub, Ruby Y. and Essertel, Gr\'{e}gory M. and Rompf, Tiark},
title = {How to Architect a Query Compiler, Revisited},
year = {2018},
isbn = {9781450347037},
publisher = {Association for Computing Machinery},
address = {New York, NY, USA},
url = {https://doi.org/10.1145/3183713.3196893},
doi = {10.1145/3183713.3196893},
booktitle = {Proceedings of the 2018 International Conference on Management of Data},
pages = {307–322},
numpages = {16},
location = {Houston, TX, USA},
series = {SIGMOD '18}
}

@inproceedings{selinger1979systemr,
author = {Selinger, P. Griffiths and Astrahan, M. M. and Chamberlin, D. D. and Lorie, R. A. and Price, T. G.},
title = {Access path selection in a relational database management system},
year = {1979},
isbn = {089791001X},
publisher = {Association for Computing Machinery},
address = {New York, NY, USA},
url = {https://doi.org/10.1145/582095.582099},
doi = {10.1145/582095.582099},
booktitle = {Proceedings of the 1979 ACM SIGMOD International Conference on Management of Data},
pages = {23–34},
numpages = {12},
location = {Boston, Massachusetts},
series = {SIGMOD '79}
}

@inbook{hellerstein1998gist,
author = {Hellerstein, Joseph M. and Naughton, Jeffrey F. and Pfeffer, Avi},
title = {Generalized search trees for database systems},
year = {1998},
isbn = {1558605231},
publisher = {Morgan Kaufmann Publishers Inc.},
address = {San Francisco, CA, USA},
booktitle = {Readings in Database Systems (3rd Ed.)},
pages = {101–112},
numpages = {12}
}

@article{meister2021bvh,
author = {Meister, Daniel and Ogaki, Shinji and Benthin, Carsten and Doyle, Michael J. and Guthe, Michael and Bittner, Jiří},
title = {A Survey on Bounding Volume Hierarchies for Ray Tracing},
journal = {Computer Graphics Forum},
volume = {40},
number = {2},
pages = {683-712},
doi = {https://doi.org/10.1111/cgf.142662},
url = {https://onlinelibrary.wiley.com/doi/abs/10.1111/cgf.142662},
eprint = {https://onlinelibrary.wiley.com/doi/pdf/10.1111/cgf.142662},
year = {2021}
}

@article{mokbel2003survey,
  author       = {Mohamed F. Mokbel and Thanaa M. Ghanem and Walid G. Aref},
  title        = {Spatio‐temporal Access Methods},
  journal      = {IEEE Data Engineering Bulletin},
  volume       = {26},
  number       = {2},
  pages        = {40--49},
  year         = {2003},
}

@article{dinh2010survey,
  author       = {Long{-}Van Nguyen{-}Dinh and
                  Walid G. Aref and
                  Mohamed F. Mokbel},
  title        = {Spatio-Temporal Access Methods: Part 2 {(2003} - 2010)},
  journal      = {{IEEE} Data Eng. Bull.},
  volume       = {33},
  number       = {2},
  pages        = {46--55},
  year         = {2010},
  url          = {http://sites.computer.org/debull/A10june/Aref.pdf},
  bibsource    = {dblp computer science bibliography, https://dblp.org}
}

@article{mahmood2019survey,
author = {Mahmood, Ahmed R. and Punni, Sri and Aref, Walid G.},
title = {Spatio-temporal access methods: a survey (2010 - 2017)},
year = {2019},
issue_date = {January   2019},
publisher = {Kluwer Academic Publishers},
address = {USA},
volume = {23},
number = {1},
issn = {1384-6175},
url = {https://doi.org/10.1007/s10707-018-0329-2},
doi = {10.1007/s10707-018-0329-2},
journal = {Geoinformatica},
month = jan,
pages = {1–36},
numpages = {36}
}

@article{gani2016survey,
	author = {Gani, Abdullah and Siddiqa, Aisha and Shamshirband, Shahaboddin and Hanum, Fariza},
	date = {2016/02/01},
	doi = {10.1007/s10115-015-0830-y},
	id = {Gani2016},
	isbn = {0219-3116},
	journal = {Knowledge and Information Systems},
	number = {2},
	pages = {241--284},
	title = {A survey on indexing techniques for big data: taxonomy and performance evaluation},
	url = {https://doi.org/10.1007/s10115-015-0830-y},
	volume = {46},
	year = {2016}}

@article{kjolstad2017taco,
author = {Kjolstad, Fredrik and Kamil, Shoaib and Chou, Stephen and Lugato, David and Amarasinghe, Saman},
title={The Tensor Algebra Compiler},
year = {2017},
issue_date = {October 2017},
publisher = {Association for Computing Machinery},
address = {New York, NY, USA},
volume = {1},
number = {OOPSLA},
url = {https://doi.org/10.1145/3133901},
doi = {10.1145/3133901},
journal = {Proc. ACM Program. Lang.},
month = oct,
articleno = {77},
numpages = {29}
}

@article{chou2018formats,
author = {Chou, Stephen and Kjolstad, Fredrik and Amarasinghe, Saman},
title = {Format Abstraction for Sparse Tensor Algebra Compilers},
year = {2018},
issue_date = {November 2018},
publisher = {Association for Computing Machinery},
address = {New York, NY, USA},
volume = {2},
number = {OOPSLA},
url = {https://doi.org/10.1145/3276493},
doi = {10.1145/3276493},
journal = {Proc. ACM Program. Lang.},
month = oct,
articleno = {123},
numpages = {30}
}

@article{henry2021,
author = {Henry, Rawn and Hsu, Olivia and Yadav, Rohan and Chou, Stephen and Olukotun, Kunle and Amarasinghe, Saman and Kjolstad, Fredrik},
title = {Compilation of Sparse Array Programming Models},
year = {2021},
issue_date = {October 2021},
publisher = {Association for Computing Machinery},
address = {New York, NY, USA},
volume = {5},
number = {OOPSLA},
url = {https://doi.org/10.1145/3485505},
doi = {10.1145/3485505},
journal = {Proc. ACM Program. Lang.},
month = oct,
articleno = {128},
numpages = {29}
}

@article{root2024burrito,
author = {Root, Alexander J and Yan, Bobby and Liu, Peiming and Gyurgyik, Christophe and Bik, Aart J.C. and Kjolstad, Fredrik},
title = {Compilation of Shape Operators on Sparse Arrays},
year = {2024},
issue_date = {October 2024},
publisher = {Association for Computing Machinery},
address = {New York, NY, USA},
volume = {8},
number = {OOPSLA2},
url = {https://doi.org/10.1145/3689752},
doi = {10.1145/3689752},
journal = {Proc. ACM Program. Lang.},
month = oct,
articleno = {312},
numpages = {27}
}

@article{liu2024stencils,
author = {Liu, Peiming and Root, Alexander J and Xu, Anlun and Li, Yinying and Kjolstad, Fredrik and Bik, Aart J.C.},
title = {Compiler Support for Sparse Tensor Convolutions},
year = {2024},
issue_date = {October 2024},
publisher = {Association for Computing Machinery},
address = {New York, NY, USA},
volume = {8},
number = {OOPSLA2},
url = {https://doi.org/10.1145/3689721},
doi = {10.1145/3689721},
journal = {Proc. ACM Program. Lang.},
month = oct,
articleno = {281},
numpages = {29}
}

@article{sundram2024recuma,
author = {Sundram, Shiv and Tariq, Muhammad Usman and Kjolstad, Fredrik},
title = {Compiling Recurrences over Dense and Sparse Arrays},
year = {2024},
issue_date = {April 2024},
publisher = {Association for Computing Machinery},
address = {New York, NY, USA},
volume = {8},
number = {OOPSLA1},
url = {https://doi.org/10.1145/3649820},
doi = {10.1145/3649820},
journal = {Proc. ACM Program. Lang.},
month = apr,
articleno = {103},
numpages = {26}
}

@article{yu2022mesh,
author = {Yu, Chang and Xu, Yi and Kuang, Ye and Hu, Yuanming and Liu, Tiantian},
title = {MeshTaichi: A Compiler for Efficient Mesh-Based Operations},
year = {2022},
issue_date = {December 2022},
publisher = {Association for Computing Machinery},
address = {New York, NY, USA},
volume = {41},
number = {6},
issn = {0730-0301},
url = {https://doi.org/10.1145/3550454.3555430},
doi = {10.1145/3550454.3555430},
journal = {ACM Trans. Graph.},
month = nov,
articleno = {252},
numpages = {17}
}

@article{zhang2018graphit,
author = {Zhang, Yunming and Yang, Mengjiao and Baghdadi, Riyadh and Kamil, Shoaib and Shun, Julian and Amarasinghe, Saman},
title = {GraphIt: a high-performance graph DSL},
year = {2018},
issue_date = {November 2018},
publisher = {Association for Computing Machinery},
address = {New York, NY, USA},
volume = {2},
number = {OOPSLA},
url = {https://doi.org/10.1145/3276491},
doi = {10.1145/3276491},
journal = {Proc. ACM Program. Lang.},
month = oct,
articleno = {121},
numpages = {30}
}

@phdthesis{bernstein2019thesis,
  author       = {Gilbert Louis Bernstein},
  title        = {Designing Languages for Parallel Portability of Physical Simulations, Using Relational Algebraic Abstractions},
  school       = {Stanford University},
  year         = {2019},
  month        = {June},
  type         = {{Ph.D.} Dissertation},
}

@article{hu2019taichi,
author = {Hu, Yuanming and Li, Tzu-Mao and Anderson, Luke and Ragan-Kelley, Jonathan and Durand, Fr\'{e}do},
title = {Taichi: a language for high-performance computation on spatially sparse data structures},
year = {2019},
issue_date = {December 2019},
publisher = {Association for Computing Machinery},
address = {New York, NY, USA},
volume = {38},
number = {6},
issn = {0730-0301},
url = {https://doi.org/10.1145/3355089.3356506},
doi = {10.1145/3355089.3356506},
journal = {ACM Trans. Graph.},
month = nov,
articleno = {201},
numpages = {16}
}

@inproceedings{perard2017ratrace,
    author = {P\'{e}rard-Gayot, Ars\`{e}ne and Weier, Martin and Membarth, Richard and Slusallek, Philipp and Lei\ss{}a, Roland and Hack, Sebastian},
    title = {RaTrace: Simple and Efficient Abstractions for BVH Ray Traversal Algorithms},
    year = {2017},
    isbn = {9781450355247},
    publisher = {Association for Computing Machinery},
    address = {New York, NY, USA},
    url = {https://doi.org/10.1145/3136040.3136044},
    doi = {10.1145/3136040.3136044},
    booktitle = {Proceedings of the 16th ACM SIGPLAN International Conference on Generative Programming: Concepts and Experiences},
    pages = {157–168},
    numpages = {12},
    location = {Vancouver, BC, Canada},
    series = {GPCE 2017}
}

@article{perard2019rodent,
    author = {P\'{e}rard-Gayot, Ars\`{e}ne and Membarth, Richard and Lei\ss{}a, Roland and Hack, Sebastian and Slusallek, Philipp},
    title = {Rodent: Generating Renderers without Writing a Generator},
    year = {2019},
    issue_date = {August 2019},
    publisher = {Association for Computing Machinery},
    address = {New York, NY, USA},
    volume = {38},
    number = {4},
    issn = {0730-0301},
    url = {https://doi.org/10.1145/3306346.3322955},
    doi = {10.1145/3306346.3322955},
    journal = {ACM Trans. Graph.},
    month = {jul},
    articleno = {40},
    numpages = {12}
}

@article{koparkar2021pargibbon,
author = {Koparkar, Chaitanya and Rainey, Mike and Vollmer, Michael and Kulkarni, Milind and Newton, Ryan R.},
title = {Efficient tree-traversals: reconciling parallelism and dense data representations},
year = {2021},
issue_date = {August 2021},
publisher = {Association for Computing Machinery},
address = {New York, NY, USA},
volume = {5},
number = {ICFP},
url = {https://doi.org/10.1145/3473596},
doi = {10.1145/3473596},
journal = {Proc. ACM Program. Lang.},
month = aug,
articleno = {91},
numpages = {29}
}

@article{singhal2024orchard,
author = {Singhal, Vidush and Sakka, Laith and Sundararajah, Kirshanthan and Newton, Ryan and Kulkarni, Milind},
title = {Orchard: Heterogeneous Parallelism and Fine-grained Fusion for Complex Tree Traversals},
year = {2024},
issue_date = {June 2024},
publisher = {Association for Computing Machinery},
address = {New York, NY, USA},
volume = {21},
number = {2},
issn = {1544-3566},
url = {https://doi.org/10.1145/3652605},
doi = {10.1145/3652605},
journal = {ACM Trans. Archit. Code Optim.},
month = may,
articleno = {41},
numpages = {25}
}

@article{ren2019simdrec,
author = {Ren, Bin and Balakrishna, Shruthi and Jo, Youngjoon and Krishnamoorthy, Sriram and Agrawal, Kunal and Kulkarni, Milind},
title = {Extracting SIMD Parallelism from Recursive Task-Parallel Programs},
year = {2019},
issue_date = {December 2019},
publisher = {Association for Computing Machinery},
address = {New York, NY, USA},
volume = {6},
number = {4},
issn = {2329-4949},
url = {https://doi.org/10.1145/3365663},
doi = {10.1145/3365663},
journal = {ACM Trans. Parallel Comput.},
month = dec,
articleno = {24},
numpages = {37}
}

@article{schardl2019tapir,
author = {Schardl, Tao B. and Moses, William S. and Leiserson, Charles E.},
title = {Tapir: Embedding Recursive Fork-join Parallelism into LLVM’s Intermediate Representation},
year = {2019},
issue_date = {December 2019},
publisher = {Association for Computing Machinery},
address = {New York, NY, USA},
volume = {6},
number = {4},
issn = {2329-4949},
url = {https://doi.org/10.1145/3365655},
doi = {10.1145/3365655},
journal = {ACM Trans. Parallel Comput.},
month = dec,
articleno = {19},
numpages = {33}
}

@article{chou2022dynamic,
author = {Chou, Stephen and Amarasinghe, Saman},
title = {Compilation of Dynamic Sparse Tensor Algebra},
year = {2022},
issue_date = {October 2022},
publisher = {Association for Computing Machinery},
address = {New York, NY, USA},
volume = {6},
number = {OOPSLA2},
url = {https://doi.org/10.1145/3563338},
doi = {10.1145/3563338},
journal = {Proc. ACM Program. Lang.},
month = oct,
articleno = {175},
numpages = {30}
}

@inproceedings{sakka2019finegraintree,
author = {Sakka, Laith and Sundararajah, Kirshanthan and Newton, Ryan R. and Kulkarni, Milind},
title = {Sound, fine-grained traversal fusion for heterogeneous trees},
year = {2019},
isbn = {9781450367127},
publisher = {Association for Computing Machinery},
address = {New York, NY, USA},
url = {https://doi.org/10.1145/3314221.3314626},
doi = {10.1145/3314221.3314626},
booktitle = {Proceedings of the 40th ACM SIGPLAN Conference on Programming Language Design and Implementation},
pages = {830–844},
numpages = {15},
location = {Phoenix, AZ, USA},
series = {PLDI 2019}
}

@article{velasquez2009boundingshaders,
author = {Vel\'{a}zquez-Armend\'{a}riz, Edgar and Zhao, Shuang and Ha\v{s}an, Milo\v{s} and Walter, Bruce and Bala, Kavita},
title = {Automatic bounding of programmable shaders for efficient global illumination},
year = {2009},
issue_date = {December 2009},
publisher = {Association for Computing Machinery},
address = {New York, NY, USA},
volume = {28},
number = {5},
issn = {0730-0301},
url = {https://doi.org/10.1145/1618452.1618488},
doi = {10.1145/1618452.1618488},
journal = {ACM Trans. Graph.},
month = dec,
pages = {1–9},
numpages = {9}
}

@inproceedings{mitchell1991interval,
  author    = {Don P. Mitchell},
  title     = {Three Applications of Interval Analysis in Computer Graphics},
  booktitle = {Frontiers of Rendering, Course Notes},
  year      = {1991},
  organization = {ACM SIGGRAPH},
  note      = {Course No. 14, SIGGRAPH '91},
}

@article{snyder1992interval,
author = {Snyder, John M.},
title = {Interval analysis for computer graphics},
year = {1992},
issue_date = {July 1992},
publisher = {Association for Computing Machinery},
address = {New York, NY, USA},
volume = {26},
number = {2},
issn = {0097-8930},
url = {https://doi.org/10.1145/142920.134024},
doi = {10.1145/142920.134024},
journal = {SIGGRAPH Comput. Graph.},
month = jul,
pages = {121–130},
numpages = {10}
}

@article{tricard2024intervalshading,
author = {Tricard, Thibault},
title = {Interval Shading: using Mesh Shaders to generate shading intervals for volume rendering},
year = {2024},
issue_date = {August 2024},
publisher = {Association for Computing Machinery},
address = {New York, NY, USA},
volume = {7},
number = {3},
url = {https://doi.org/10.1145/3675380},
doi = {10.1145/3675380},
journal = {Proc. ACM Comput. Graph. Interact. Tech.},
month = aug,
articleno = {43},
numpages = {11}
}

@article{keeter2020fidgit,
author = {Keeter, Matthew J.},
title = {Massively parallel rendering of complex closed-form implicit surfaces},
year = {2020},
issue_date = {August 2020},
publisher = {Association for Computing Machinery},
address = {New York, NY, USA},
volume = {39},
number = {4},
issn = {0730-0301},
url = {https://doi.org/10.1145/3386569.3392429},
doi = {10.1145/3386569.3392429},
journal = {ACM Trans. Graph.},
month = aug,
articleno = {141},
numpages = {10}
}

@article{graefe1993query,
author = {Graefe, Goetz},
title = {Query evaluation techniques for large databases},
year = {1993},
issue_date = {June 1993},
publisher = {Association for Computing Machinery},
address = {New York, NY, USA},
volume = {25},
number = {2},
issn = {0360-0300},
url = {https://doi.org/10.1145/152610.152611},
doi = {10.1145/152610.152611},
journal = {ACM Comput. Surv.},
month = jun,
pages = {73–169},
numpages = {97}
}

@book{samet2005multi,
author = {Samet, Hanan},
title = {Foundations of Multidimensional and Metric Data Structures (The Morgan Kaufmann Series in Computer Graphics and Geometric Modeling)},
year = {2005},
isbn = {0123694469},
publisher = {Morgan Kaufmann Publishers Inc.},
address = {San Francisco, CA, USA}
}

@inproceedings{chen2022traversalsynthesis,
author = {Chen, Yanju and Liu, Junrui and Feng, Yu and Bodik, Rastislav},
title = {Tree traversal synthesis using domain-specific symbolic compilation},
year = {2022},
isbn = {9781450392051},
publisher = {Association for Computing Machinery},
address = {New York, NY, USA},
url = {https://doi.org/10.1145/3503222.3507751},
doi = {10.1145/3503222.3507751},
booktitle = {Proceedings of the 27th ACM International Conference on Architectural Support for Programming Languages and Operating Systems},
pages = {1030–1042},
numpages = {13},
location = {Lausanne, Switzerland},
series = {ASPLOS '22}
}

@inproceedings{graefe2009zones,
	address = {Berlin, Heidelberg},
	author = {Graefe, Goetz},
	booktitle = {Data Warehousing and Knowledge Discovery},
	editor = {Pedersen, Torben Bach and Mohania, Mukesh K. and Tjoa, A. Min},
	isbn = {978-3-642-03730-6},
	pages = {111--124},
	publisher = {Springer Berlin Heidelberg},
	title = {Fast Loads and Fast Queries},
	year = {2009}}

@inproceedings{zimmerer2025snowflake,
author = {Zimmerer, Andreas and Dam, Damien and Kossmann, Jan and Waack, Juliane and Oukid, Ismail and Kipf, Andreas},
title = {Pruning in Snowflake: Working Smarter, Not Harder},
year = {2025},
isbn = {9798400715648},
publisher = {Association for Computing Machinery},
address = {New York, NY, USA},
url = {https://doi.org/10.1145/3722212.3724447},
doi = {10.1145/3722212.3724447},
booktitle = {Companion of the 2025 International Conference on Management of Data},
pages = {757–770},
numpages = {14},
location = {Berlin, Germany},
series = {SIGMOD/PODS '25}
}

@article{sudhir2023pando,
author = {Sudhir, Sivaprasad and Tao, Wenbo and Laptev, Nikolay and Habis, Cyrille and Cafarella, Michael and Madden, Samuel},
title = {Pando: Enhanced Data Skipping with Logical Data Partitioning},
year = {2023},
issue_date = {May 2023},
publisher = {VLDB Endowment},
volume = {16},
number = {9},
issn = {2150-8097},
url = {https://doi.org/10.14778/3598581.3598601},
doi = {10.14778/3598581.3598601},
journal = {Proc. VLDB Endow.},
month = may,
pages = {2316–2329},
numpages = {14}
}

@inproceedings{sun2014dataskip,
author = {Sun, Liwen and Franklin, Michael J. and Krishnan, Sanjay and Xin, Reynold S.},
title = {Fine-grained partitioning for aggressive data skipping},
year = {2014},
isbn = {9781450323765},
publisher = {Association for Computing Machinery},
address = {New York, NY, USA},
url = {https://doi.org/10.1145/2588555.2610515},
doi = {10.1145/2588555.2610515},
booktitle = {Proceedings of the 2014 ACM SIGMOD International Conference on Management of Data},
pages = {1115–1126},
numpages = {12},
location = {Snowbird, Utah, USA},
series = {SIGMOD '14}
}

@inproceedings{root2023pitchfork,
author = {Root, Alexander J and Ahmad, Maaz Bin Safeer and Sharlet, Dillon and Adams, Andrew and Kamil, Shoaib and Ragan-Kelley, Jonathan},
title = {Fast Instruction Selection for Fast Digital Signal Processing},
year = {2024},
isbn = {9798400703942},
publisher = {Association for Computing Machinery},
address = {New York, NY, USA},
url = {https://doi.org/10.1145/3623278.3624768},
doi = {10.1145/3623278.3624768},
booktitle = {Proceedings of the 28th ACM International Conference on Architectural Support for Programming Languages and Operating Systems, Volume 4},
pages = {125–137},
numpages = {13},
location = {Vancouver, BC, Canada},
series = {ASPLOS '23}
}

@article{newcomb2020trs,
author = {Newcomb, Julie L. and Adams, Andrew and Johnson, Steven and Bodik, Rastislav and Kamil, Shoaib},
title = {Verifying and improving Halide’s term rewriting system with program synthesis},
year = {2020},
issue_date = {November 2020},
publisher = {Association for Computing Machinery},
address = {New York, NY, USA},
volume = {4},
number = {OOPSLA},
url = {https://doi.org/10.1145/3428234},
doi = {10.1145/3428234},
journal = {Proc. ACM Program. Lang.},
month = nov,
articleno = {166},
numpages = {28}
}

@article{willsey2021egg,
author = {Willsey, Max and Nandi, Chandrakana and Wang, Yisu Remy and Flatt, Oliver and Tatlock, Zachary and Panchekha, Pavel},
title = {egg: Fast and extensible equality saturation},
year = {2021},
issue_date = {January 2021},
publisher = {Association for Computing Machinery},
address = {New York, NY, USA},
volume = {5},
number = {POPL},
url = {https://doi.org/10.1145/3434304},
doi = {10.1145/3434304},
journal = {Proc. ACM Program. Lang.},
month = jan,
articleno = {23},
numpages = {29}
}

@article{nandi2021ruler,
author = {Nandi, Chandrakana and Willsey, Max and Zhu, Amy and Wang, Yisu Remy and Saiki, Brett and Anderson, Adam and Schulz, Adriana and Grossman, Dan and Tatlock, Zachary},
title = {Rewrite rule inference using equality saturation},
year = {2021},
issue_date = {October 2021},
publisher = {Association for Computing Machinery},
address = {New York, NY, USA},
volume = {5},
number = {OOPSLA},
url = {https://doi.org/10.1145/3485496},
doi = {10.1145/3485496},
journal = {Proc. ACM Program. Lang.},
month = oct,
articleno = {119},
numpages = {28}
}

@InProceedings{zhang2024dbt,
  author =	{Zhang, Yihong and Suciu, Dan and Wang, Yisu Remy and Willsey, Max},
  title =	{{Database Theory in Action: Search-Based Program Optimization}},
  booktitle =	{28th International Conference on Database Theory (ICDT 2025)},
  pages =	{34:1--34:6},
  series =	{Leibniz International Proceedings in Informatics (LIPIcs)},
  ISBN =	{978-3-95977-364-5},
  ISSN =	{1868-8969},
  year =	{2025},
  volume =	{328},
  editor =	{Roy, Sudeepa and Kara, Ahmet},
  publisher =	{Schloss Dagstuhl -- Leibniz-Zentrum f{\"u}r Informatik},
  address =	{Dagstuhl, Germany},
  URL =		{https://drops.dagstuhl.de/entities/document/10.4230/LIPIcs.ICDT.2025.34},
  URN =		{urn:nbn:de:0030-drops-229759},
  doi =		{10.4230/LIPIcs.ICDT.2025.34}
}

@article{zhang2022relational,
  author = {
    Zhang, Yihong and 
    Wang, Yisu Remy and 
    Willsey, Max and 
    Tatlock, Zachary
  },
  title = {Relational E-Matching},
  year = {2022},
  issue_date = {January 2022},
  publisher = {Association for Computing Machinery},
  address = {New York, NY, USA},
  volume = {6},
  number = {POPL},
  url = {https://doi.org/10.1145/3498696},
  doi = {10.1145/3498696},
  journal = {Proc. ACM Program. Lang.},
  month = {jan},
  articleno = {35},
  numpages = {22}
}

@inproceedings{yang2021equality,
	author = {Yang, Yichen and Phothilimthana, Phitchaya and Wang, Yisu and Willsey, Max and Roy, Sudip and Pienaar, Jacques},
	booktitle = {Proceedings of Machine Learning and Systems},
	editor = {A. Smola and A. Dimakis and I. Stoica},
	pages = {255--268},
	title = {Equality Saturation for Tensor Graph Superoptimization},
	url = {https://proceedings.mlsys.org/paper_files/paper/2021/file/cc427d934a7f6c0663e5923f49eba531-Paper.pdf},
	volume = {3},
	year = {2021}}

@article{howard2019quantized,
  title={Quantized bounding volume hierarchies for neighbor search in molecular simulations on graphics processing units},
  author={Howard, Michael P and Statt, Antonia and Madutsa, Felix and Truskett, Thomas M and Panagiotopoulos, Athanassios Z},
  journal={Computational Materials Science},
  volume={164},
  pages={139--146},
  year={2019},
  publisher={Elsevier}
}

@misc{sqlite,
  title        = {SQLite},
  author       = {Hipp, Richard D. and The SQLite Development Team},
  howpublished = {Available at \url{https://www.sqlite.org/}},
  year         = {2025},
  note         = {Accessed: October 19, 2025}
}

@article{khayyat2015iejoin,
author = {Khayyat, Zuhair and Lucia, William and Singh, Meghna and Ouzzani, Mourad and Papotti, Paolo and Quian\'{e}-Ruiz, Jorge-Arnulfo and Tang, Nan and Kalnis, Panos},
title = {Lightning fast and space efficient inequality joins},
year = {2015},
issue_date = {September 2015},
publisher = {VLDB Endowment},
volume = {8},
number = {13},
issn = {2150-8097},
url = {https://doi.org/10.14778/2831360.2831362},
doi = {10.14778/2831360.2831362},
journal = {Proc. VLDB Endow.},
month = sep,
pages = {2074–2085},
numpages = {12}
}

@inproceedings{raasveldt2019duckdb,
author = {Raasveldt, Mark and M\"{u}hleisen, Hannes},
title = {DuckDB: an Embeddable Analytical Database},
year = {2019},
isbn = {9781450356435},
publisher = {Association for Computing Machinery},
address = {New York, NY, USA},
url = {https://doi.org/10.1145/3299869.3320212},
doi = {10.1145/3299869.3320212},
booktitle = {Proceedings of the 2019 International Conference on Management of Data},
pages = {1981–1984},
numpages = {4},
location = {Amsterdam, Netherlands},
series = {SIGMOD '19}
}

@article{sawhney2020wos,
author = {Sawhney, Rohan and Crane, Keenan},
title = {Monte Carlo geometry processing: a grid-free approach to PDE-based methods on volumetric domains},
year = {2020},
issue_date = {August 2020},
publisher = {Association for Computing Machinery},
address = {New York, NY, USA},
volume = {39},
number = {4},
issn = {0730-0301},
url = {https://doi.org/10.1145/3386569.3392374},
doi = {10.1145/3386569.3392374},
journal = {ACM Trans. Graph.},
month = aug,
articleno = {123},
numpages = {18}
}

@INPROCEEDINGS{pan2012fcl,
  author={Pan, Jia and Chitta, Sachin and Manocha, Dinesh},
  booktitle={2012 IEEE International Conference on Robotics and Automation}, 
  title={FCL: A general purpose library for collision and proximity queries}, 
  year={2012},
  volume={},
  number={},
  pages={3859-3866},
  doi={10.1109/ICRA.2012.6225337}}

@article{emre2025cones,
author = {Emre, U. and Kanak, A. and Steinberg, S.},
title = {High-Performance Elliptical Cone Tracing},
journal = {Computer Graphics Forum},
volume = {44},
number = {7},
pages = {e70230},
doi = {https://doi.org/10.1111/cgf.70230},
url = {https://onlinelibrary.wiley.com/doi/abs/10.1111/cgf.70230},
eprint = {https://onlinelibrary.wiley.com/doi/pdf/10.1111/cgf.70230},
year = {2025}
}

@misc{McGuire2017Data,
   title = {Computer Graphics Archive},
   author = {Morgan McGuire},
   year = {2017},
   month = {July},
   note = {\small \texttt{https://casual-effects.com/data}},
   url = {https://casual-effects.com/data}
}

@article{codd1970relational,
author = {Codd, E. F.},
title = {A relational model of data for large shared data banks},
year = {1970},
issue_date = {June 1970},
publisher = {Association for Computing Machinery},
address = {New York, NY, USA},
volume = {13},
number = {6},
issn = {0001-0782},
url = {https://doi.org/10.1145/362384.362685},
doi = {10.1145/362384.362685},
journal = {Commun. ACM},
month = jun,
pages = {377–387},
numpages = {11}
}

@MISC{eigenweb,
  author = {Ga\"{e}l Guennebaud and Beno\^{i}t Jacob and others},
  title = {Eigen v3},
  howpublished = {http://eigen.tuxfamily.org},
  year = {2010}
 }

@article{wald2001interactive,
author = {Wald, Ingo and Slusallek, Philipp and Benthin, Carsten and Wagner, Markus},
title = {Interactive Rendering with Coherent Ray Tracing},
journal = {Computer Graphics Forum},
volume = {20},
number = {3},
pages = {153-165},
doi = {https://doi.org/10.1111/1467-8659.00508},
url = {https://onlinelibrary.wiley.com/doi/abs/10.1111/1467-8659.00508},
eprint = {https://onlinelibrary.wiley.com/doi/pdf/10.1111/1467-8659.00508},
year = {2001}
}

@inproceedings{aila2009understanding,
author = {Aila, Timo and Laine, Samuli},
title = {Understanding the efficiency of ray traversal on GPUs},
year = {2009},
isbn = {9781605586038},
publisher = {Association for Computing Machinery},
address = {New York, NY, USA},
url = {https://doi.org/10.1145/1572769.1572792},
doi = {10.1145/1572769.1572792},
booktitle = {Proceedings of the Conference on High Performance Graphics 2009},
pages = {145–149},
numpages = {5},
location = {New Orleans, Louisiana},
series = {HPG '09}
}

@inproceedings{aila2010treelets,
author = {Aila, Timo and Karras, Tero},
title = {Architecture considerations for tracing incoherent rays},
year = {2010},
publisher = {Eurographics Association},
address = {Goslar, DEU},
booktitle = {Proceedings of the Conference on High Performance Graphics},
pages = {113–122},
numpages = {10},
location = {Saarbrucken, Germany},
series = {HPG '10}
}

@article{bentley1975kdtree,
author = {Bentley, Jon Louis},
title = {Multidimensional binary search trees used for associative searching},
year = {1975},
issue_date = {Sept. 1975},
publisher = {Association for Computing Machinery},
address = {New York, NY, USA},
volume = {18},
number = {9},
issn = {0001-0782},
url = {https://doi.org/10.1145/361002.361007},
doi = {10.1145/361002.361007},
journal = {Commun. ACM},
month = sep,
pages = {509–517},
numpages = {9}
}

@INPROCEEDINGS{eltabakh2007deduplication,
author = { Eltabakh, M. Y. and Aref, Walid G. and Ouzzani, Mourad },
booktitle = { Scientific and Statistical Database Management, International Conference on },
title = {{ Duplicate Elimination in Space-partitioning Tree Indexes }},
year = {2007},
volume = {},
ISSN = {1551-6393},
pages = {18},
doi = {10.1109/SSDBM.2007.10},
url = {https://doi.ieeecomputersociety.org/10.1109/SSDBM.2007.10},
publisher = {IEEE Computer Society},
address = {Los Alamitos, CA, USA},
month =Jul}

@article{lauterbach2009fast,
author = {Lauterbach, C. and Garland, M. and Sengupta, S. and Luebke, D. and Manocha, D.},
title = {Fast BVH Construction on GPUs},
journal = {Computer Graphics Forum},
volume = {28},
number = {2},
pages = {375-384},
doi = {https://doi.org/10.1111/j.1467-8659.2009.01377.x},
url = {https://onlinelibrary.wiley.com/doi/abs/10.1111/j.1467-8659.2009.01377.x},
eprint = {https://onlinelibrary.wiley.com/doi/pdf/10.1111/j.1467-8659.2009.01377.x},
year = {2009}
}

@inproceedings{pantaleoni2010hlbvh,
author = {Pantaleoni, J. and Luebke, D.},
title = {HLBVH: hierarchical LBVH construction for real-time ray tracing of dynamic geometry},
year = {2010},
publisher = {Eurographics Association},
address = {Goslar, DEU},
booktitle = {Proceedings of the Conference on High Performance Graphics},
pages = {87–95},
numpages = {9},
location = {Saarbrucken, Germany},
series = {HPG '10}
}

@inproceedings{gu2013efficient,
author = {Gu, Yan and He, Yong and Fatahalian, Kayvon and Blelloch, Guy},
title = {Efficient BVH construction via approximate agglomerative clustering},
year = {2013},
isbn = {9781450321358},
publisher = {Association for Computing Machinery},
address = {New York, NY, USA},
url = {https://doi.org/10.1145/2492045.2492054},
doi = {10.1145/2492045.2492054},
booktitle = {Proceedings of the 5th High-Performance Graphics Conference},
pages = {81–88},
numpages = {8},
location = {Anaheim, California},
series = {HPG '13}
}

@inproceedings{karras2012maximizing,
author = {Karras, Tero},
title = {Maximizing parallelism in the construction of BVHs, octrees, and k-d trees},
year = {2012},
isbn = {9783905674415},
publisher = {Eurographics Association},
address = {Goslar, DEU},
booktitle = {Proceedings of the Fourth ACM SIGGRAPH / Eurographics Conference on High-Performance Graphics},
pages = {33–37},
numpages = {5},
location = {Paris, France},
series = {EGGH-HPG'12}
}

@inproceedings{apetrei2014fast,
booktitle = {Computer Graphics and Visual Computing (CGVC)},
editor = {Rita Borgo and Wen Tang},
title = {{Fast and Simple Agglomerative LBVH Construction}},
author = {Apetrei, Ciprian},
year = {2014},
publisher = {The Eurographics Association},
ISBN = {978-3-905674-70-5},
DOI = {10.2312/cgvc.20141206},
pages = {41–44},
}

@article{meister2017parallel,
  title={Parallel locally-ordered clustering for bounding volume hierarchy construction},
  author={Meister, Daniel and Bittner, Ji{\v{r}}{\'\i}},
  journal={IEEE transactions on visualization and computer graphics},
  volume={24},
  number={3},
  pages={1345--1353},
  year={2017},
  publisher={IEEE}
}

@article{benthin2024hploc,
  title={H-PLOC: Hierarchical Parallel Locally-Ordered Clustering for Bounding Volume Hierarchy Construction},
  author={Benthin, Carsten and Meister, Daniel and Barczak, Joshua and Mehalwal, Rohan and Tsakok, John and Kensler, Andrew},
  journal={Proceedings of the ACM on Computer Graphics and Interactive Techniques},
  volume={7},
  number={3},
  pages={1--14},
  year={2024},
  publisher={ACM New York, NY, USA}
}

@article{gyurgyik2026scion,
  author  = {Christophe Gyurgyik and Alexander J Root and Fredrik Kjolstad},
  title   = {Decoupling Data Layouts from Bounding Volume Hierarchies},
  journal = {Proceedings of the ACM on Programming Languages},
  volume  = {10},
  number  = {PLDI},
  articleno = {175},
  month   = jun,
  year    = {2026},
  doi     = {10.1145/3808253},
  url     = {https://doi.org/10.1145/3808253},
  publisher = {Association for Computing Machinery}
}

@software{root2026zenodo,
	author = {Root, Alexander J and Christophe, Gyurgyik and Purvi, Goel and Fatahalian, Kayvon and Ragan-Kelley, Jonathan and Andrew, Adams and Kjolstad, Fredrik Berg},
	doi = {10.5281/zenodo.19091467},
	month = mar,
	publisher = {Zenodo},
	title = {Artifact for PLDI 2026 Paper: Bonsai: Compiling Queries into Work-Efficient Tree Traversals},
	url = {https://doi.org/10.5281/zenodo.19091467},
	year = 2026}

\newpage
\appendix
\section{Scalar Symbolic Interval Analysis}
\label{sec:appendix-interval-analysis}

Interval analysis is a recursive bottom-up technique. For symbolic interval analysis, the compiler builds an AST that represents the lower bound and upper bound of a symbolic expression, where varying parameters are replaced with their interval (or volumes) and uniform parameters are singular-valued intervals. To support additional operations, one needs to define the semantics of how an operator operates on an interval instead of a scalar value; this is generally done by reasoning about the monotonicity of an operation. \Cref{sec:scalar-interval} describes how such reasoning can be done for boolean combinators and comparison operations; here, we document how intervals can be computed numerically, allowing for analyzing predicates containing computation.

\paragraph{Numerical operations}
Addition is monotonically increasing in both arguments; thus, the upper bound is the sum of the upper bound of its arguments. Subtraction, however, is monotonically increasing in its first argument, but monotonically decreasing in the second argument. These properties result in the following bounds:
\[
    \lceil x + y \rceil \mapsto \lceil x \rceil + \lceil y \rceil
    \qquad
    \lceil x - y \rceil \mapsto \lceil x \rceil - \lfloor y \rfloor
\]
\[
    \lfloor x + y \rfloor \mapsto \lfloor x \rfloor + \lfloor y \rfloor
    \qquad
    \lfloor x - y \rfloor \mapsto \lfloor x \rfloor - \lceil y \rceil
\]
Multiplication is non-linear, and therefore requires evaluating all interval end-points\footnote{Inlining all multiplications blows up the AST; \code{let} statements avoid this.}:
\[
    \text{let } S = \{ \lceil x \rceil * \lceil y \rceil, \lceil x \rceil * \lfloor y \rfloor, \lfloor x \rfloor * \lceil y \rceil, \lfloor x \rfloor * \lfloor y \rfloor \} \text{ in}
\]%
\[
    \lceil x * y \rceil \mapsto \max(S)
    \qquad
    \lfloor x * y \rfloor \mapsto \min(S)
\]
Thus far, no numerical operations have been type-specific.\footnote{A practical concern here is integer overflow, under which these rules are incorrect. This is most easily avoided by defining numerical operations to saturate.} However, the bounds of division are quite different for floating-point versus integer types. Computing the bounds of a floating-point division is a well-studied but complex problem~\cite{fpbounds2017joldes}, and we do not attempt to tersely express accurate symbolic bounds for floating-point division here. Integer division and modulo can be reasoned about somewhat more easily, but require a significant amount of control flow to handle reasoning about the sign of each operand. For brevity, we do not include the upper and lower bounds of these operators, but do walk through the (symbolic) casework needed for the upper bound\footnote{The lower bound does not exist when the interval of the denominator includes zero.} of integer (Euclidean) division below.
\[
    \lceil x / y \rceil \mapsto \begin{cases}
\lceil x \rceil / \lfloor y \rfloor & \lceil x \rceil > 0 \land \lfloor y \rfloor > 0\\
\lceil x \rceil / \lceil y \rceil & \lceil x \rceil < 0 \land \lfloor y \rfloor > 0\\
\lfloor x \rfloor / \lceil y \rceil & \lfloor x \rfloor < 0 \land \lceil y \rceil < 0\\
\lfloor x \rfloor / \lfloor y \rfloor & \lfloor x \rfloor > 0 \land \lceil y \rceil < 0\\
\max(-\lfloor x \rfloor, \lceil x \rceil) & \text{otherwise}

\end{cases}
\]
In order, these cases correspond to: $x$ can be positive and $y$ must be positive; $x$ must be negative and $y$ must be positive; $x$ can be negative and $y$ must be negative; $x$ must be positive and $y$ must be negative; and lastly, $y$ can be positive or negative. In each of these cases, if the result must be negative, the lowest-magnitude negative result is computed, and if the result can be positive, the highest-magnitude positive result is computed.

\paragraph{Monotonic Functions} The bounds of monotonic functions, such as \texttt{min}, \texttt{max}, \texttt{ceil}, \texttt{floor}, \texttt{exp}, \texttt{sqrt}, or \texttt{ln}, are simply the function applied to the corresponding bounds of its argument(s).

\paragraph{Conditionals} Boolean bounds can also be used to bound conditional statements. We denote the ternary conditional operator (AKA if-then-else) as \texttt{ite}, and provide the following expression for the upper bound:
\[
    \lceil \texttt{ite}(a, x, y) \rceil \mapsto \max(\texttt{ite}(\lfloor a \rfloor, \lceil x \rceil , \lceil y \rceil), \texttt{ite}(\lceil a \rceil, \lceil x \rceil , \lceil y \rceil))
\]
Logically, if the condition $a$ \textit{must be true} ($\lfloor a \rfloor = \lceil a \rceil = true$), then the upper bound of \texttt{ite} is just the upper bound of $x$, $\lceil x \rceil$. If $a$ must be false ($\lfloor a \rfloor = \lceil a \rceil = false$), then the upper bound is just $\lceil y \rceil$. Otherwise, $a$ is unbounded, and the upper bound is just the maximum of the two possible upper bounds. The same reasoning is applied to produce the expression for the lower bound:
\[
    \lfloor \texttt{ite}(a, x, y) \rfloor \mapsto \min(\texttt{ite}(\lfloor a \rfloor, \lfloor x \rfloor , \lfloor y \rfloor), \texttt{ite}(\lceil a \rceil, \lfloor x \rfloor , \lfloor y \rfloor))
\]

\paragraph{Non-monotonic Functions} There are many interesting non-monotonic functions as well, which require slightly more complex reasoning to bound their values. Piecewise-monotonic (monotonic on certain intervals) functions can be bounded relatively easily. For example, consider the \texttt{trunc} function from the SQL standard~\cite{date1989sql}, which accepts a floating-point value and an integer number representing the number of digits past 0 to round to. If the rounding integer is uniform (constant), this function is monotonic in the floating-point argument, but if not (e.g., the rounding integer is a varying piece of indexed data), then we require more control-flow to be generated:
\[
    \lceil \texttt{trunc}(x, y) \rceil \mapsto \texttt{ite}(\lceil x \rceil > 0, \texttt{trunc}(\lceil x \rceil, \lceil y \rceil), \texttt{trunc}(\lceil x \rceil, \lfloor y \rfloor))
\]
If $x$ can be positive, the upper bound of $\texttt{trunc}(x, y)$ is the upper bound of $x$ with the most precision (with $\lceil y \rceil$ decimals), but if $x$ is strictly negative, then the upper bound is the upper bound of $x$ with the \textit{least} precision (with $\lfloor y \rfloor$ decimals). Similar reasoning can be applied to the lower bound of \texttt{trunc}, as well as other functions like \texttt{pow} and \texttt{round}.

\end{document}